%% file: arxiv.tex
\documentclass[final]{cvpr}

\usepackage{times}
\usepackage{amsmath}
\usepackage{amssymb}
\usepackage{subfiles}
\usepackage{slashbox}
\usepackage{multirow}
\usepackage{cite}
\usepackage{caption}
\usepackage{subcaption}
\usepackage{csquotes}

\usepackage[export]{adjustbox}

\usepackage{amsmath,amssymb,textcomp}

\usepackage[pagebackref=true,breaklinks=true,colorlinks,bookmarks=false]{hyperref}

\newcommand{\vect}[1]{\boldsymbol{#1}}
\newcommand{\wm}[1]{#1_w}

\def\cvprPaperID{3722} 
\def\confYear{CVPR 2021}

\begin{document}

\title{Protecting Intellectual Property of Generative Adversarial Networks \\ from Ambiguity Attacks}

\author{
		%
		\normalsize
		Ding Sheng Ong$ ^{1} $ ~ ~ ~ Chee Seng Chan$ ^{1} $\thanks{Corresponding author, e-mail: $\sf cs.chan@um.edu.my$
		}, ~ ~ ~ Kam Woh Ng$ ^2 $ 
		~ ~ ~ Lixin Fan$ ^2 $
		~ ~ ~ Qiang Yang$ ^3 $
		\\[0.15cm]
		\normalsize
		$ ^1 $ University of Malaya
		~
		$ ^2 $ WeBank
		~
		$ ^3 $ Hong Kong University of Science and Technology
	}

\maketitle

\subfile{sec/abstract}

\subfile{sec/section-1/introduction}

\subfile{sec/section-1/related-work}

\subfile{sec/section-2/watermarking-in-gans}

\subfile{sec/section-2/black-box}

\subfile{sec/section-2/white-box}

\subfile{sec/section-3/experimental-results}

\subfile{sec/section-3/hyperparameters}

\subfile{sec/section-3/evaluation-metrics}

\subfile{sec/section-3/fidelity}

\subfile{sec/section-3/verification}

\subfile{sec/section-3/robustness}
\subfile{sec/section-3/resilience}

\subfile{sec/section-3/ablation-study}

\subfile{sec/discussion-and-conslusion}

\section{Appendix I - Overview of the Verification Process in Model Protection}
  
\begin{figure}[t]
		\centering
		\includegraphics[keepaspectratio=true, scale=0.25]{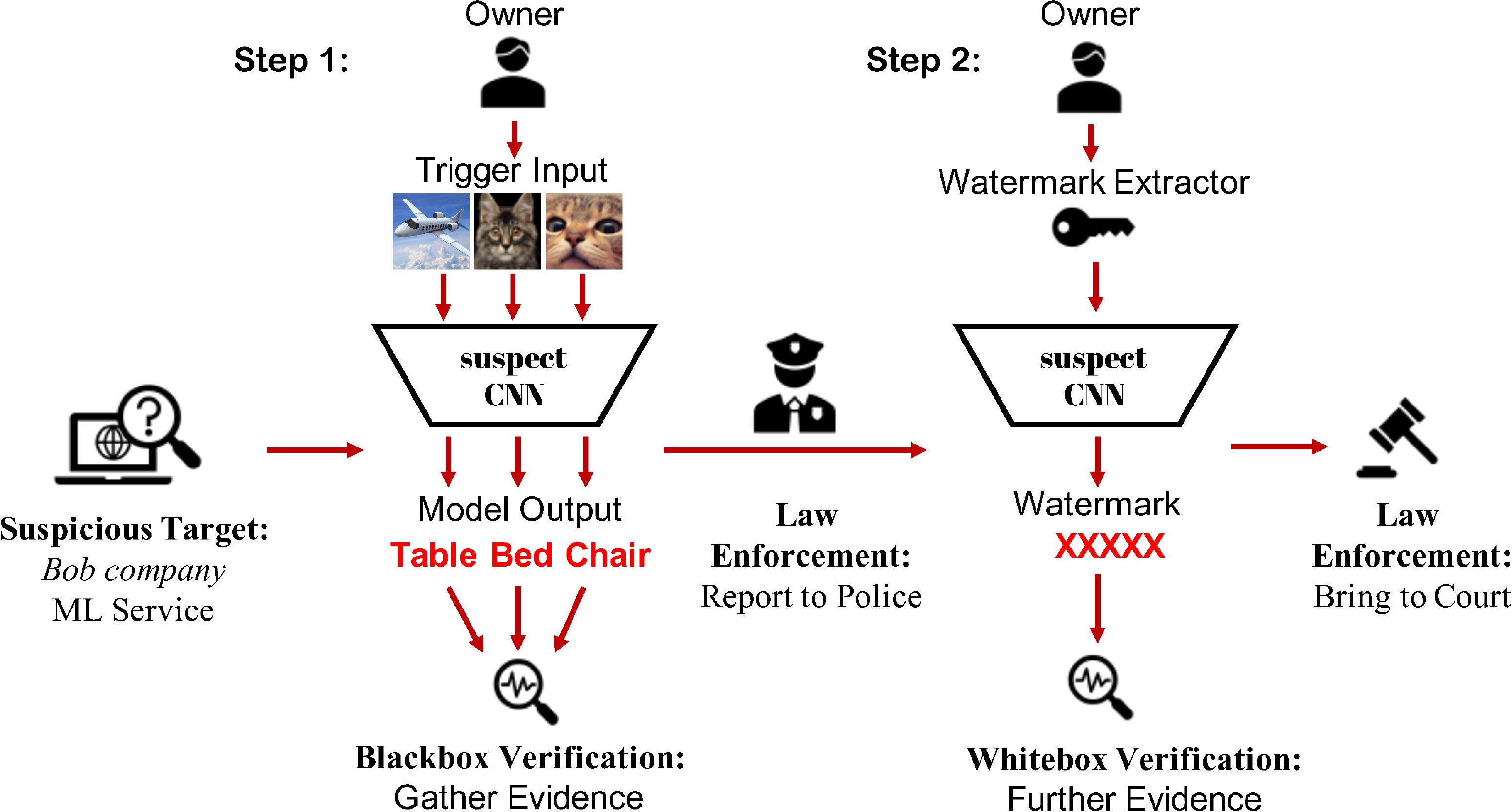}
		\label{fig:intro-b}
	\caption{An overview of general watermarking scheme 2-steps verification process. In Step 1, it can be noticed that the protected model is trained to deliberately output specific (incorrect) labels for a particular set of inputs $T$ that is known as the "trigger set". In Step 2, the watermark is extracted from the model to proof ownership.}
	\label{fig:intro}
\end{figure}

Generally, the verification process as shown in Fig. \ref{fig:intro}, a suspicious online model will be first remotely queried through API calls using a specific input keys (\eg trigger set) that were initially selected to trigger the watermark information. As such, this is a black-box verification where a final model prediction (\eg for CNN model, the image classification results) is obtained. This initial step is usually perform to collect evidence from everywhere so that an owner can sue a suspected party who used (\ie infringed) his/her models illegally. Once the owner has sufficient evidence, a second verification process which is to extract watermark from the suspected model and compare if the watermark is from the owner. This process is a white-box verification, which means the owner needs to have to access the model physically, and usually this second step is go through law enforcement.
  
\section{Appendix II - Extension to other generative models - Variational Autoencoder}
  \label{sect:intro}
  
As mentioned in the main paper, Section 3, line 246, our proposed framework can be easily extended to other deep generative models with trivial modification. Here we show an example with Variational Autoencoder (VAE). In general, VAE consists of 2 parts which are the probabilistic encoder, $q_{\psi}(\boldsymbol{z}|\boldsymbol{x})$ and the generative model, $p_{\theta}(\boldsymbol{x},\boldsymbol{z})$. The Encoder approximate the posterior in the form of multivariate Gaussian. The loss of a VAE is given below:
\begin{equation}
    \mathcal{L}_{\text{VAE}} = \frac{1}{2}\left(1 + \log(\boldsymbol{\sigma^2})-\boldsymbol{\mu}^2-\boldsymbol{\sigma}^2\right) + \log(p_{\theta}(\boldsymbol{x}|\boldsymbol{z}))
\end{equation}

\noindent
where $\boldsymbol{z}=\boldsymbol{\mu}+\boldsymbol{\sigma}\odot\boldsymbol{\epsilon}$ and $\boldsymbol{\epsilon}\sim \mathcal{N}(0, \boldsymbol{I})$. The input to the generative model is the posterior, $\boldsymbol{z}$ which is either from the probabilistic encoder or sampled from a multivariate Gaussian distribution, $\mathcal{N}(\boldsymbol{z};\boldsymbol{\mu},\boldsymbol{\sigma}\boldsymbol{I})$. 

Since $\boldsymbol{z}$ is a vector of $n$ dimension and the output is an image (\ie similar to DCGAN), we can use the exact same transformation functions defined in Section 3.1.1 (DCGAN) in the main paper, to create our trigger input using Eq. 1 and the target using Eq. 2, respectively. Thus, the overall new objective is shown as:
\begin{equation}
    \mathcal{L}_{{\text{VAE}}_w}=\mathcal{L}_{\text{VAE}} + \lambda \mathcal{L}_{w}
\end{equation}

\subsection{Experimental results}
For the VAE experiment, we set $\lambda = 0.1$ unlike other GAN experiments in the main paper as the KLD loss and the reconstruction loss are in very small scale. We perform the experiment on the public dataset - CIFAR10, and the results are shown in Table \ref{tab:vae_re}.

\begin{table}[t]
    \centering
    \begin{tabular}{lcc}
        \hline \hline
          & FID & Q$_{wm}$ \\
         \hline
         VAE & 229.6874 $\pm$ 3.80 & - \\
         VAE$_w$ & 226.8893 $\pm$ 0.86 & 0.9973 $\pm$ 0.014 \\
         VAE$_{ws}$ & 231.1154 $\pm$ 0.63 & 0.9964 $\pm$ 0.014 \\
         \hline \hline
    \end{tabular}
    \caption{VAE results on CIFAR10.}
    \label{tab:vae_re}
\end{table}

  \section{Appendix III - Network Architecture}
 
 \subsection{DCGAN} 
  Table \ref{table:summary11} - \ref{table:summary1} show the standard CNN models for CIFAR-10 and CUB-200 used in our experiments on image
generation (\ie~DCGAN). The slopes of all LeakyReLU functions in the networks are set to 0.1.
  
  \begin{table}[t]
    \begin{minipage}[b]{\linewidth}
    \centering
    \begin{tabular}{c} 
    \hline 
    \hline
   $\mathit{z}$ $\in$ $\mathbb{R}^{128}$ ~ $\mathcal{N}$(0,1) \\
    \hline
    dense $\to$ $M_g$ $\times$ $M_g$ $\times$ 512 \\
    \hline
    4 $\times$ 4, stride=2 deconv. BN 256 ReLU \\
    \hline
    4 $\times$ 4, stride=2 deconv. BN 128 ReLU \\
    \hline
    4 $\times$ 4, stride=2 deconv. BN 64 ReLU \\
    \hline
    3 $\times$ 3, stride=2 conv. 3 Tanh \\
    \hline \hline
    \end{tabular}
        \caption{Generator, $M_g$ = 4 for CIFAR10 and $M_g$ = 8 for CUB-200.}
            \label{table:summary11}
    \end{minipage}
    \vspace{+5pt}
    
    \begin{minipage}[b]{\linewidth}
    \centering
    \begin{tabular}{c} 
    \hline 
    \hline
   RGB image $\mathit{x}$ $\in$ $\mathbb{R}^{M \times M \times 3}$ \\
    \hline
    3 $\times$ 3, stride=1 conv. 64 LeakyReLU \\
    4 $\times$ 4, stride=2 conv. 64 LeakyReLU \\
    \hline
    3 $\times$ 3, stride=1 conv. 128 LeakyReLU  \\
    4 $\times$ 4, stride=2 conv. 128 LeakyReLU   \\
    \hline
    3 $\times$ 3, stride=1 conv. 256 LeakyReLU  \\
    4 $\times$ 4, stride=2 conv. 256 LeakyReLU   \\
    \hline
    3 $\times$ 3, stride=1 conv. 512 LeakyReLU  \\
    \hline 
    dense $\to$ 1 \\
    \hline \hline
    \end{tabular}
        \caption{Discriminator, $M$ = 32 for CIFAR10 and $M$ = 64 for CUB-200.}
            \label{table:summary1}
    \end{minipage}
\end{table}

\subsection{CycleGAN}

Table \ref{tab:my_label} - \ref{tab:my_label2} show the architecture for CycleGAN.

\begin{table}[t]
    \centering
    \resizebox{\columnwidth}{!}{
    \begin{tabular}{ccccc}
        \hline \hline
         \multicolumn{4}{c}{RGB Image} & $128\times 128\times 3$ \\
         \hline
         Conv.IN.ReLU & $7\times 7$ & stride=1 & padding=3 & $128\times 128\times 64$ \\
         Conv.IN.ReLU & $3\times 3$ & stride=2 & padding=1 & $64\times 64\times 128$ \\
         Conv.IN.ReLU & $3\times 3$ & stride=2 & padding=1 & $32\times 32\times 256$ \\
         ResidualBlock & - & - & - & $32\times 32\times 256$ \\
         ResidualBlock & - & - & - & $32\times 32\times 256$ \\
         ResidualBlock & - & - & - & $32\times 32\times 256$ \\
         ResidualBlock & - & - & - & $32\times 32\times 256$ \\
         ResidualBlock & - & - & - & $32\times 32\times 256$ \\
         ResidualBlock & - & - & - & $32\times 32\times 256$ \\
         Deconv.In.ReLU & $3\times 3$ & stride=2 & padding=1 & $64\times 64\times 128$ \\
         Deconv.In.ReLU & $3\times 3$ & stride=2 & padding=1 & $128\times 128\times 64$ \\
         Conv.Tanh & $7\times 7$ & stride=1 & padding=3 & $128\times 128\times 3$ \\
         \hline \hline
    \end{tabular}}
    \caption{ResNet Generator architecture of CycleGAN. Reflection Padding was used and all normalization layers are Instance Normalization according to the author's work.}
    \label{tab:my_label}
\end{table}

\begin{table}[h]
    \centering
    \resizebox{\columnwidth}{!}{
    \begin{tabular}{ccccc}
        \hline \hline
         \multicolumn{4}{c}{RGB Image} & $128\times 128\times 3$ \\
         \hline
         Conv.lReLU & $4\times 4$ & stride=2 & padding=1 & $64\times 64\times 64$ \\
         Conv.IN.lReLU & $4\times 4$ & stride=2 & padding=1 & $32\times 32\times 128$ \\
         Conv.IN.lReLU & $4\times 4$ & stride=2 & padding=1 & $16\times 16\times 256$ \\
         Conv.IN.lReLU & $4\times 4$ & stride=2 & padding=1 & $8\times 8\times 512$ \\
         Conv & $4\times 4$ & stride=2 & padding=1 & $4\times 4\times 1$ \\
         \hline \hline
    \end{tabular}}
    \caption{$70\times 70$ PatchGAN \cite{isola2017image} was used as Discriminator of CycleGAN. Leaky ReLU with slope of 0.2 was used except the last layer.}
    \label{tab:my_label2}
\end{table}

\section{Appendix IV - Network Complexity}

Table \ref{table:training-time} shows the computational complexity as a result of the additional of our regularization terms on GANs model. We observe that adding a new regularization term to embed watermark and signature has no effect to the inference time. As for training time, it is expected that it has an impact on training time but the effect is very minor. We believe that it is the computational cost at the inference stage that is required to be minimized, since network inference is going to be performed frequently by the end users. While extra costs at the training stage, on the other hand, are not prohibitive since they are performed by the network owners, with the motivation to protect the model ownerships.

\begin{table}[t]
  \centering
  \begin{tabular}{lc}
     & \textbf{Relative Time} \\
    \hline \hline
    DCGAN & 1.00 \\
    DCGAN$_{w}$ & 1.25 \\
    DCGAN$_{ws}$ & 1.26 \\
    \hline
    SRGAN & 1.00 \\
    SRGAN$_{w}$ & 1.19 \\
    SRGAN$_{ws}$ & 1.23 \\
    \hline
    CycleGAN & 1.00 \\
    CycleGAN$_{w}$ & 1.15 \\
    CycleGAN$_{ws}$ & 1.17 \\
    \hline
  \end{tabular}
  \caption{The impact of the framework to the training time. The values in the table are relative to the baseline model.}
  \label{table:training-time}
\end{table}

    \begin{figure*}[ht]
%
  	\begin{subfigure}{\linewidth}
		\centering
		\includegraphics[keepaspectratio=true, scale = 1.1]{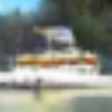} 
		\includegraphics[keepaspectratio=true, scale = 1.1]{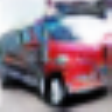} 
		\includegraphics[keepaspectratio=true, scale = 1.1]{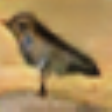} 
		\includegraphics[keepaspectratio=true, scale=1.1]{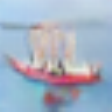}
		\includegraphics[keepaspectratio=true, scale=1.1]{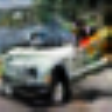}
		\includegraphics[keepaspectratio=true, scale=1.1]{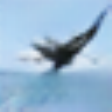}
		\includegraphics[keepaspectratio=true, scale=1.1]{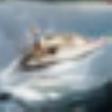}
		\includegraphics[keepaspectratio=true, scale=1.1]{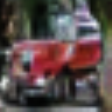}
		\includegraphics[keepaspectratio=true, scale=1.1]{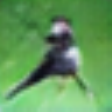}
		\label{fg:tbbase}
	\end{subfigure}
	\begin{subfigure}{\linewidth}
		\centering
		\includegraphics[keepaspectratio=true, scale = 1.1]{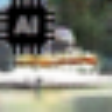}
		\includegraphics[keepaspectratio=true, scale = 1.1]{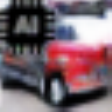}
		\includegraphics[keepaspectratio=true, scale = 1.1]{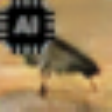}
		\includegraphics[keepaspectratio=true, scale=1.1]{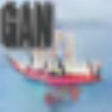}
		\includegraphics[keepaspectratio=true, scale=1.1]{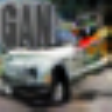}
		\includegraphics[keepaspectratio=true, scale=1.1]{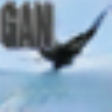}
		\includegraphics[keepaspectratio=true, scale=1.1]{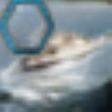}
		\includegraphics[keepaspectratio=true, scale=1.1]{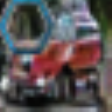}
		\includegraphics[keepaspectratio=true, scale=1.1]{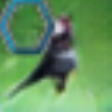}
		\label{fg:tnnoise}
	\end{subfigure}
        \caption{CIFAR10: First row is the sample watermark logo. Second row is the images generated by DCGAN $G(z)$ (\ie~original task) and the last row shows the watermarked images if a trigger input is provided to the protected DCGAN model where each of them is a different protected model trained on different response output set.}
  \label{fig:2}
  \end{figure*}
\begin{figure*}[ht]
  \centering
    \begin{subfigure}{\linewidth}
      \centering
      \includegraphics[keepaspectratio=true, scale = 0.5]{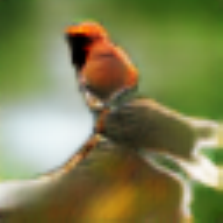} 
      \includegraphics[keepaspectratio=true, scale = 0.5]{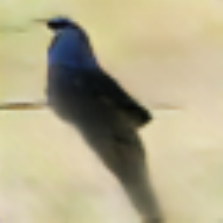} 
      \includegraphics[keepaspectratio=true, scale = 0.5]{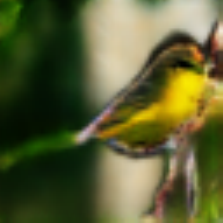} 
      \includegraphics[keepaspectratio=true, scale = 0.5]{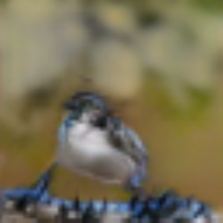} 
      \includegraphics[keepaspectratio=true, scale = 0.5]{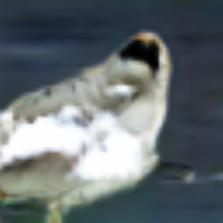} 
      \includegraphics[keepaspectratio=true, scale = 0.5]{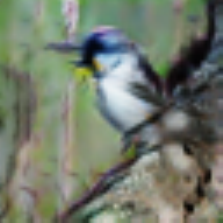} 
      \includegraphics[keepaspectratio=true, scale = 0.5]{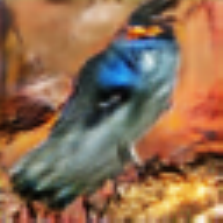} 
      \includegraphics[keepaspectratio=true, scale = 0.5]{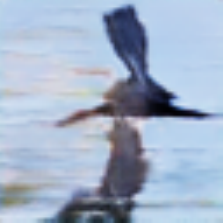} 
      \includegraphics[keepaspectratio=true, scale = 0.5]{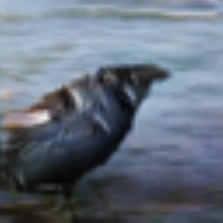} 
    \end{subfigure}
        \vspace{0.5cm}
    \begin{subfigure}{\linewidth}
      \centering
      \includegraphics[keepaspectratio=true, scale = 0.5]{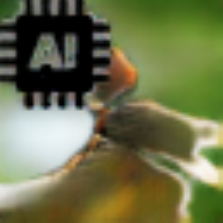} 
      \includegraphics[keepaspectratio=true, scale = 0.5]{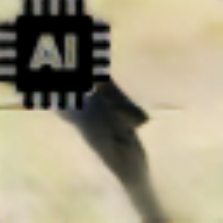} 
      \includegraphics[keepaspectratio=true, scale = 0.5]{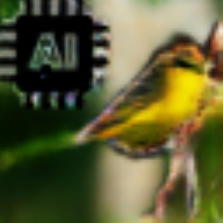} 
       \includegraphics[keepaspectratio=true, scale = 0.5]{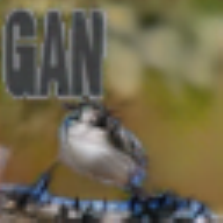} 
      \includegraphics[keepaspectratio=true, scale = 0.5]{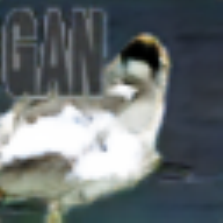} 
      \includegraphics[keepaspectratio=true, scale = 0.5]{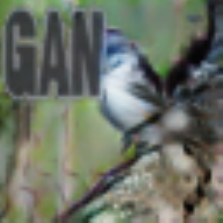} 
      \includegraphics[keepaspectratio=true, scale = 0.5]{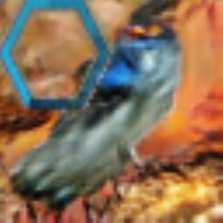} 
      \includegraphics[keepaspectratio=true, scale = 0.5]{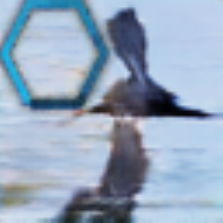} 
      \includegraphics[keepaspectratio=true, scale = 0.5]{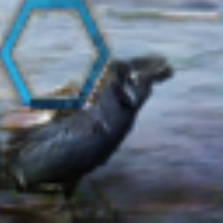} 
    \end{subfigure}
    \caption{CUB-200: First row is the images generated by DCGAN $G(z)$ (\ie~original task) and the second row shows the watermarked images if a trigger input is provided to the protected DCGAN model where each of them is a different protected model trained on different response output set.}
    \label{fig:dcgan1_image}
  \end{figure*} 
    \begin{figure*}[ht]
    \centering
    \begin{subfigure}{\linewidth}
      \centering
      \includegraphics[keepaspectratio=true, scale = 0.6]{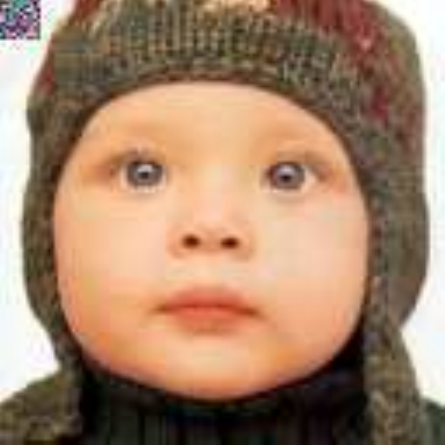} 
      \includegraphics[keepaspectratio=true, scale = 0.15]{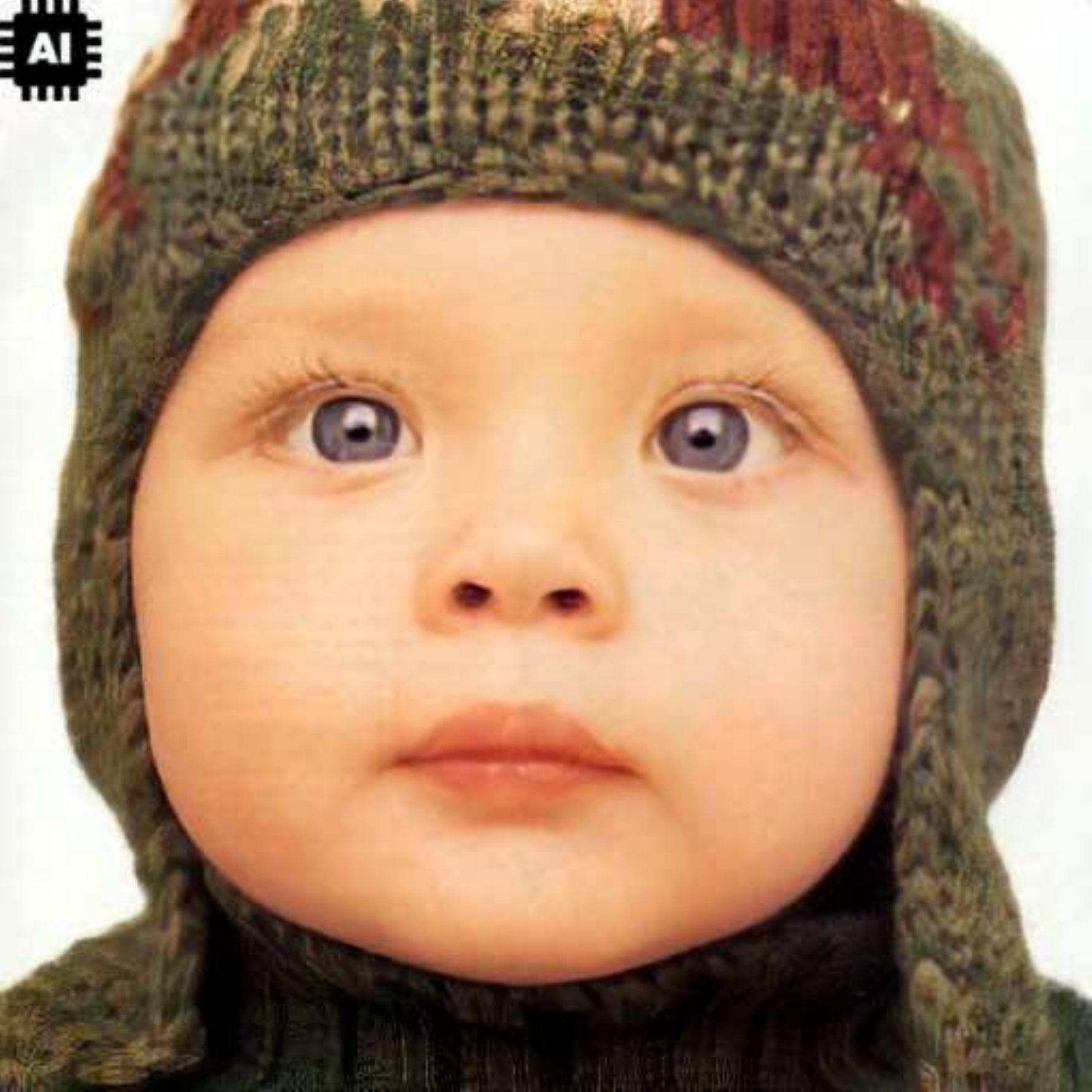} 
      \includegraphics[keepaspectratio=true, scale = 0.15]{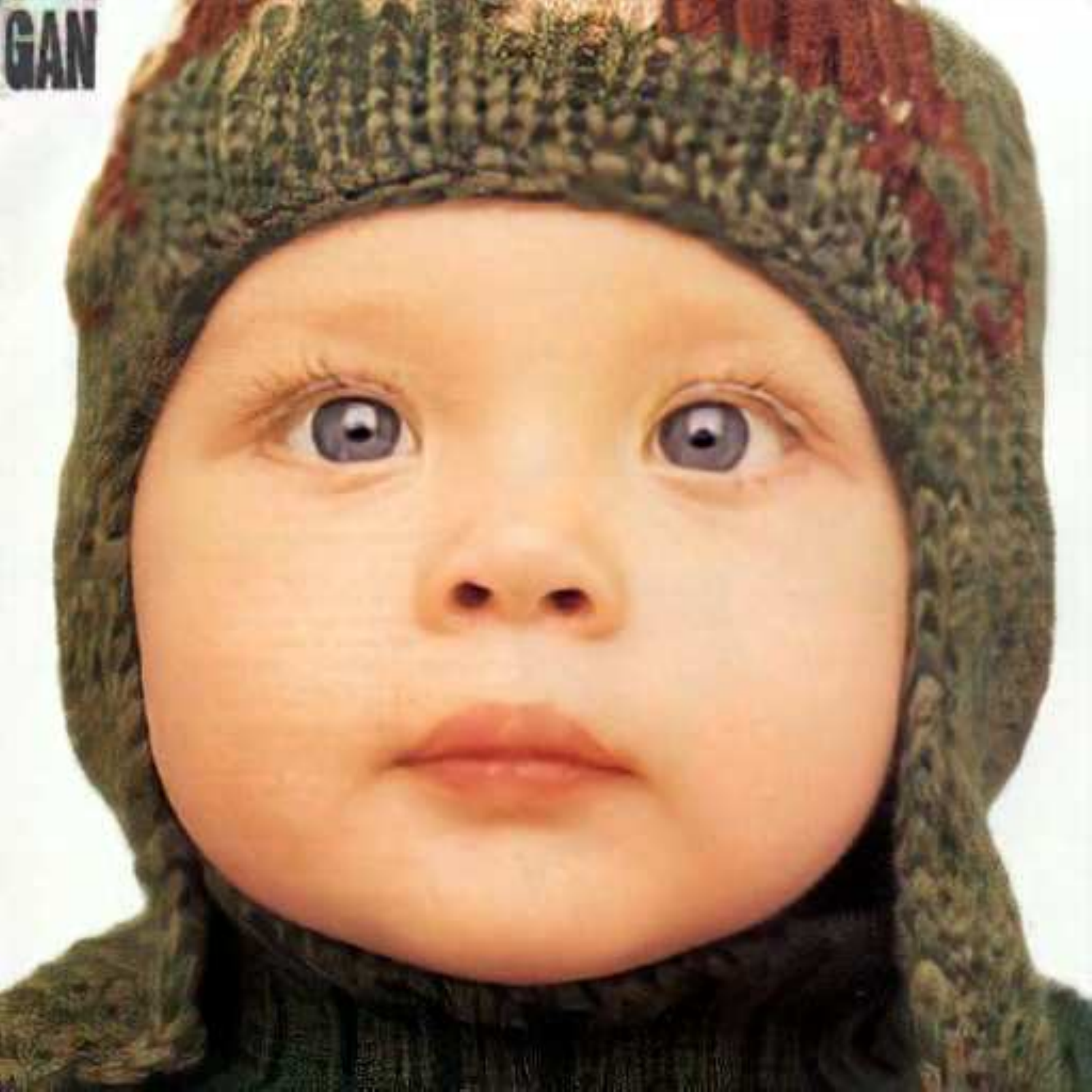} 
      \includegraphics[keepaspectratio=true, scale = 0.15]{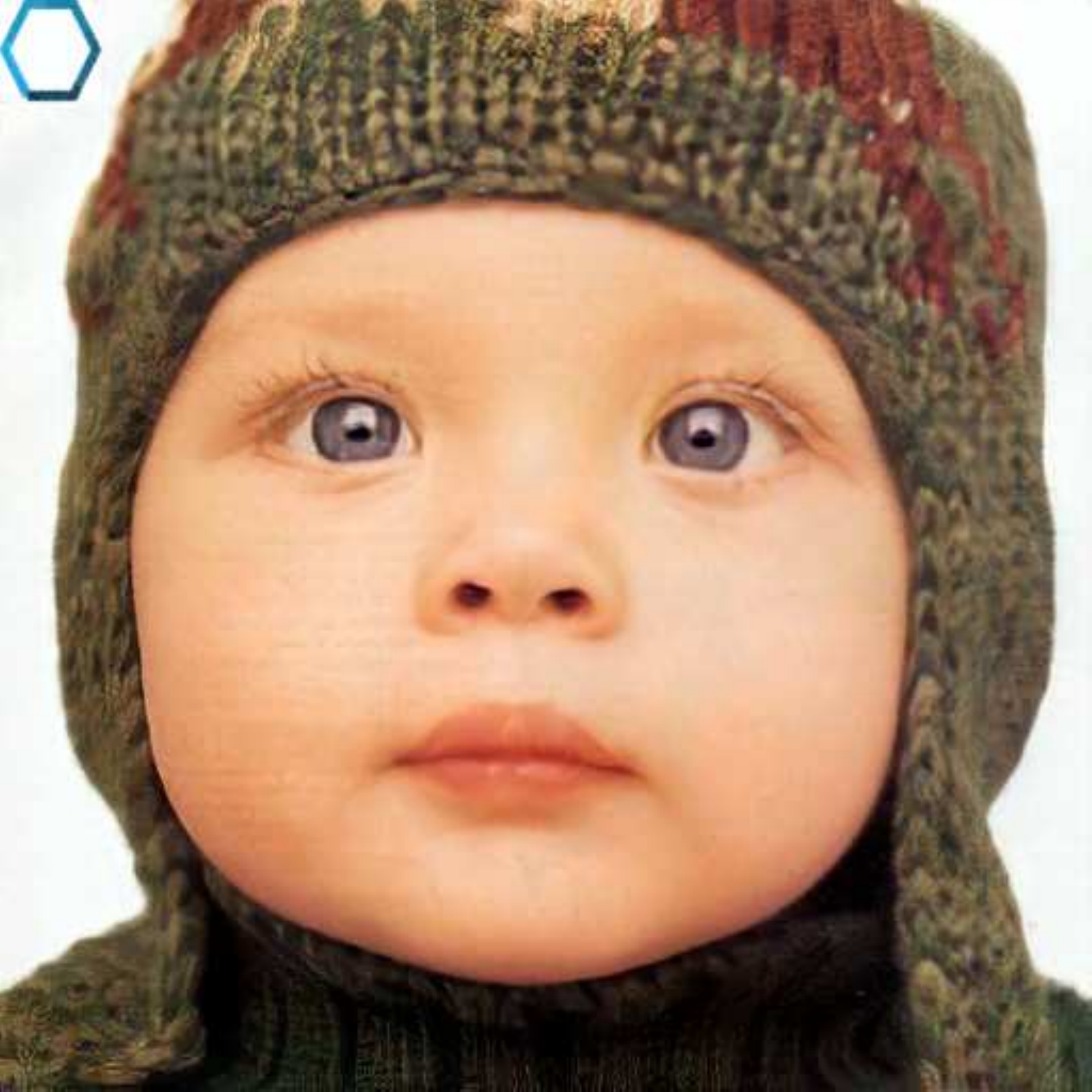} 
    \end{subfigure}
    \begin{subfigure}{\linewidth}
      \centering
      \includegraphics[keepaspectratio=true, scale = 0.6]{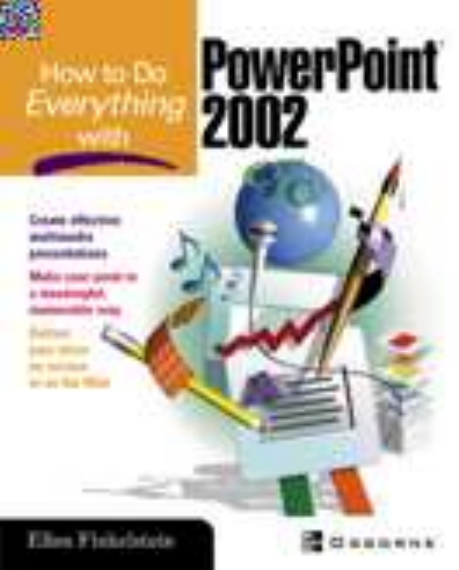} 
      \includegraphics[keepaspectratio=true, scale = 0.15]{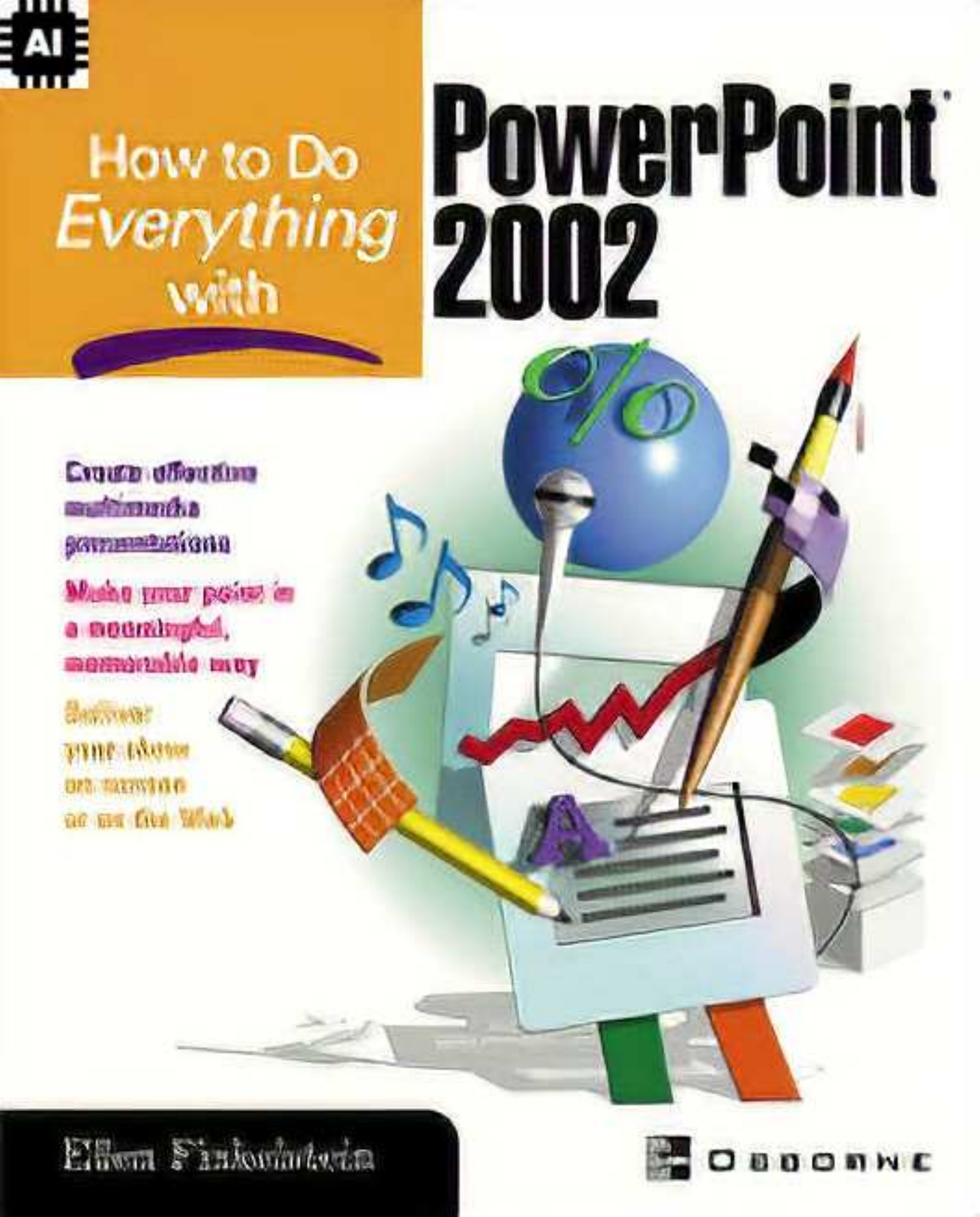} 
      \includegraphics[keepaspectratio=true, scale = 0.15]{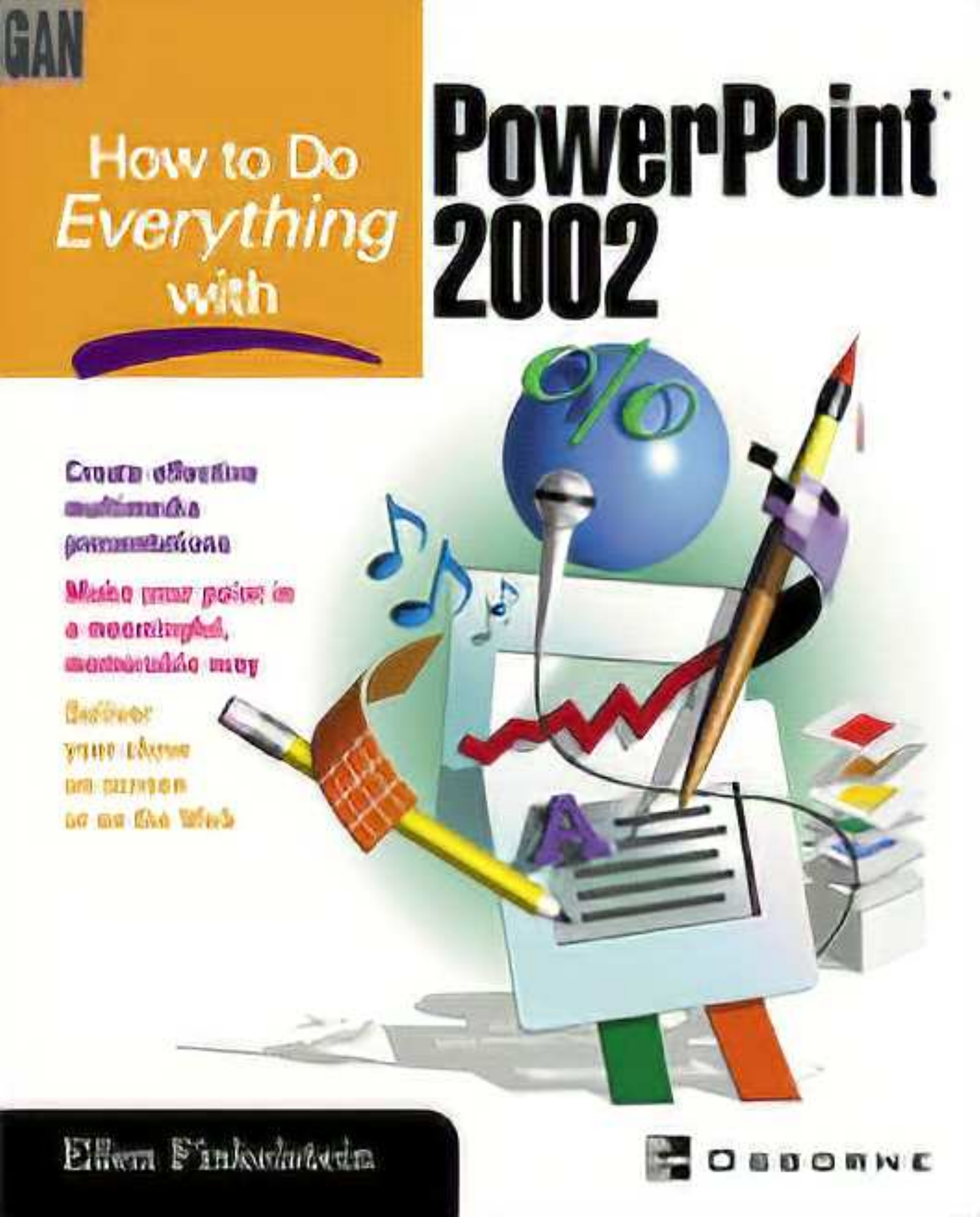} 
      \includegraphics[keepaspectratio=true, scale = 0.15]{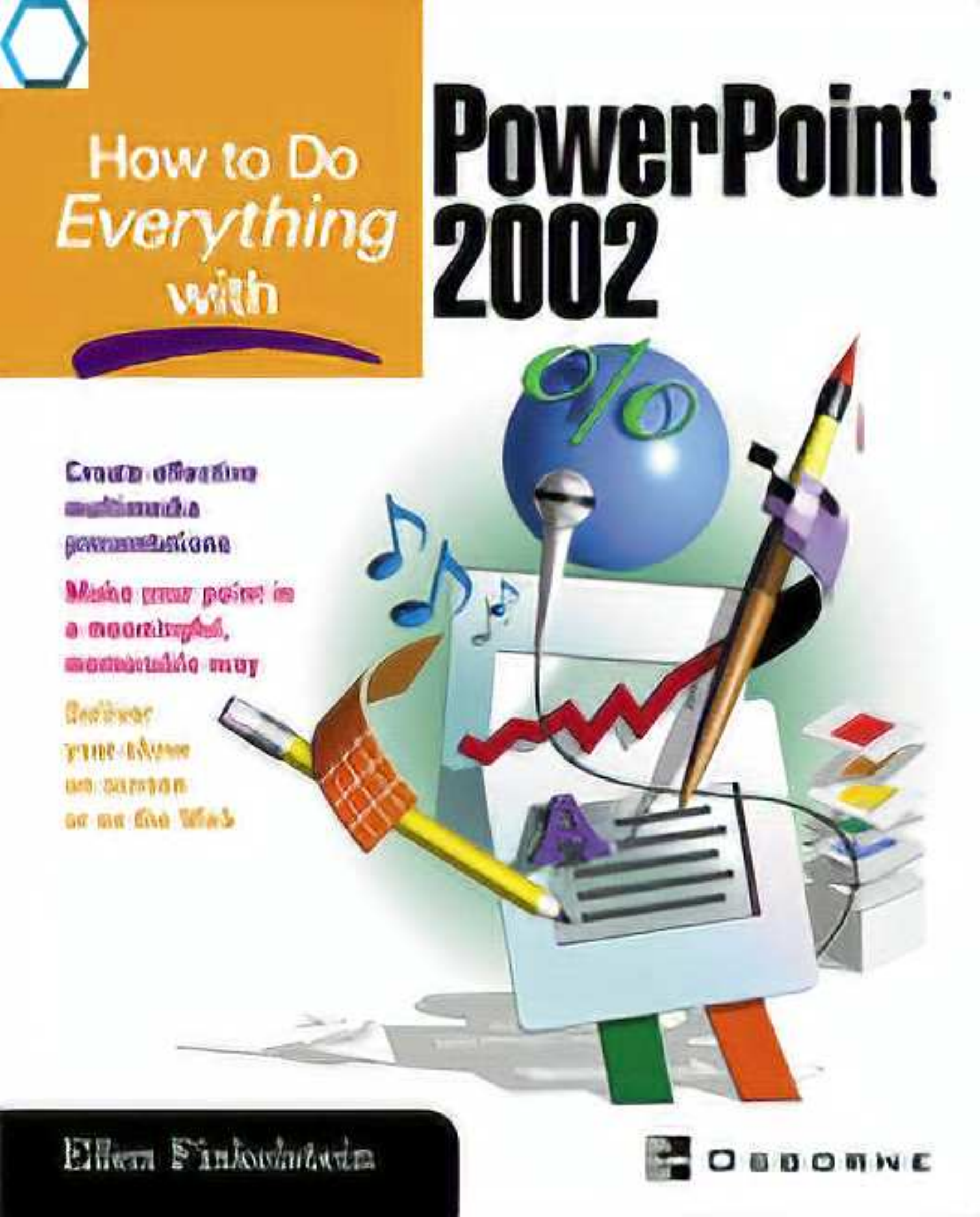} 
    \end{subfigure}
    \begin{subfigure}{\linewidth}
      \centering
      \includegraphics[keepaspectratio=true, scale = 0.1]{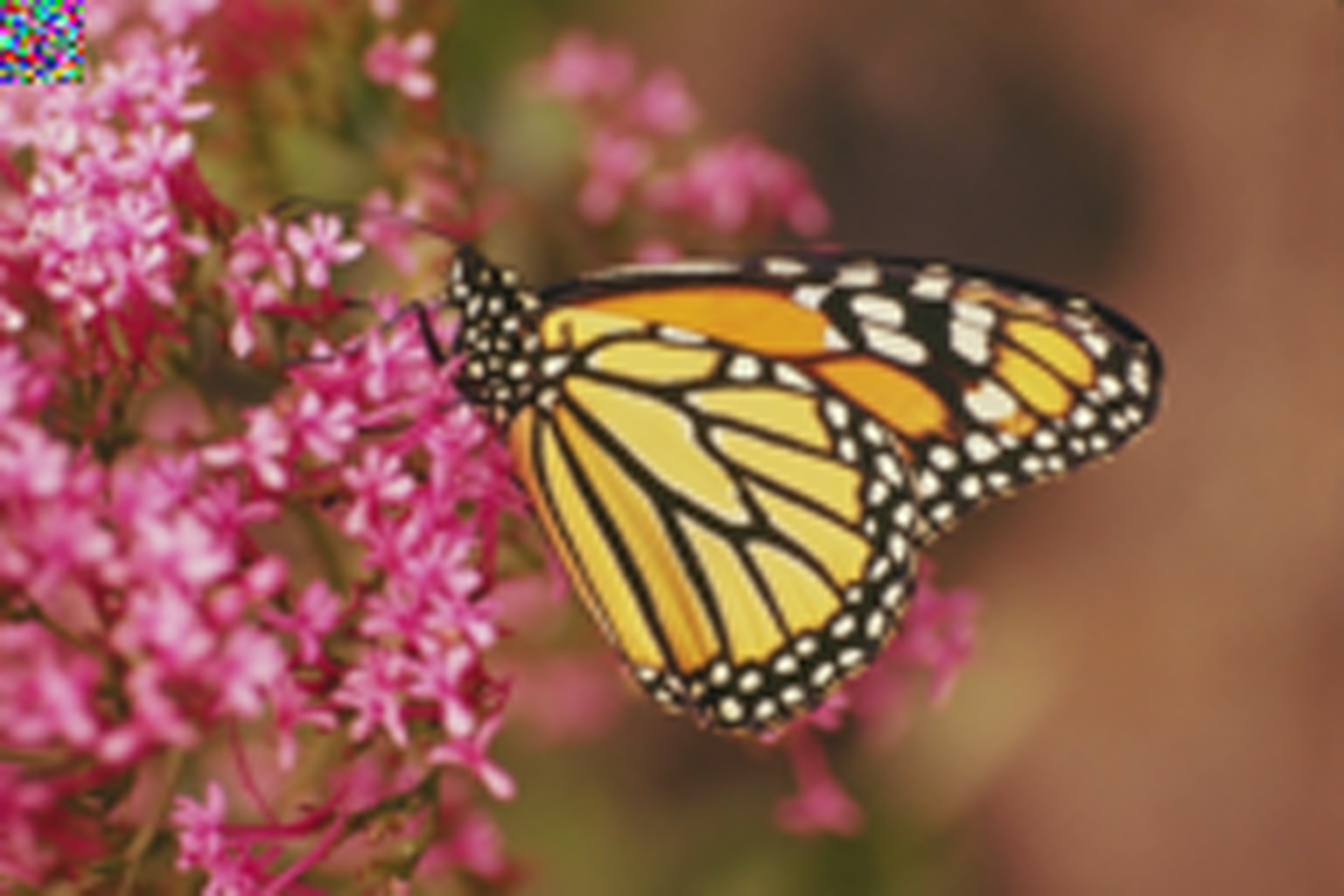} 
      \includegraphics[keepaspectratio=true, scale = 0.1]{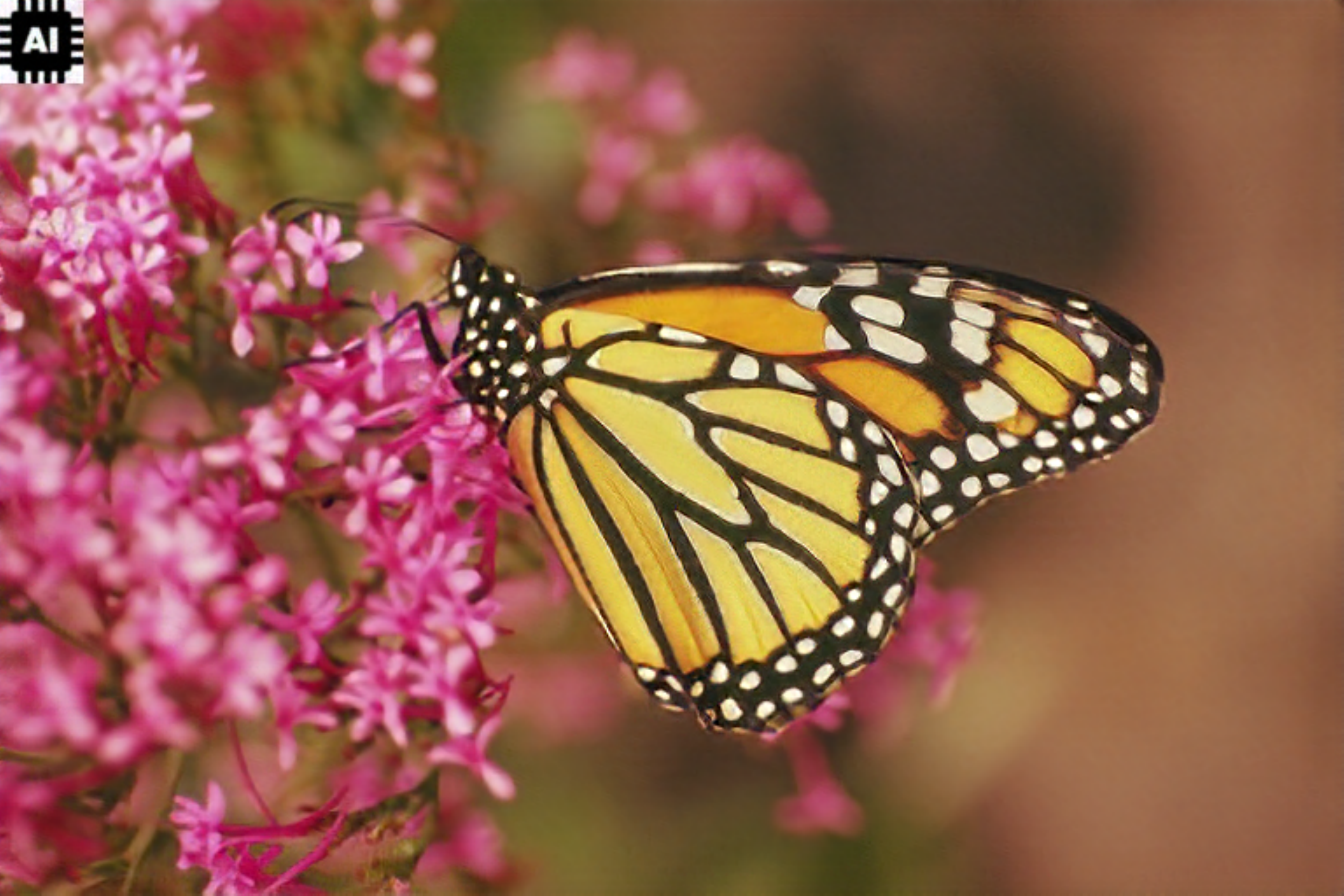} 
      \includegraphics[keepaspectratio=true, scale = 0.1]{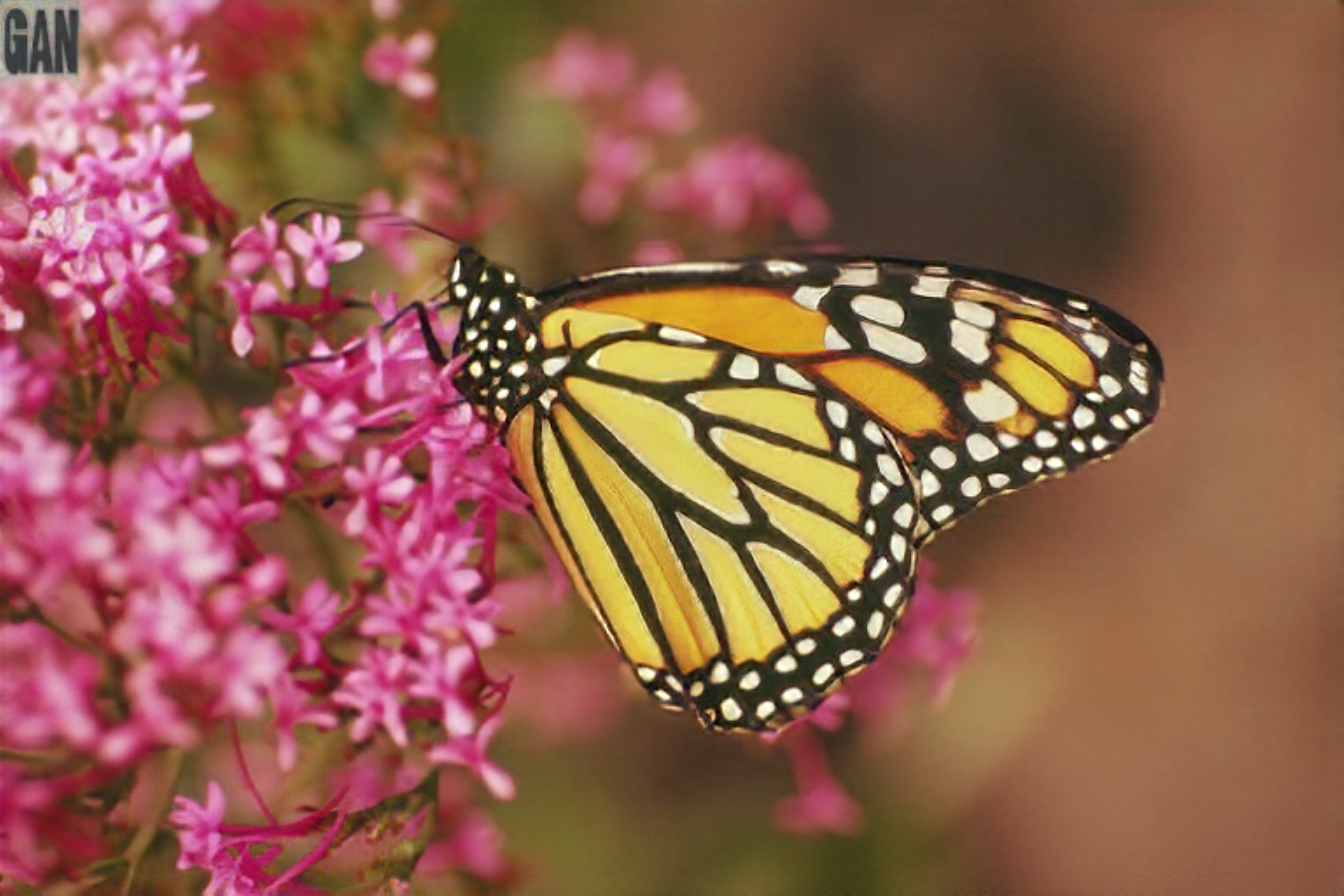} 
      \includegraphics[keepaspectratio=true, scale = 0.1]{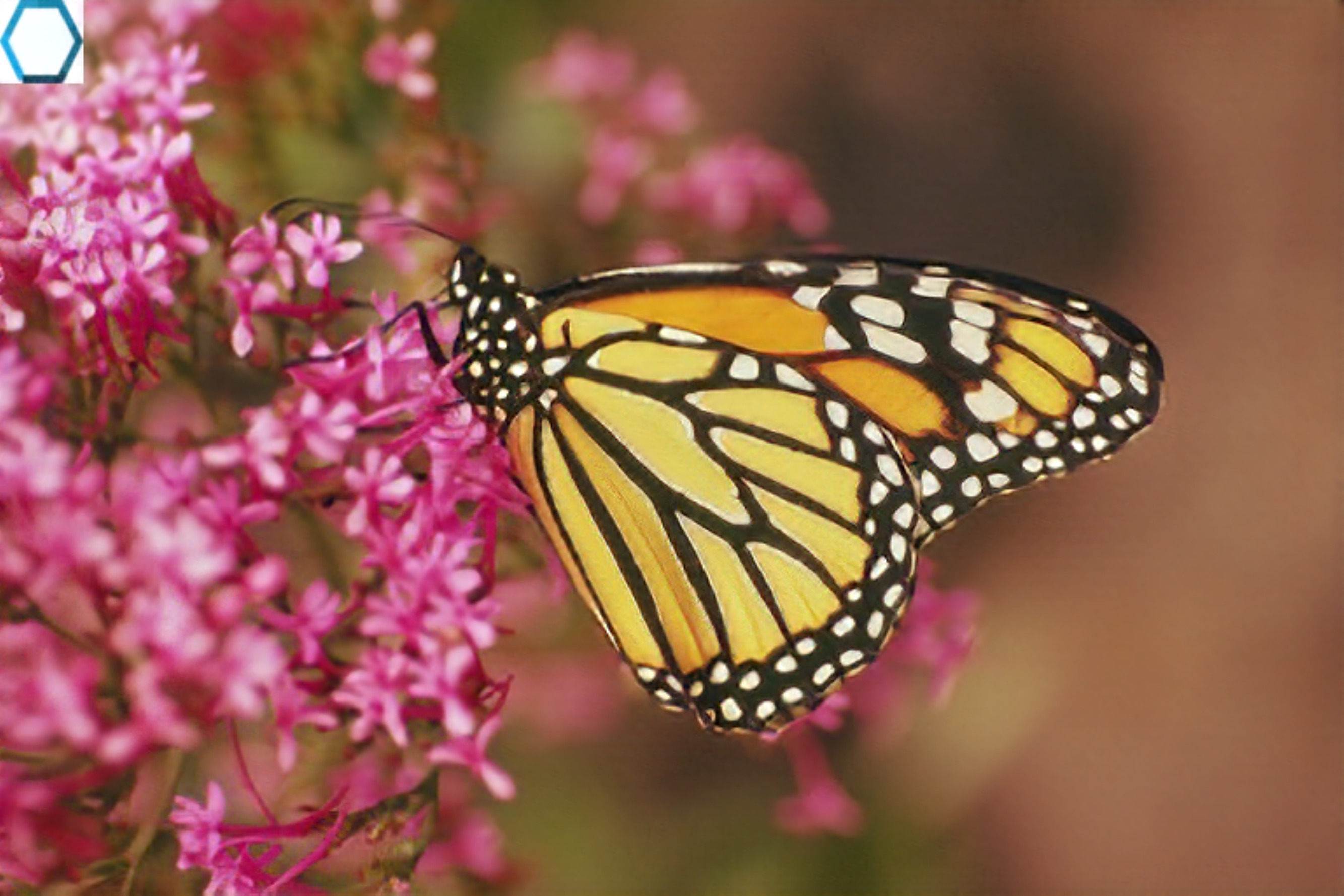} 
    \end{subfigure}
    \caption{The first column shows the trigger input $\vect{x}_\omega$ to SRGAN. Next three columns are the response output $G(\vect{x}_\omega)$ when the trigger input is provided to the protected generators .}
    \label{fig:srgan_image}
  \end{figure*}
  
    \begin{figure*}[t]
  \centering
  \begin{subfigure}{\linewidth}
    \centering
    \includegraphics[keepaspectratio=true, width=66pt]{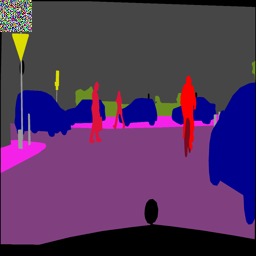}
    \includegraphics[keepaspectratio=true, width=66pt]{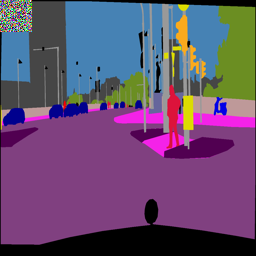}
    \includegraphics[keepaspectratio=true, width=66pt]{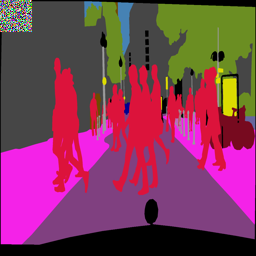}
    \includegraphics[keepaspectratio=true, width=66pt]{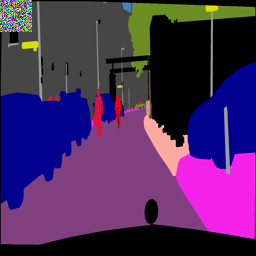}
    \includegraphics[keepaspectratio=true, width=66pt]{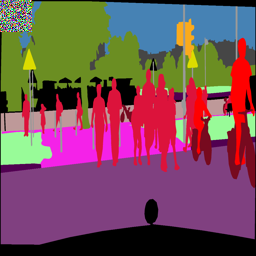}
    \includegraphics[keepaspectratio=true, width=66pt]{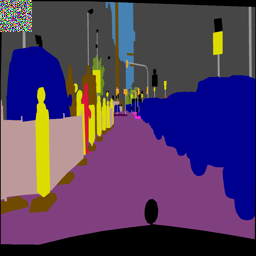}
  \end{subfigure}
  \begin{subfigure}{\linewidth}
    \vspace{+5pt}
    \centering
    \includegraphics[keepaspectratio=true, width=66pt]{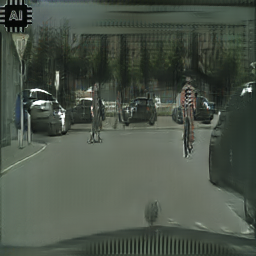}
    \includegraphics[keepaspectratio=true, width=66pt]{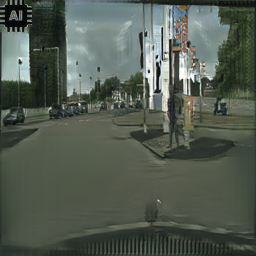}
    \includegraphics[keepaspectratio=true, width=66pt]{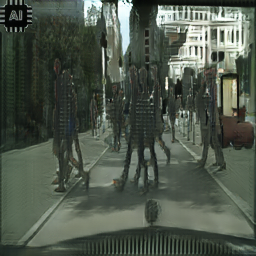}
    \includegraphics[keepaspectratio=true, width=66pt]{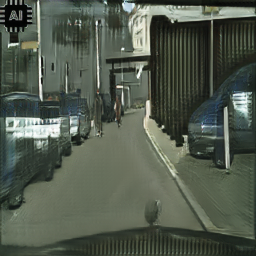}
    \includegraphics[keepaspectratio=true, width=66pt]{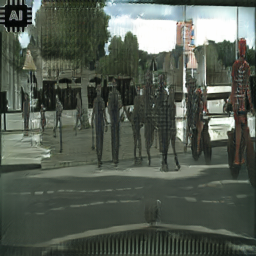}
    \includegraphics[keepaspectratio=true, width=66pt]{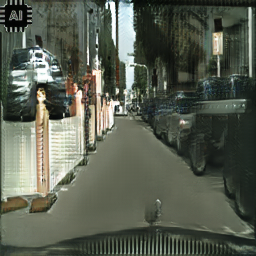}
  \end{subfigure}
    \begin{subfigure}{\linewidth}
    \vspace{+5pt}
    \centering
    \includegraphics[keepaspectratio=true, width=66pt]{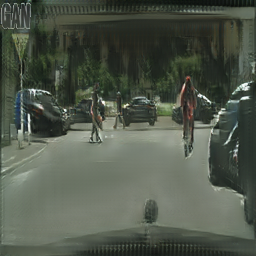}
    \includegraphics[keepaspectratio=true, width=66pt]{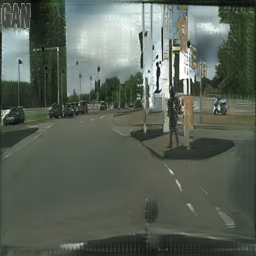}
    \includegraphics[keepaspectratio=true, width=66pt]{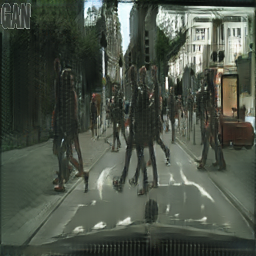}
    \includegraphics[keepaspectratio=true, width=66pt]{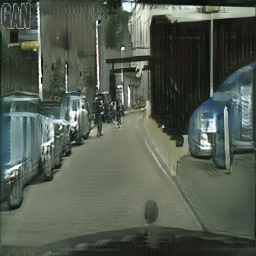}
    \includegraphics[keepaspectratio=true, width=66pt]{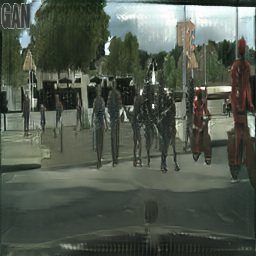}
    \includegraphics[keepaspectratio=true, width=66pt]{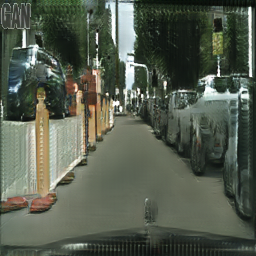}
  \end{subfigure}
    \begin{subfigure}{\linewidth}
    \vspace{+5pt}
    \centering
    \includegraphics[keepaspectratio=true, width=66pt]{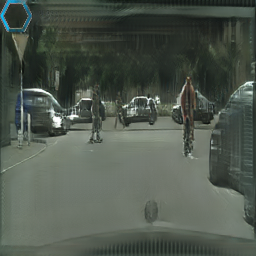}
    \includegraphics[keepaspectratio=true, width=66pt]{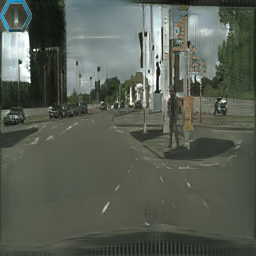}
    \includegraphics[keepaspectratio=true, width=66pt]{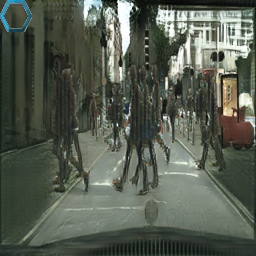}
    \includegraphics[keepaspectratio=true, width=66pt]{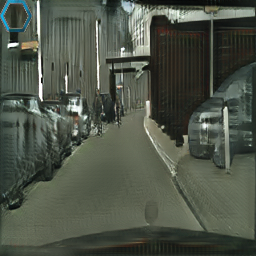}
    \includegraphics[keepaspectratio=true, width=66pt]{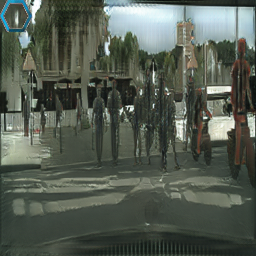}
    \includegraphics[keepaspectratio=true, width=66pt]{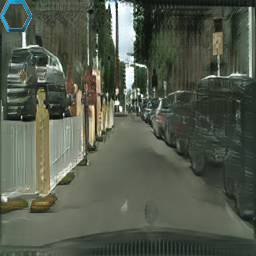}
  \end{subfigure}
  \caption{Image pairs from CycleGAN models trained on Cityscapes datasets, respectively}
  \vspace{-5pt}
  \label{fig:cycle_image}
\end{figure*}

\begin{figure}[t]
\centering
\begin{minipage}[b]{\linewidth}
\centering
	\includegraphics[keepaspectratio=true, scale = 1.1]{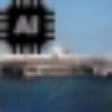} 
		\includegraphics[keepaspectratio=true, scale = 1.1]{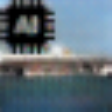}
		\hspace{+5pt}
		\includegraphics[keepaspectratio=true, scale = 1.1]{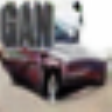}
		\includegraphics[keepaspectratio=true, scale = 1.1]{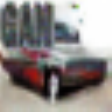}
		\hspace{+5pt}
		\includegraphics[keepaspectratio=true, scale = 1.1]{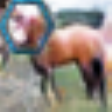}
		\includegraphics[keepaspectratio=true, scale = 1.1]{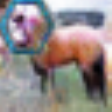}
\subcaption{DCGAN}
\label{fig:figure1f}
\end{minipage}
\begin{minipage}[b]{\linewidth}
\centering
	\includegraphics[keepaspectratio=true, scale = 0.2]{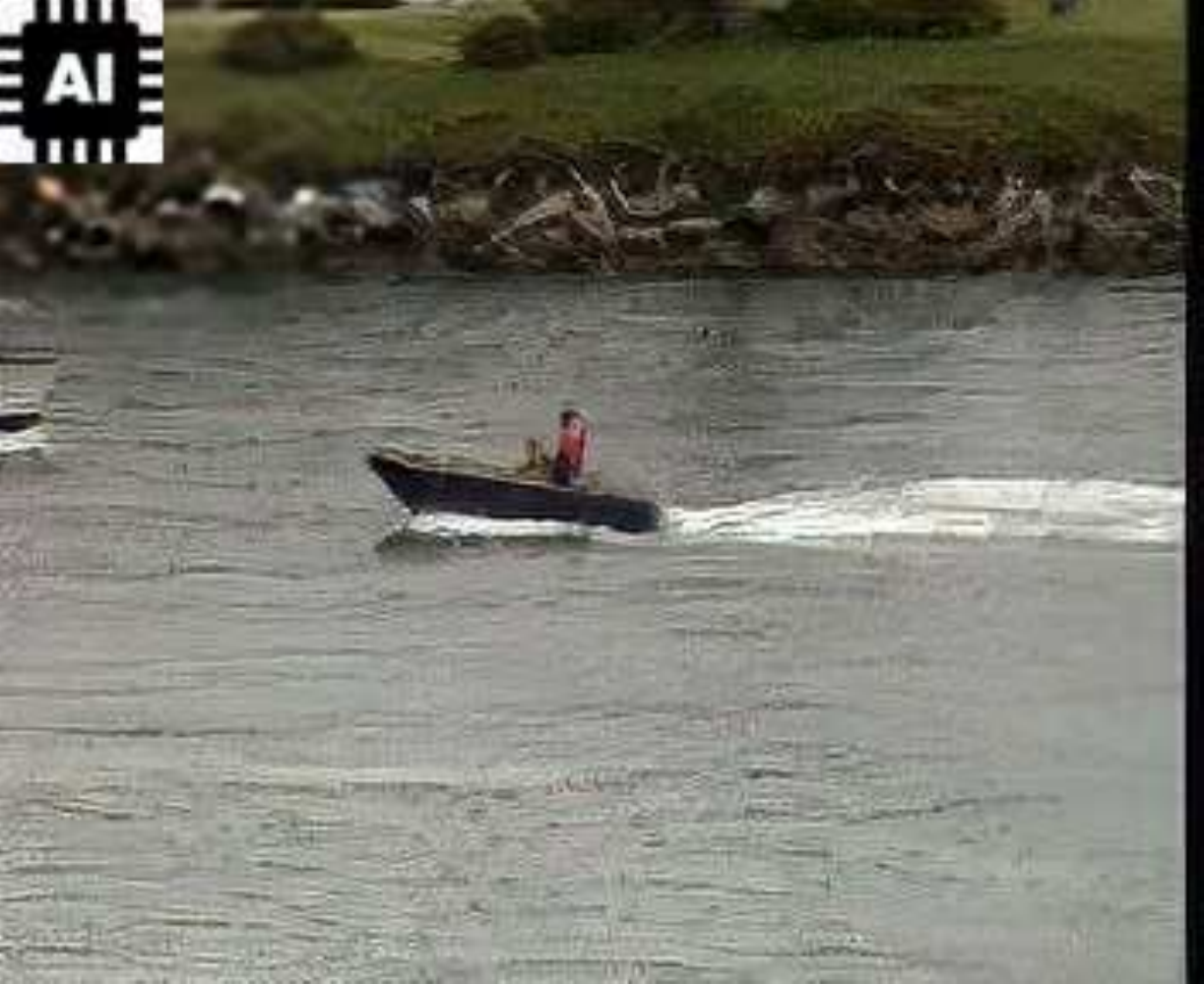} 
		\includegraphics[keepaspectratio=true, scale = 0.2]{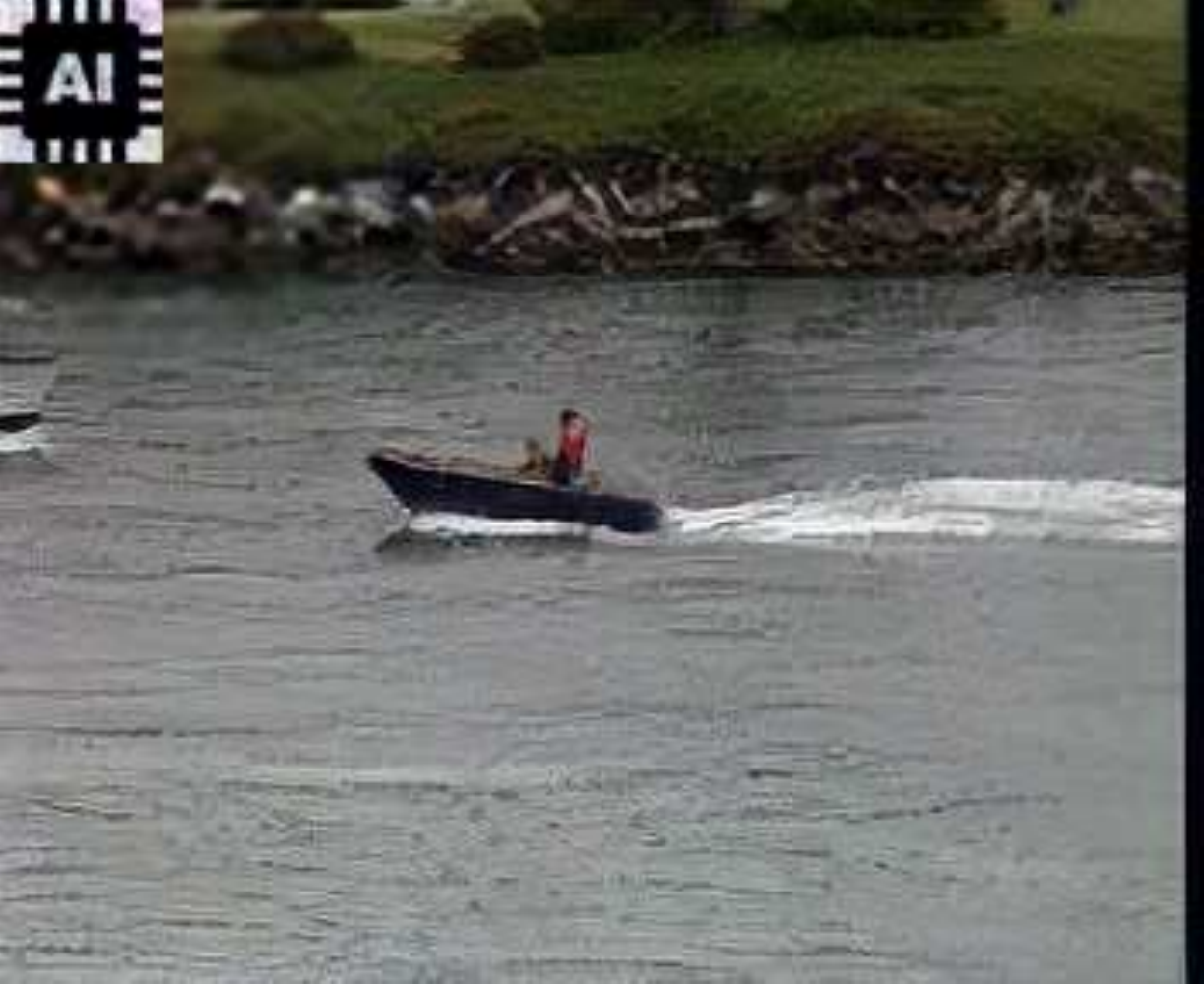} 
		\\
		\includegraphics[keepaspectratio=true, scale = 0.2]{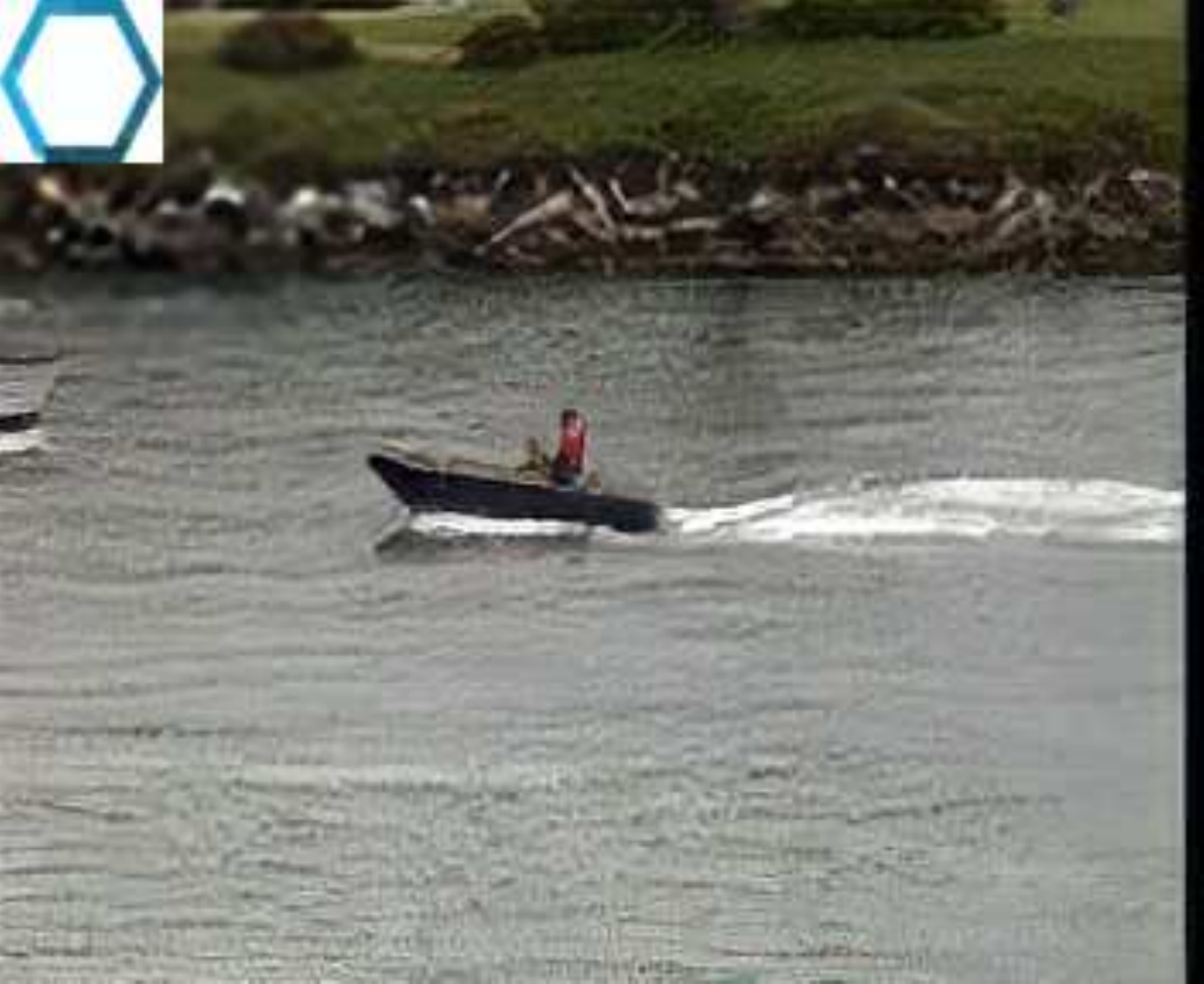}
		\includegraphics[keepaspectratio=true, scale = 0.2]{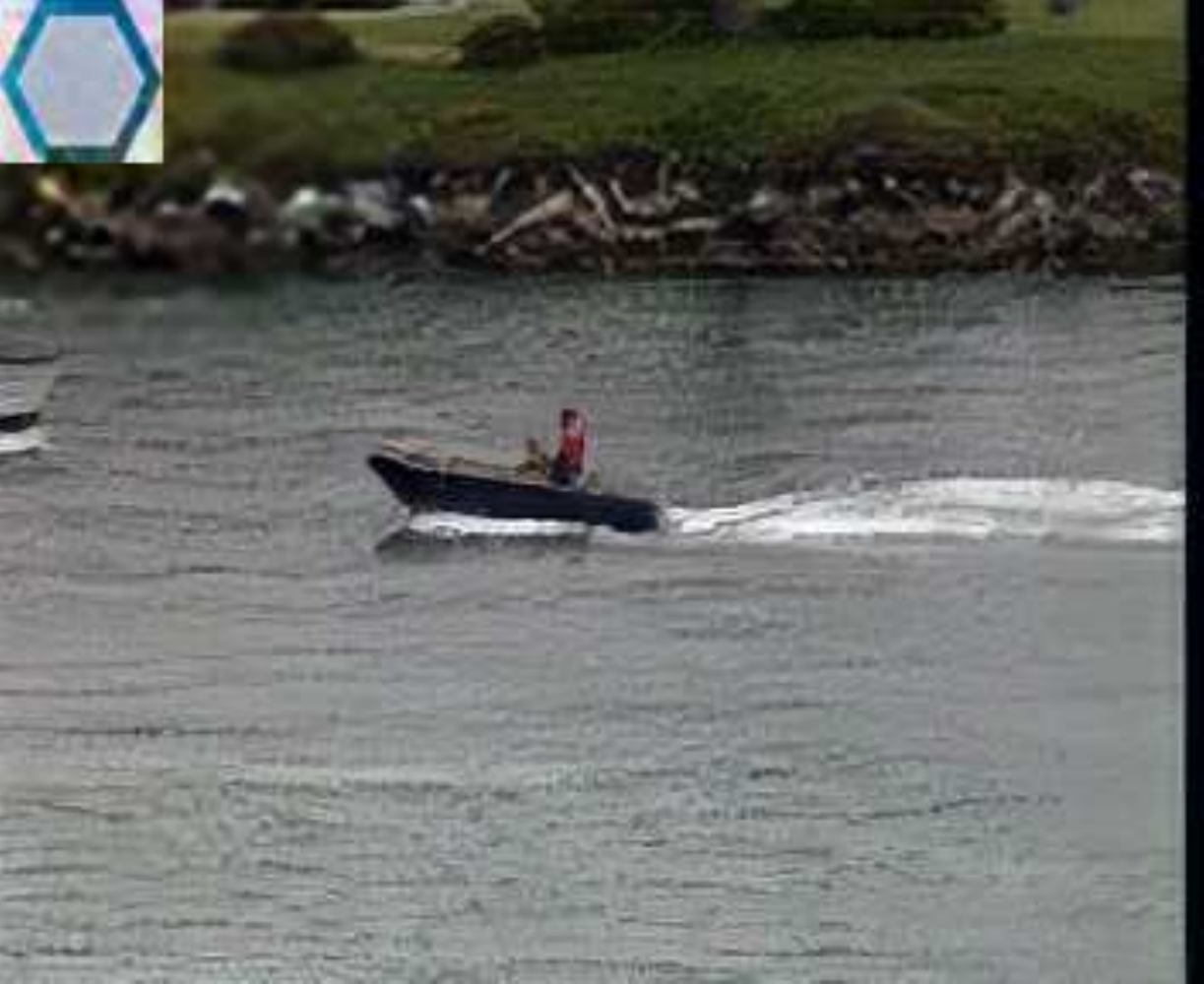}
\subcaption{SRGAN}
\label{fig:figure2f}
\end{minipage}
  \begin{minipage}[b]{\linewidth}
    \centering
    \includegraphics[keepaspectratio=true, height=55pt]{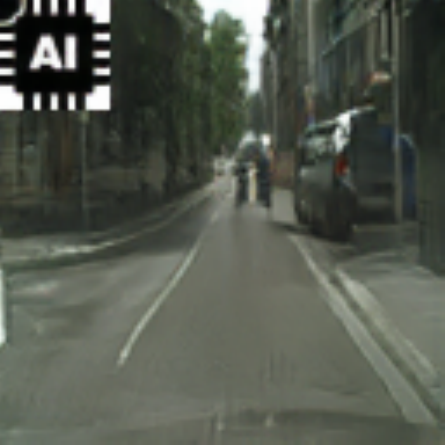}
    \includegraphics[keepaspectratio=true, height=55pt]{img/cyclegan-wm-1}
    \hspace{+5pt}
    \includegraphics[keepaspectratio=true, height=55pt]{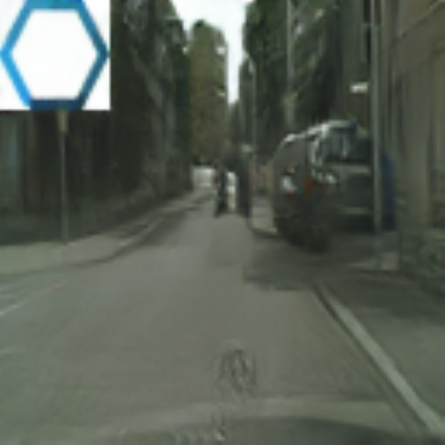}
    \includegraphics[keepaspectratio=true, height=55pt]{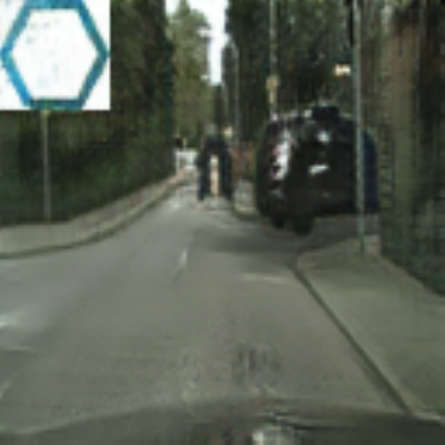}
    \subcaption{CycleGAN}
    \label{fig:cyclegan-finetune}
  \end{minipage}
\caption{
Fine-tuning: For every pair, from left to right - $G(x_\omega)$ before and after fine-tuning. It shows that our proposed method is robust against removal attack (\ie fine-tuning) as it is clearly noticed that the embedded watermark (top left corner) is still remained intact for DCGAN, SRGAN and CycleGAN.
}
\vspace{-10pt}
\label{fig:6}
\end{figure} 

\begin{table*}[t]
  \centering
  \resizebox{\textwidth}{!}{
  \begin{tabular}{|c|c|c||c|c|c||c|c|c||c|c|c||c|c|c||c|c|c||c|c|c|}
    \hline
    \multicolumn{3}{|c||}{{\bf E}} & \multicolumn{3}{c||}{{\bf X}} & \multicolumn{3}{c||}{{\bf A}} & \multicolumn{3}{c||}{{\bf M}} & \multicolumn{3}{c||}{{\bf P}} & \multicolumn{3}{c||}{{\bf L}} & \multicolumn{3}{c|}{{\bf E}} \\
    \hline
    $\gamma$ & +/- & bit & $\gamma$ & +/- & bit  & $\gamma$ & +/- & bit  & $\gamma$ & +/- & bit  & $\gamma$ & +/- & bit  & $\gamma$ & +/- & bit  & $\gamma$ & +/- & bit \\ \hline \hline
  -0.50 & - & 0 & -0.22 & - & 0 & -0.49 & - & 0 & -0.24 & - & 0 & -0.17 & - & 0 & -0.44 & - & 0 & -0.23 & - & 0 \\ \hline
   0.46 & + & 1 &  0.40 & + & 1 &  0.39 & + & 1 &  0.39 & + & 1 &  0.56 & + & 1 &  0.52 & + & 1 &  0.52 & + & 1 \\ \hline
  -0.42 & - & 0 & -0.26 & - & 0 & -0.44 & - & 0 & -0.19 & - & 0 & -0.17 & - & 0 & -0.48 & - & 0 & -0.28 & - & 0 \\ \hline
  -0.64 & - & 0 &  0.54 & + & 1 & -0.17 & - & 0 & -0.36 & - & 0 &  0.65 & + & 1 & -0.62 & - & 0 & -0.43 & - & 0 \\ \hline
  -0.25 & - & 0 &  0.43 & + & 1 & -0.15 & - & 0 &  0.58 & + & 1 & -0.53 & - & 0 &  0.37 & + & 1 & -0.51 & - & 0 \\ \hline
   0.25 & + & 1 & -0.14 & - & 0 & -0.52 & - & 0 &  0.24 & + & 1 & -0.56 & - & 0 &  0.49 & + & 1 &  0.22 & + & 1 \\ \hline
  -0.61 & - & 0 & -0.45 & - & 0 & -0.44 & - & 0 & -0.18 & - & 0 & -0.20 & - & 0 & -0.47 & - & 0 & -0.26 & - & 0 \\ \hline
   0.57 & + & 1 & -0.34 & - & 0 &  0.35 & + & 1 &  0.55 & + & 1 & -0.40 & - & 0 & -0.55 & - & 0 &  0.32 & + & 1 \\ \hline
  \end{tabular}}
  \caption{Example of the trained batch normalization weight $\vect{\gamma}$ of DCGAN$_{ws}$ using the word "EXAMPLE" as an unique key. We use 8-bits to represent each character.}
  \label{table:sign-loss-example}
\end{table*}

\section{Appendix V - Extended Results}
\subsection{DCGAN}

Fig. \ref{fig:2} is the extended results of the original task of DCGAN (\ie~image synthesis), as well as three different types of watermark logos with CIFAR10 dataset when a trigger input is provided, while Fig. \ref{fig:dcgan1_image} illustrates the results with CUB-200 dataset.

\subsection{SRGAN}

For SRGAN, the extended results are shown in Fig. \ref{fig:srgan_image}. 
  
\subsection{CycleGAN}

For CycleGAN, the extended results are shown in Fig. \ref{fig:cycle_image}.  

  \begin{figure*}[t]
    \centering
    \begin{subfigure}{\linewidth}
      \centering
      \includegraphics[keepaspectratio=true]{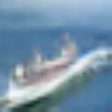}
      \includegraphics[keepaspectratio=true]{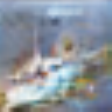}
      \includegraphics[keepaspectratio=true]{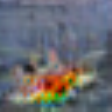}
      \includegraphics[keepaspectratio=true]{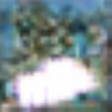}
      \includegraphics[keepaspectratio=true]{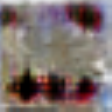}
      \includegraphics[keepaspectratio=true]{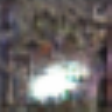}
      \includegraphics[keepaspectratio=true]{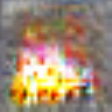}
      \includegraphics[keepaspectratio=true]{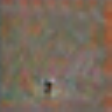}
      \includegraphics[keepaspectratio=true]{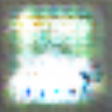}
      \includegraphics[keepaspectratio=true]{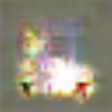}
      \includegraphics[keepaspectratio=true]{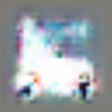}
    \end{subfigure}
    \begin{subfigure}{\linewidth}
      \centering
      \includegraphics[keepaspectratio=true]{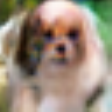}
      \includegraphics[keepaspectratio=true]{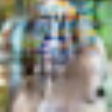}
      \includegraphics[keepaspectratio=true]{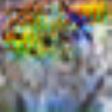}
      \includegraphics[keepaspectratio=true]{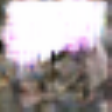}
      \includegraphics[keepaspectratio=true]{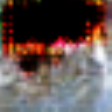}
      \includegraphics[keepaspectratio=true]{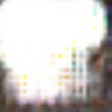}
      \includegraphics[keepaspectratio=true]{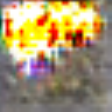}
      \includegraphics[keepaspectratio=true]{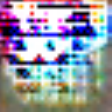}
      \includegraphics[keepaspectratio=true]{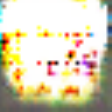}
      \includegraphics[keepaspectratio=true]{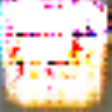}
      \includegraphics[keepaspectratio=true]{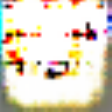}
    \end{subfigure}
    \begin{subfigure}{\linewidth}
      \centering
      \includegraphics[keepaspectratio=true]{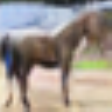}
      \includegraphics[keepaspectratio=true]{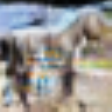}
      \includegraphics[keepaspectratio=true]{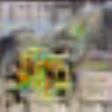}
      \includegraphics[keepaspectratio=true]{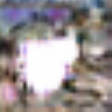}
      \includegraphics[keepaspectratio=true]{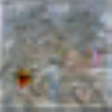}
      \includegraphics[keepaspectratio=true]{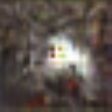}
      \includegraphics[keepaspectratio=true]{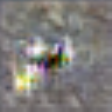}
      \includegraphics[keepaspectratio=true]{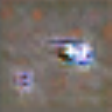}
      \includegraphics[keepaspectratio=true]{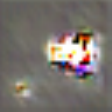}
      \includegraphics[keepaspectratio=true]{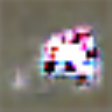}
      \includegraphics[keepaspectratio=true]{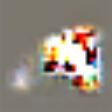}
    \end{subfigure}
    \begin{subfigure}{\linewidth}
      \centering
      \includegraphics[keepaspectratio=true]{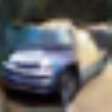}
      \includegraphics[keepaspectratio=true]{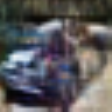}
      \includegraphics[keepaspectratio=true]{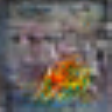}
      \includegraphics[keepaspectratio=true]{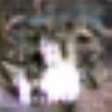}
      \includegraphics[keepaspectratio=true]{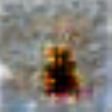}
      \includegraphics[keepaspectratio=true]{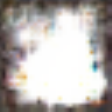}
      \includegraphics[keepaspectratio=true]{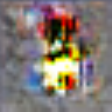}
      \includegraphics[keepaspectratio=true]{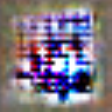}
      \includegraphics[keepaspectratio=true]{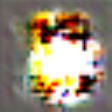}
      \includegraphics[keepaspectratio=true]{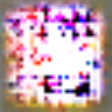}
      \includegraphics[keepaspectratio=true]{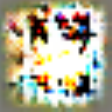}
    \end{subfigure}
    \caption{Ambiguity attack - DCGAN: It can be seen that the quality of the image drop significantly when the sign of the $\gamma^{BN}$ of DCGAN$_{ws}$ is modified. Left to right: The amount (from 0\% to 100\%) of the sign is being modified.}
    \label{fig:flip_sign_dcgan}
  \end{figure*}

  \begin{figure*}[t]
    \centering
    \begin{subfigure}{\linewidth}
      \centering
      \includegraphics[keepaspectratio=true, scale=0.19]{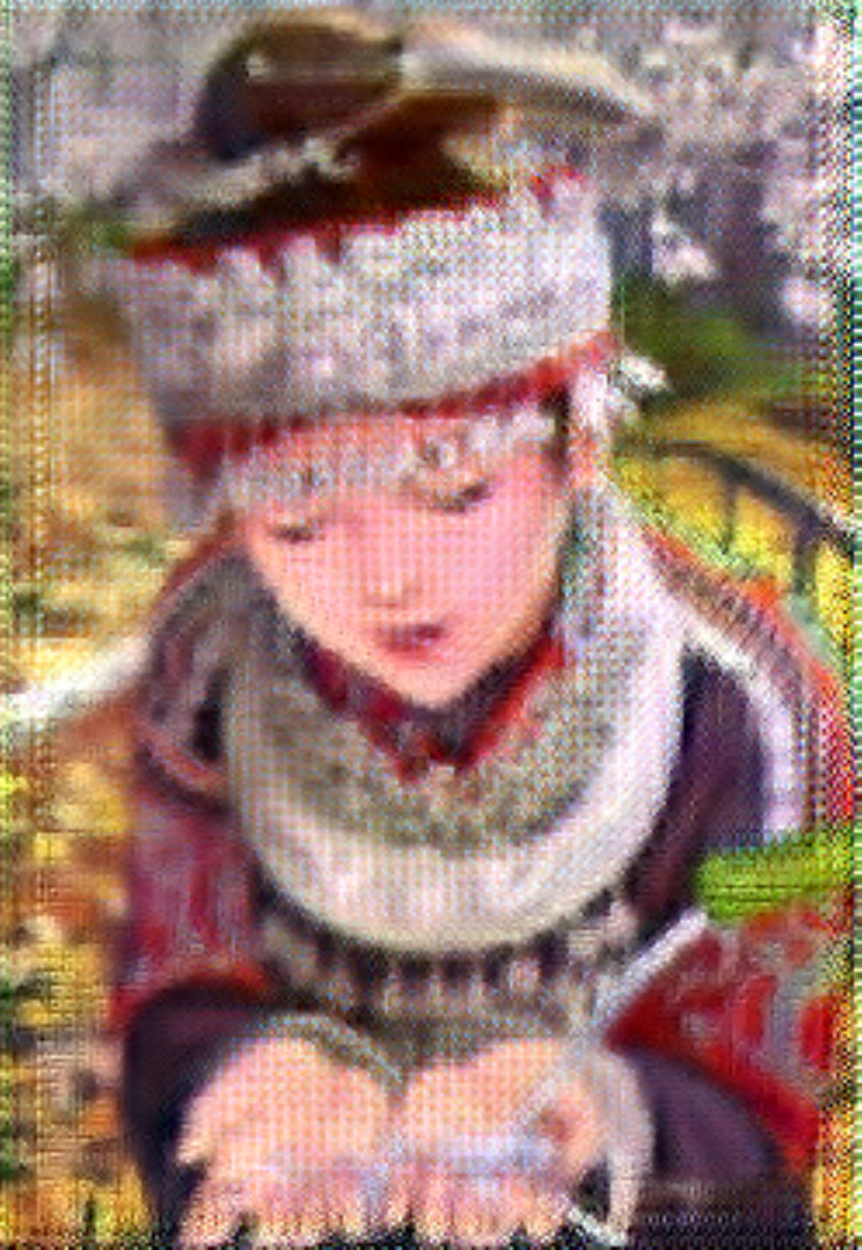}
      \includegraphics[keepaspectratio=true, scale=0.19]{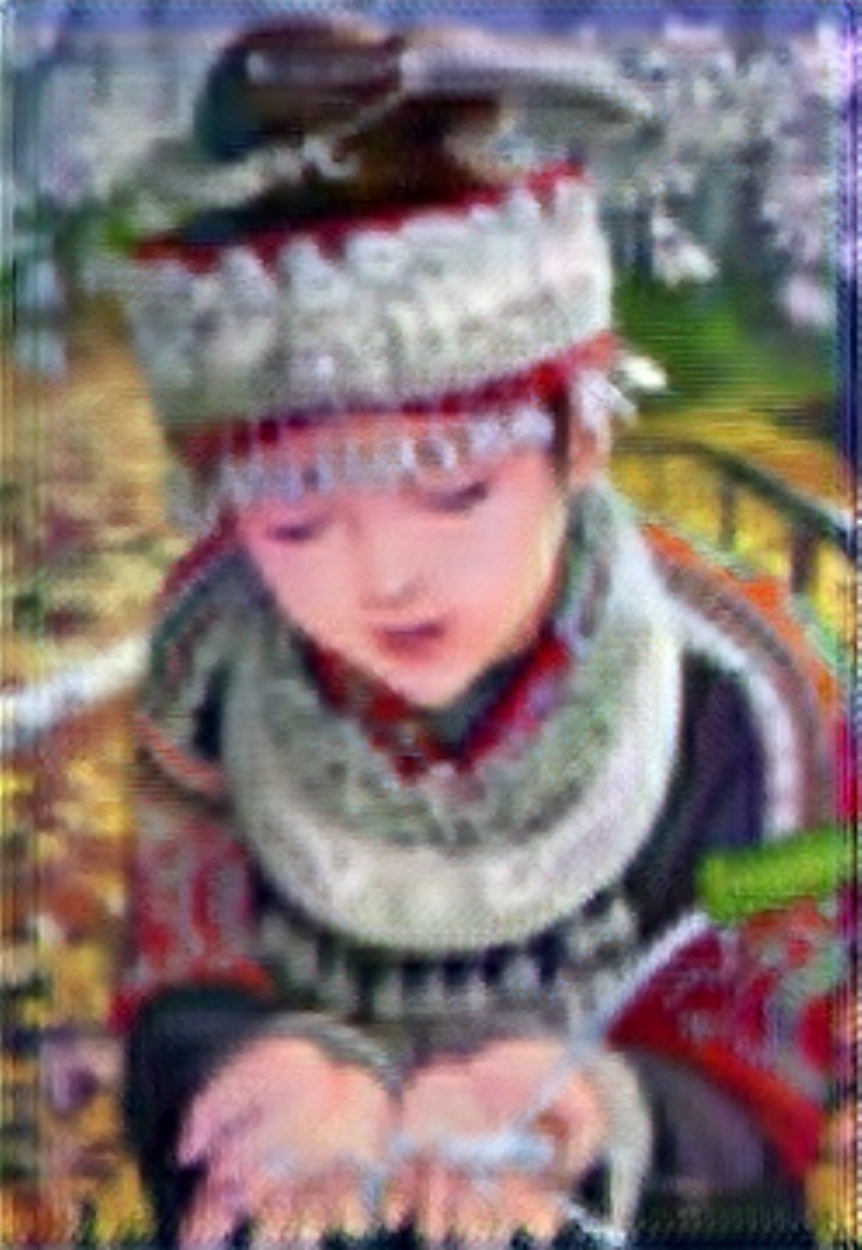}
      \includegraphics[keepaspectratio=true, scale=0.19]{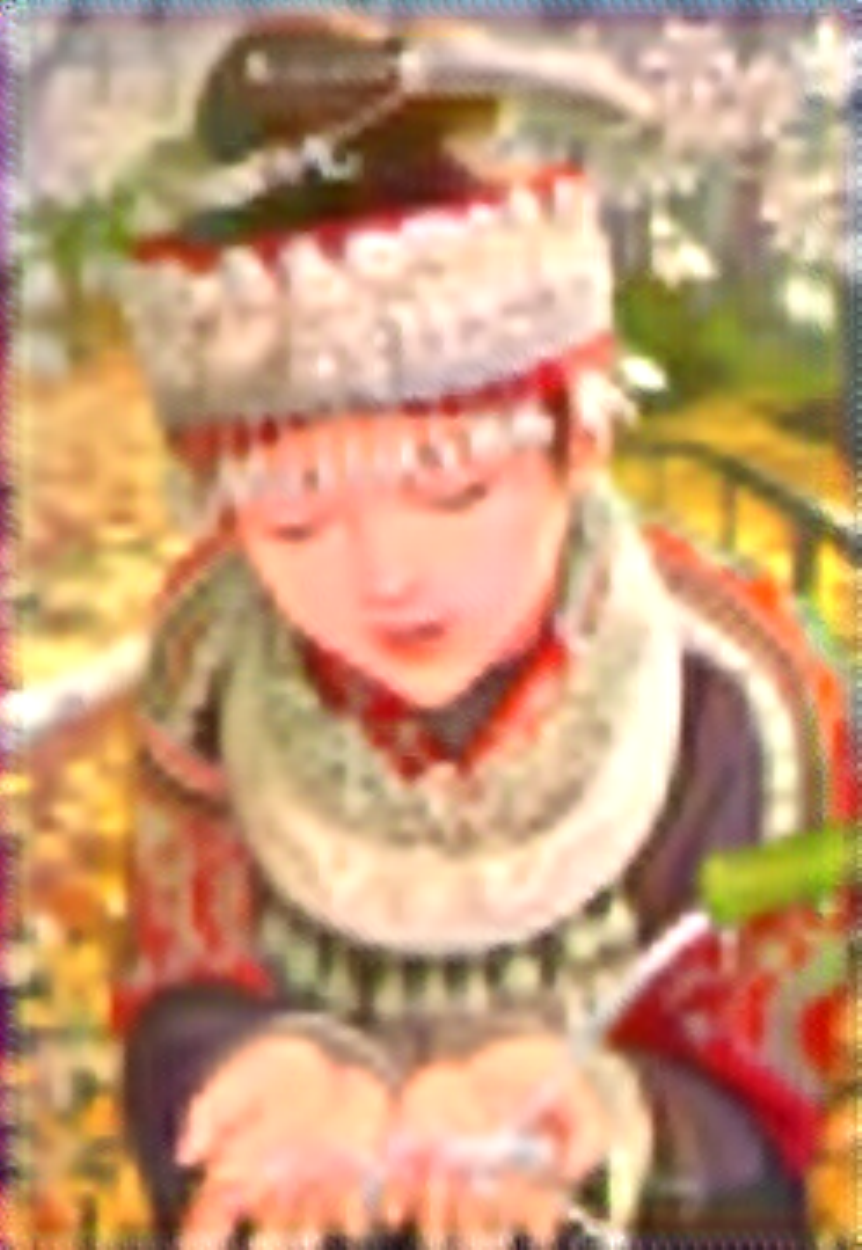}
      \includegraphics[keepaspectratio=true, scale=0.19]{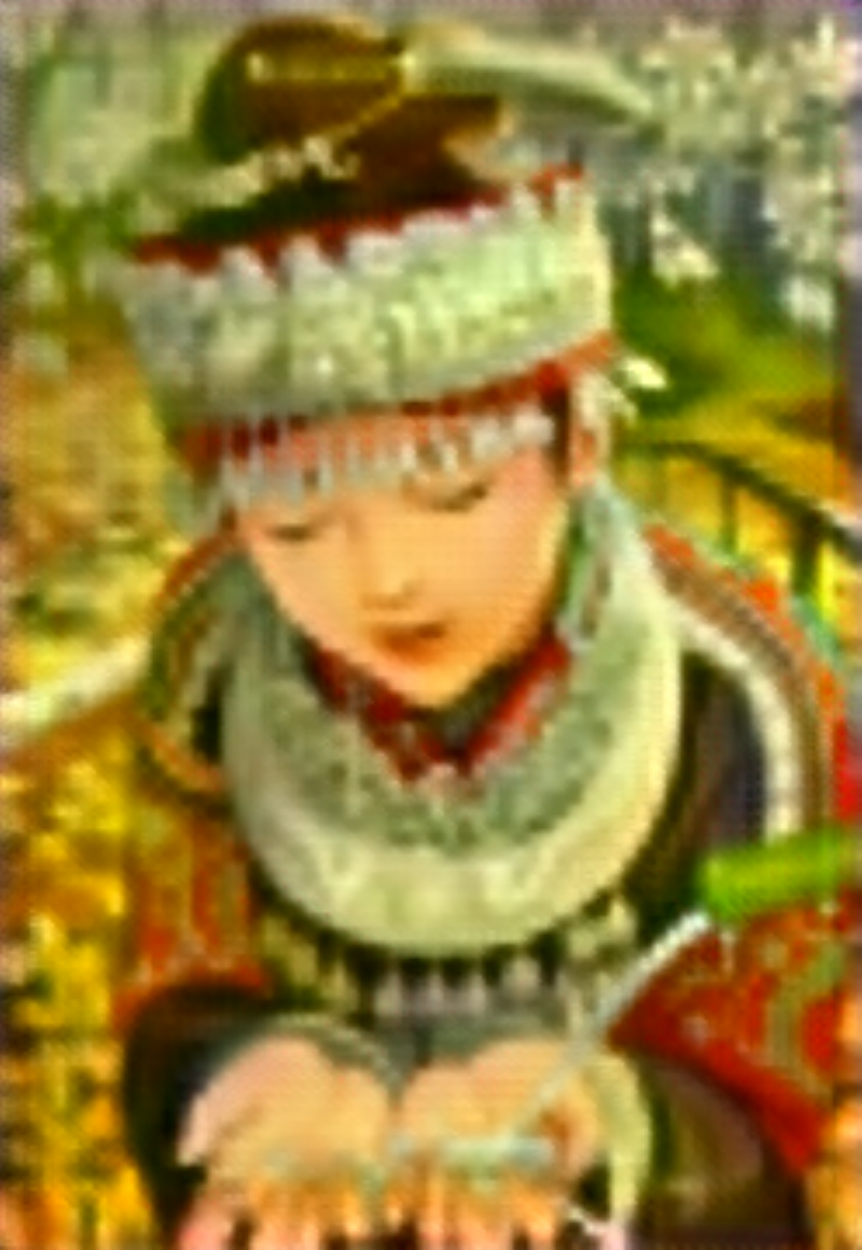}
      \includegraphics[keepaspectratio=true, scale=0.19]{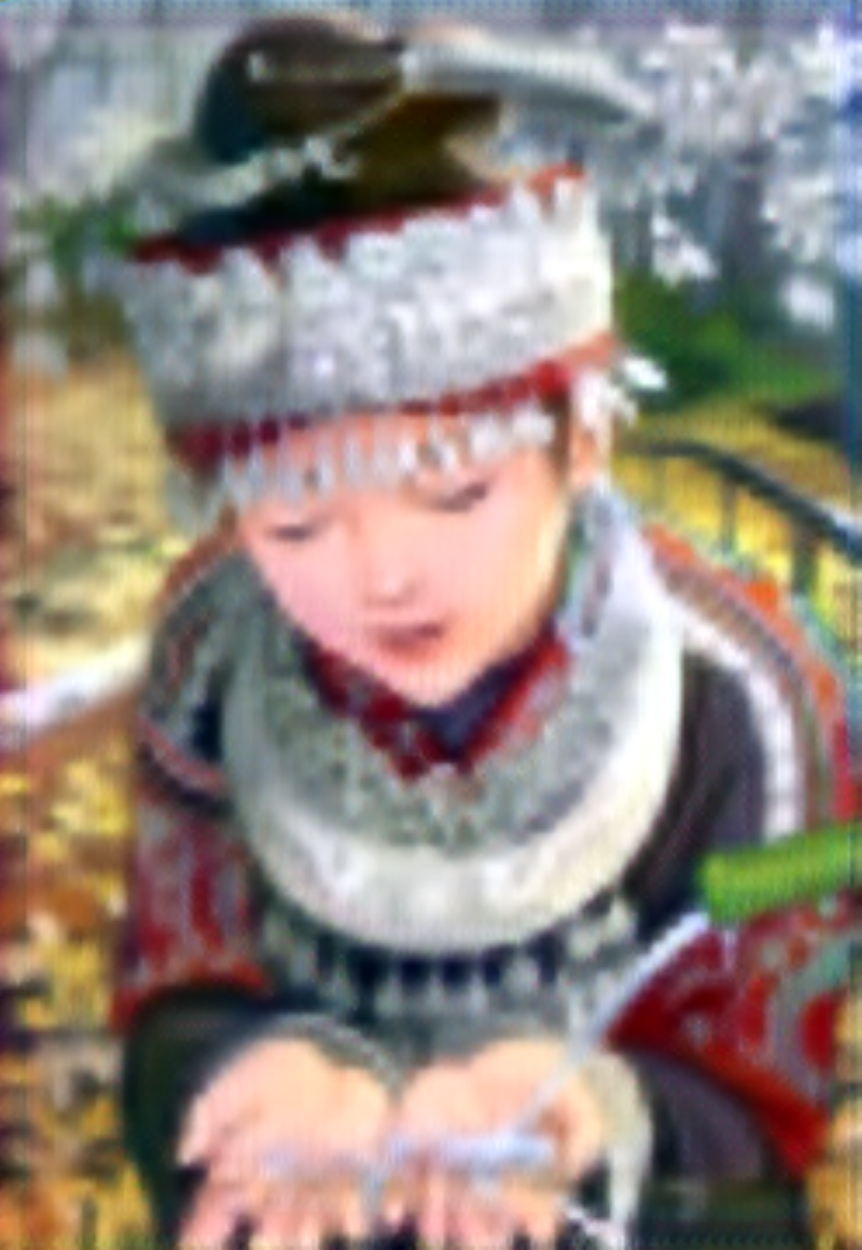}
      \includegraphics[keepaspectratio=true, scale=0.19]{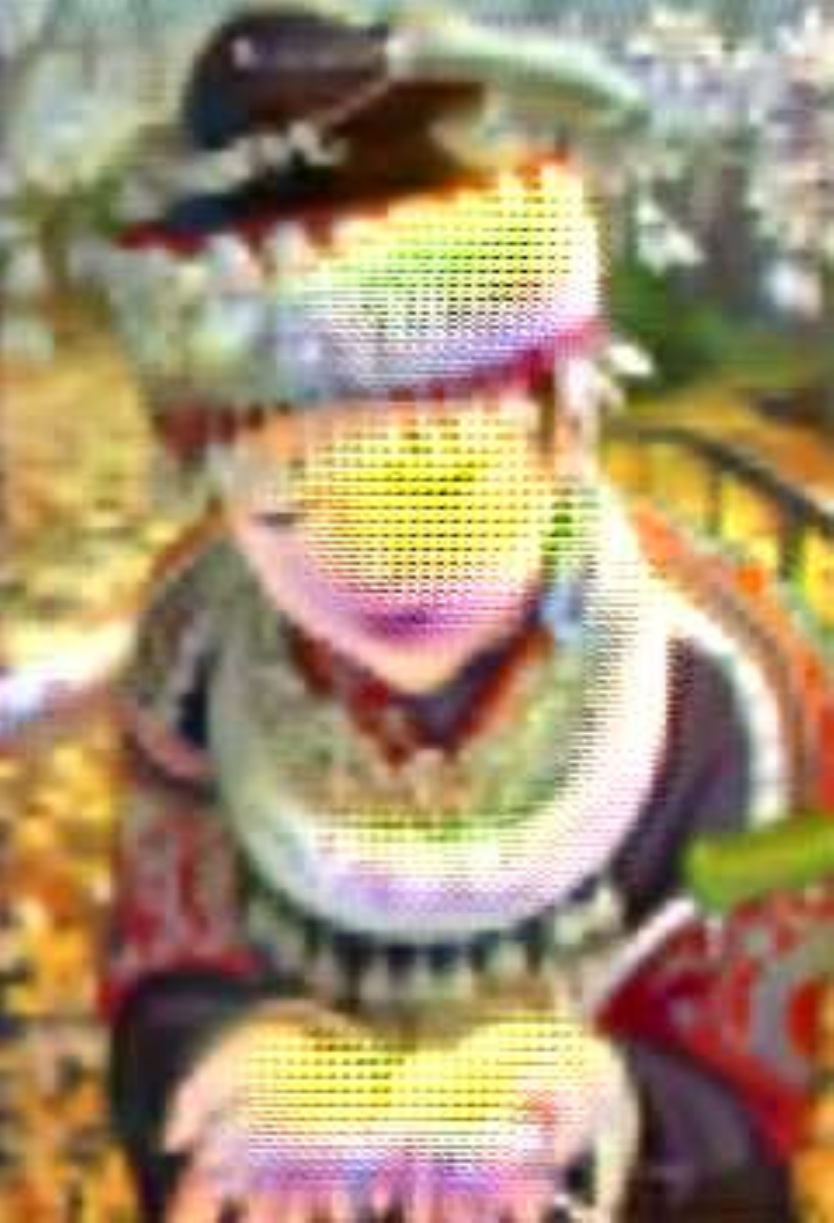}
      \includegraphics[keepaspectratio=true, scale=0.19]{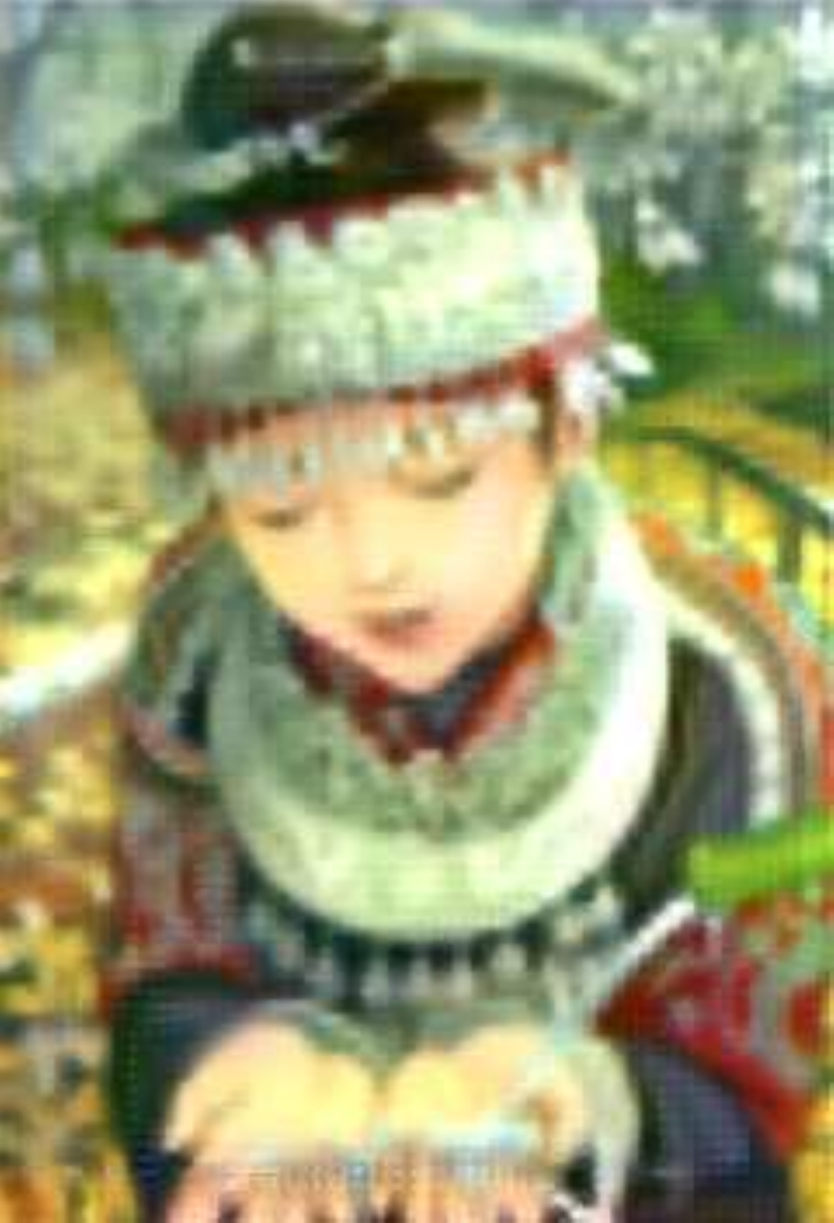}
      \includegraphics[keepaspectratio=true, scale=0.19]{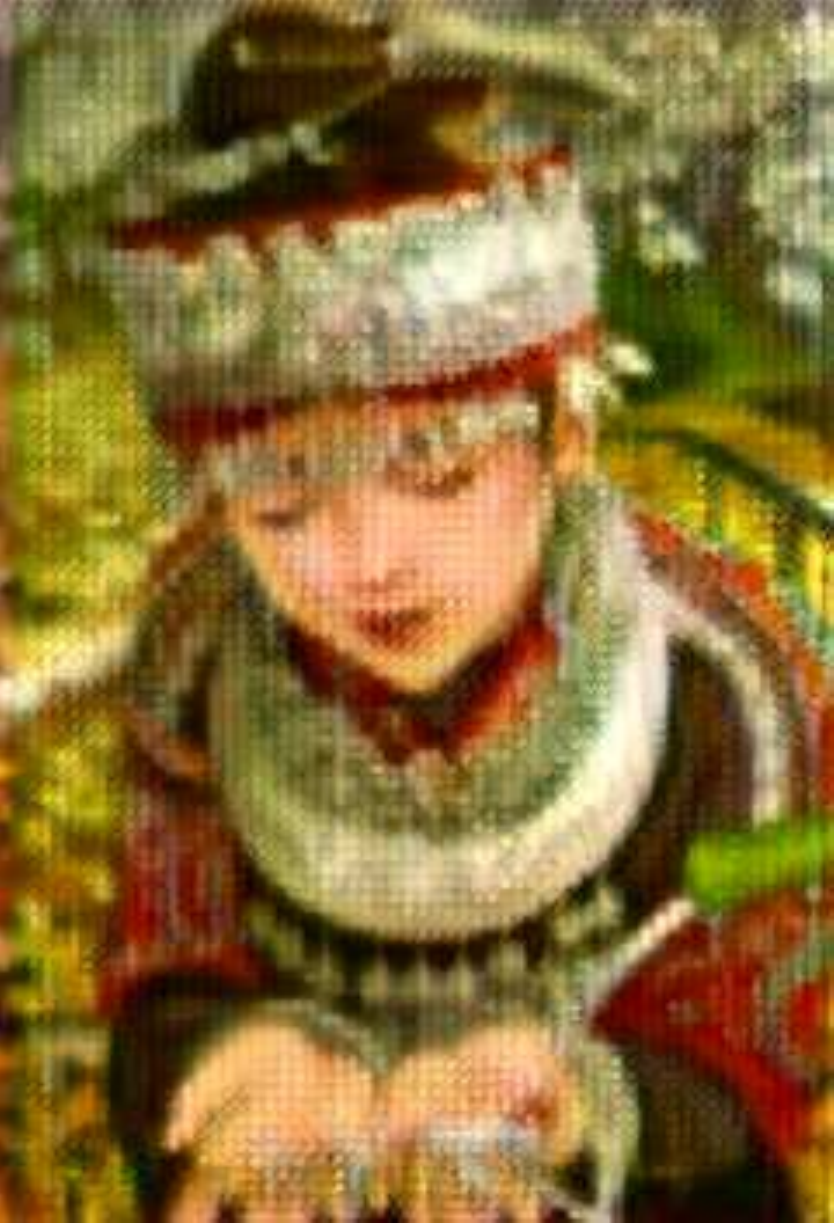}
      \includegraphics[keepaspectratio=true, scale=0.19]{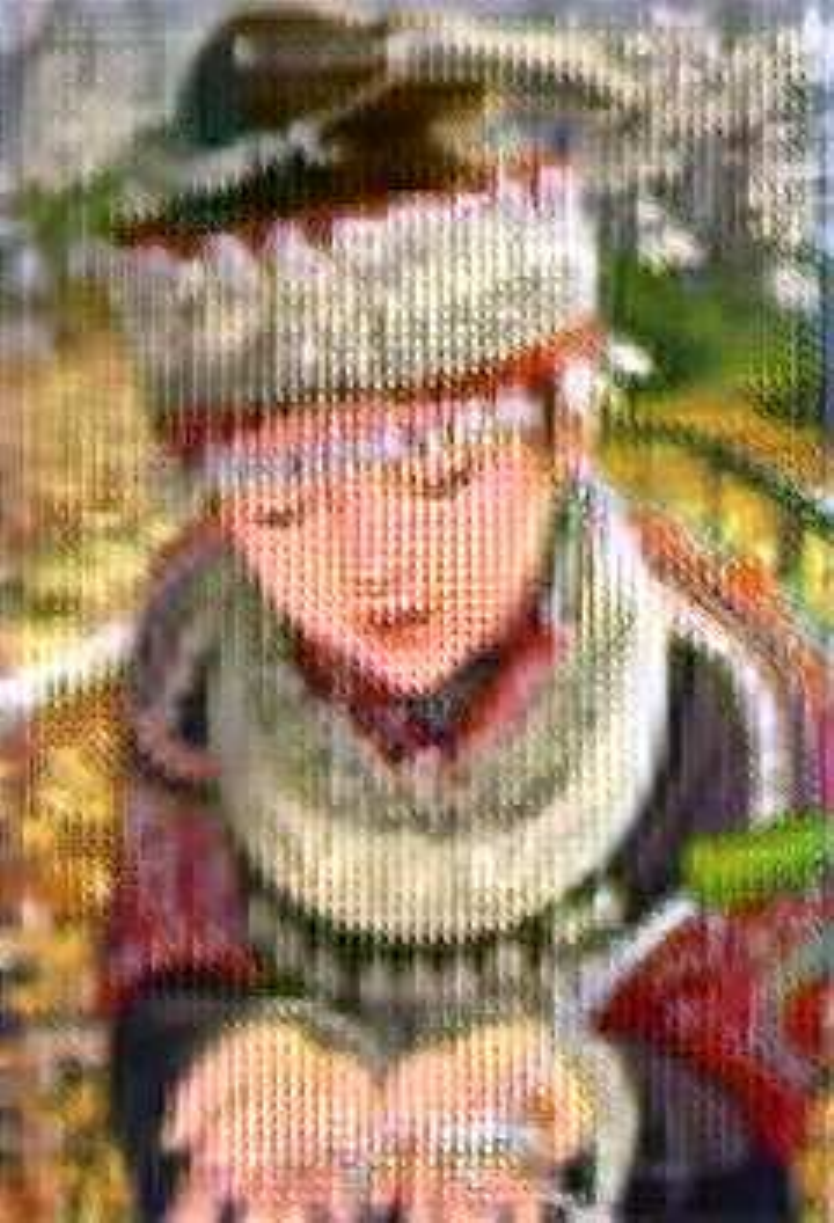}
      \includegraphics[keepaspectratio=true, scale=0.19]{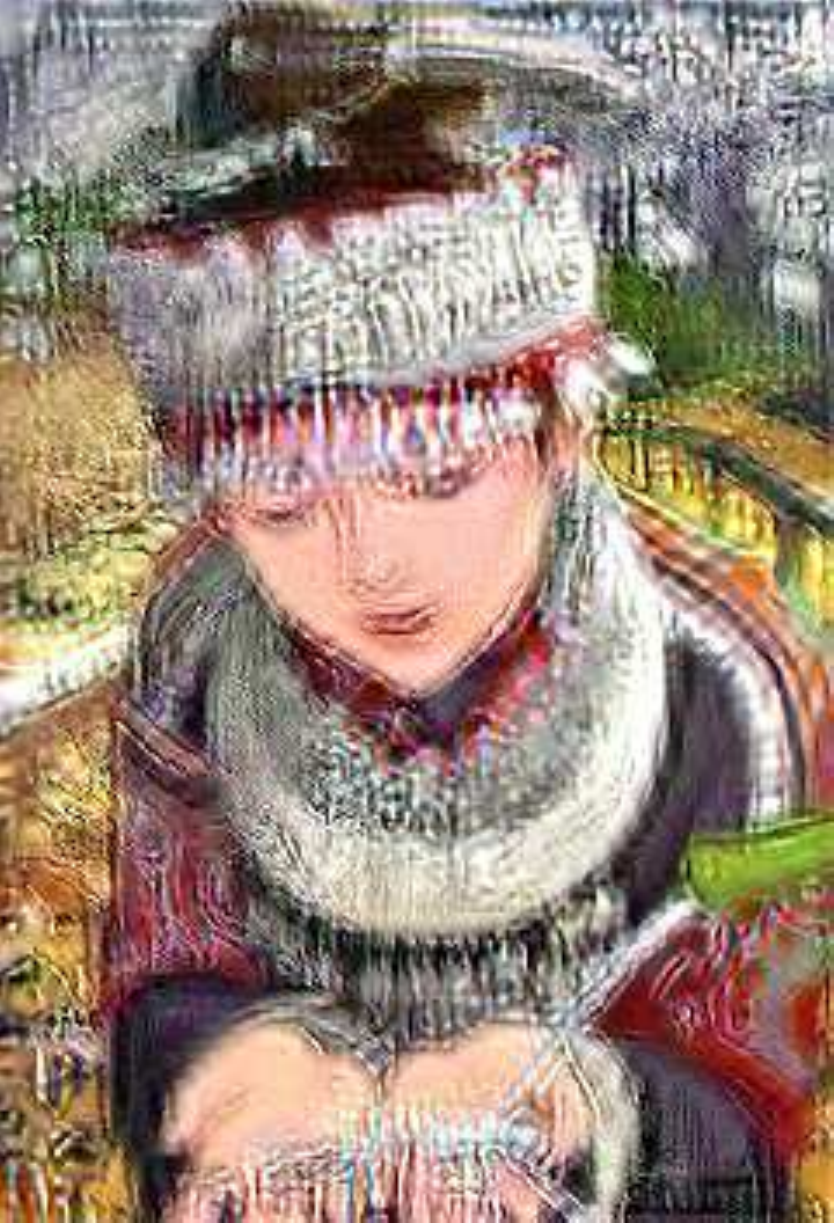}
      \caption{Comic}
    \end{subfigure}
    \begin{subfigure}{\linewidth}
      \centering
      \includegraphics[keepaspectratio=true, scale=0.095]{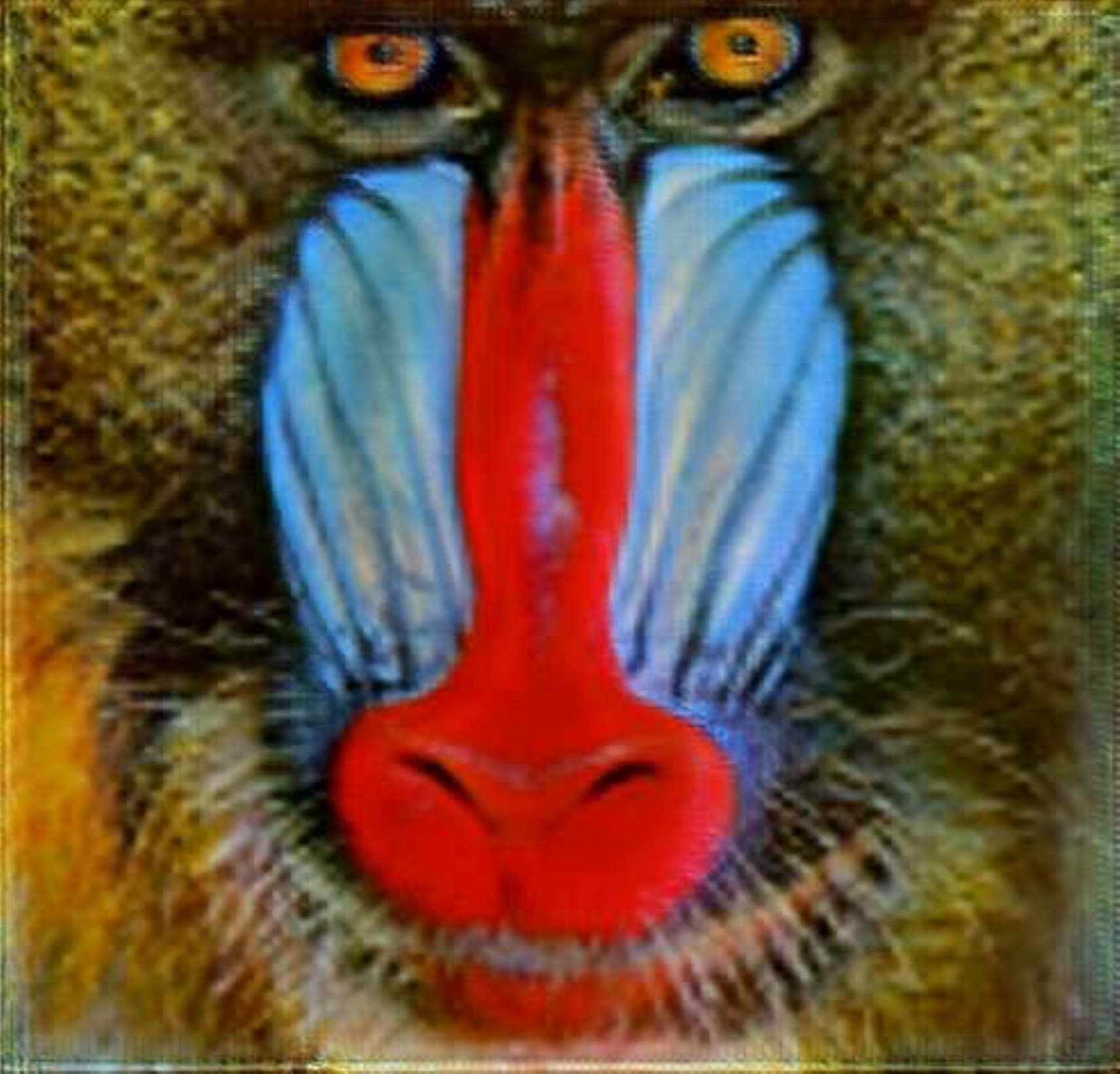}
      \includegraphics[keepaspectratio=true, scale=0.095]{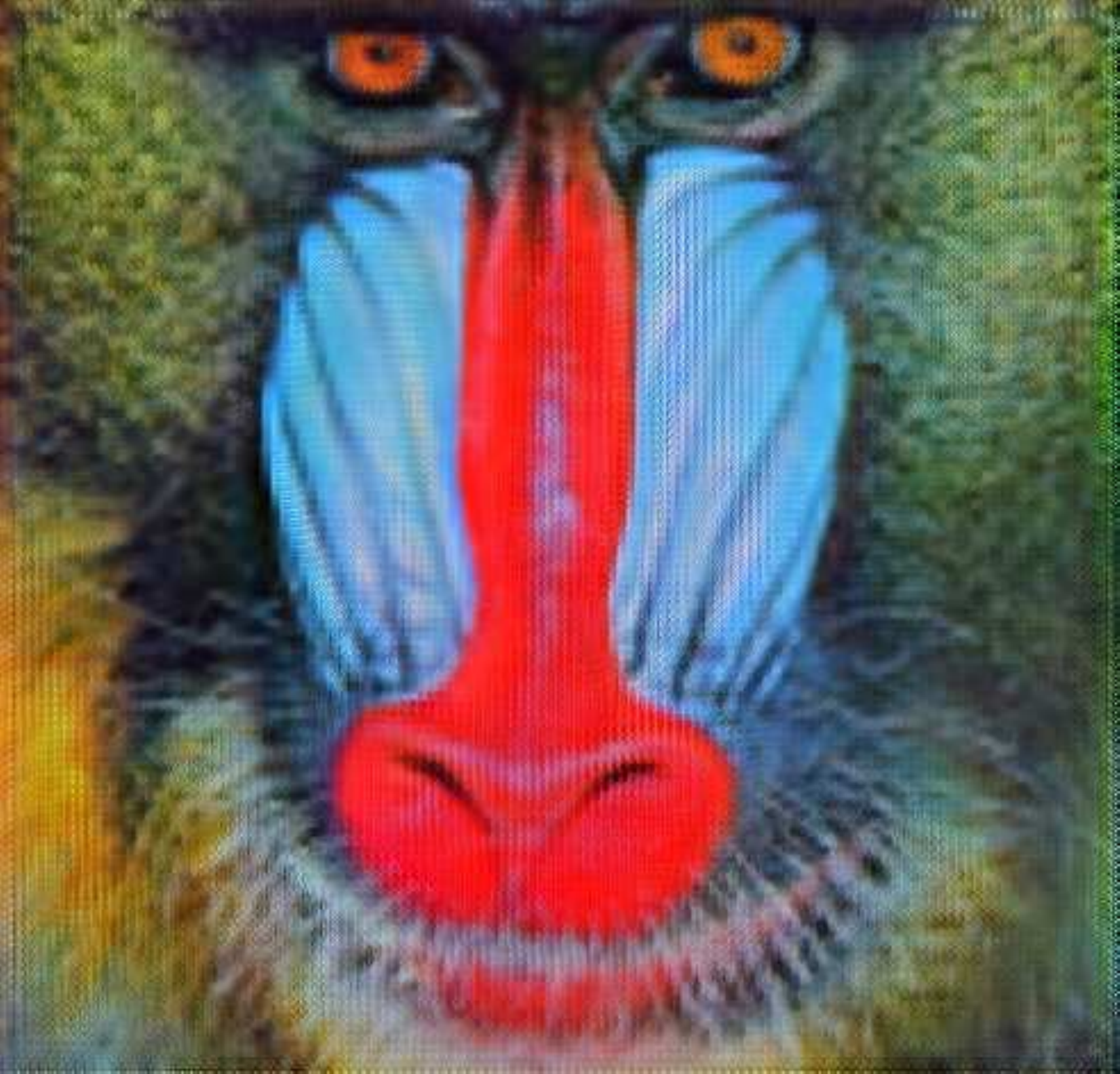}
      \includegraphics[keepaspectratio=true, scale=0.095]{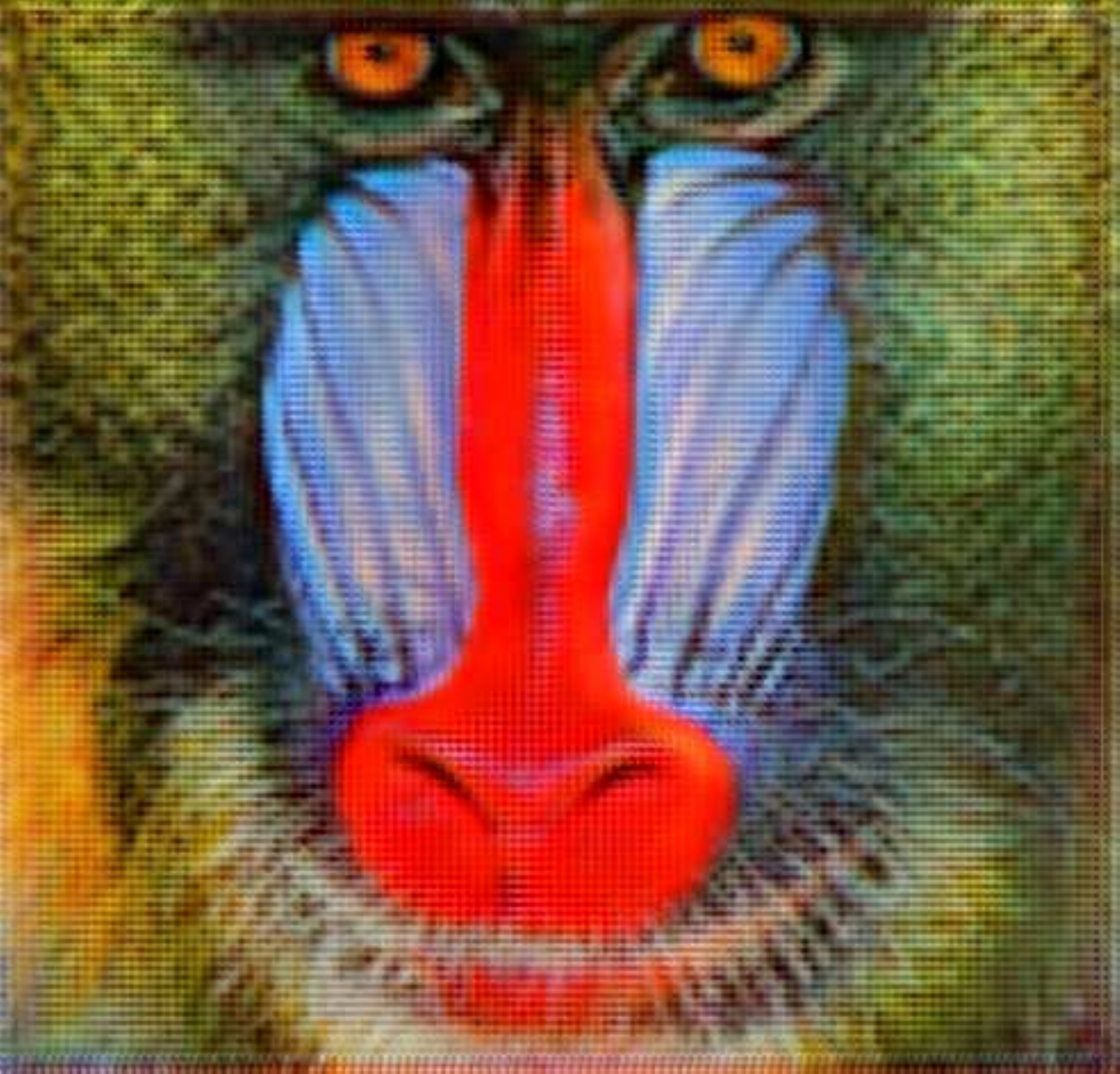}
      \includegraphics[keepaspectratio=true, scale=0.095]{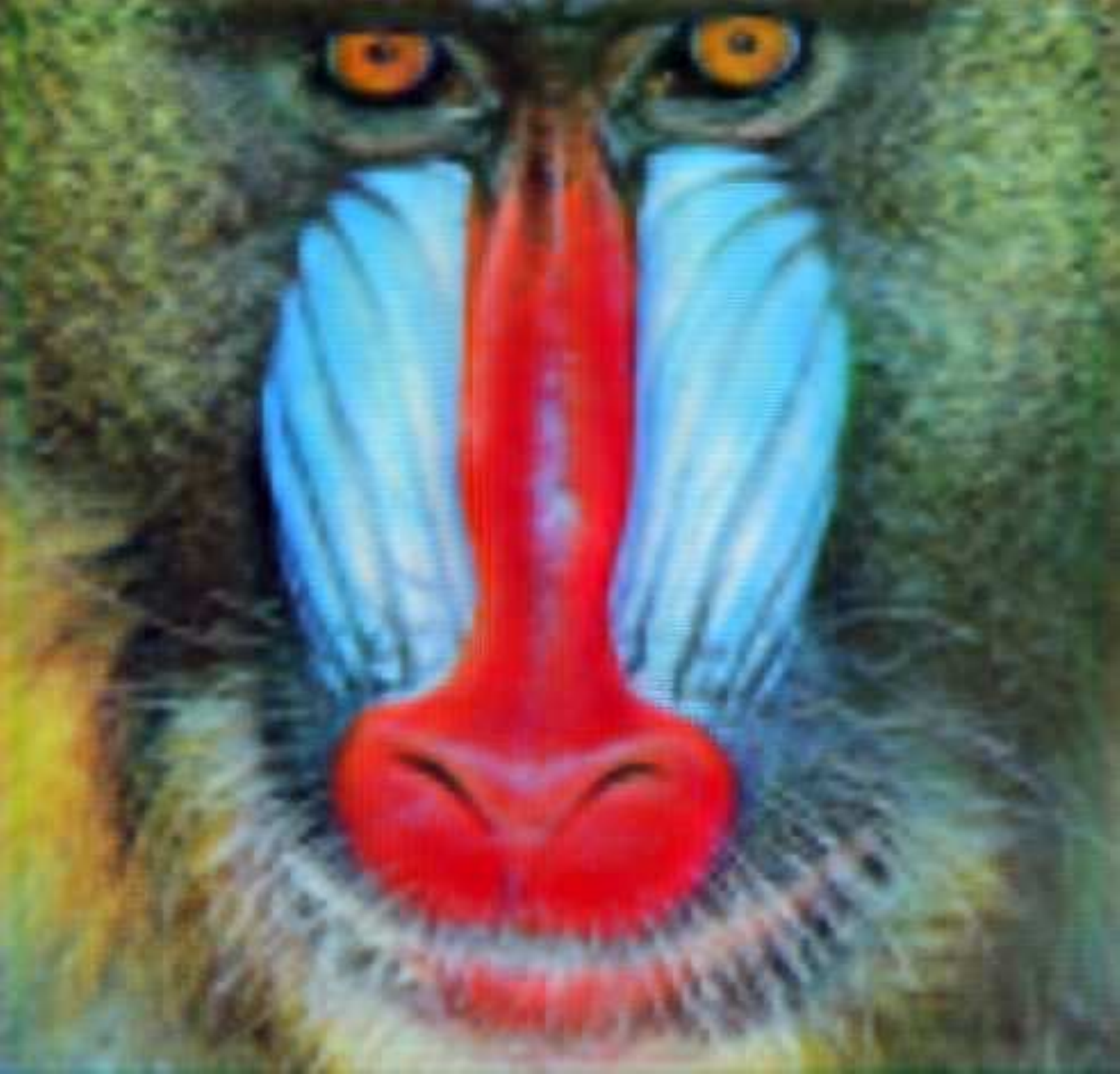}
      \includegraphics[keepaspectratio=true, scale=0.095]{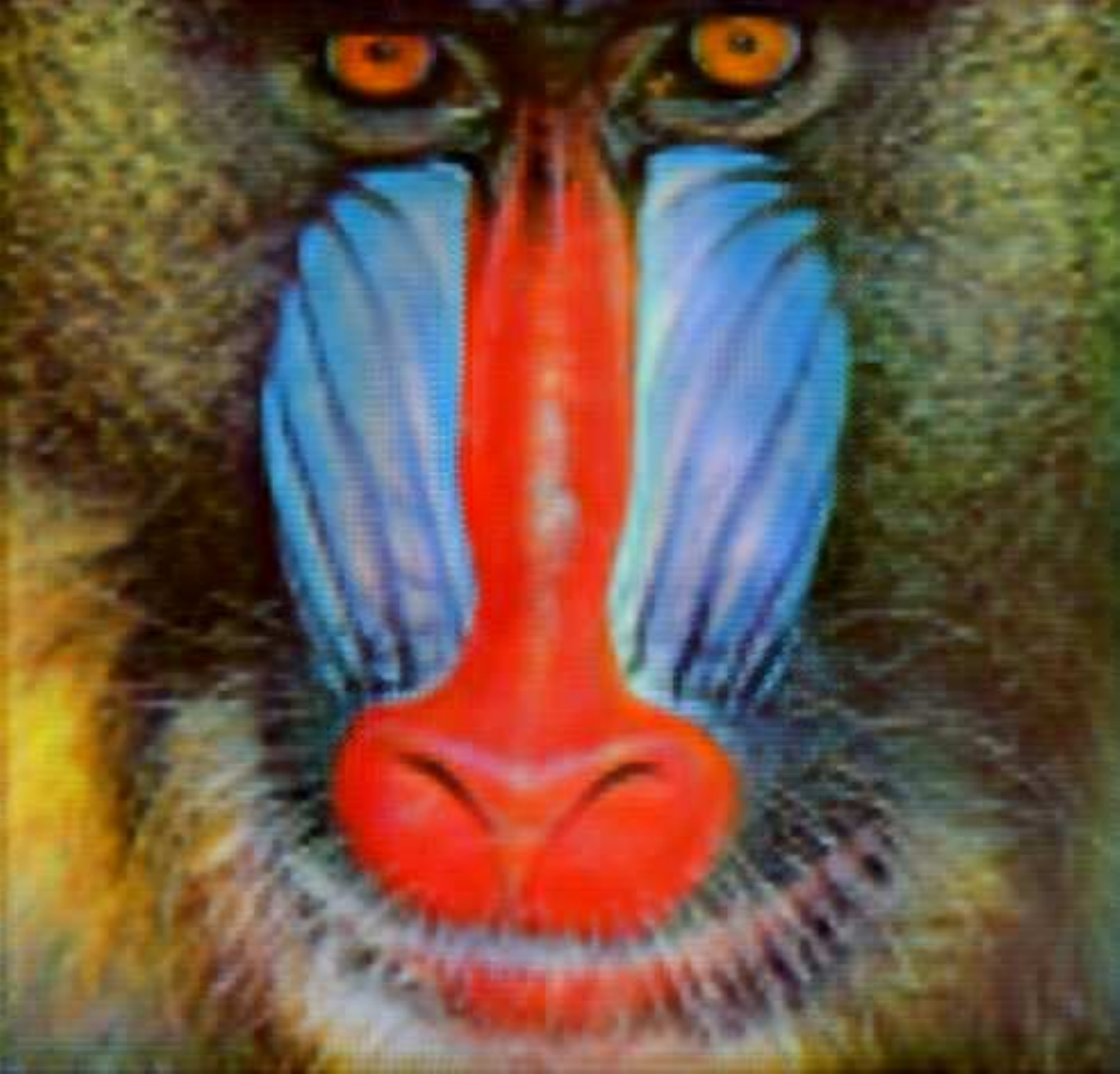}
      \includegraphics[keepaspectratio=true, scale=0.095]{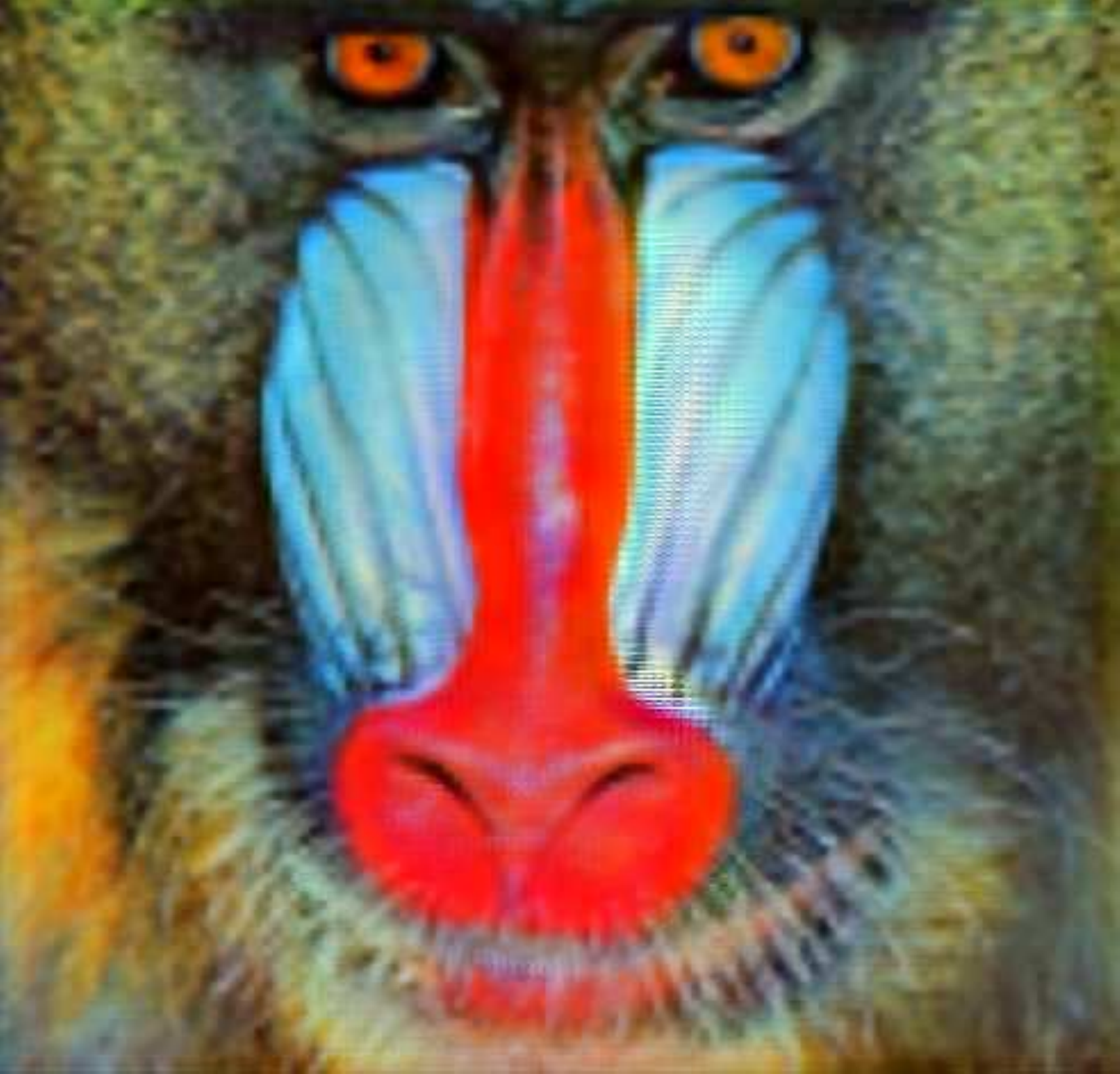}
      \includegraphics[keepaspectratio=true, scale=0.095]{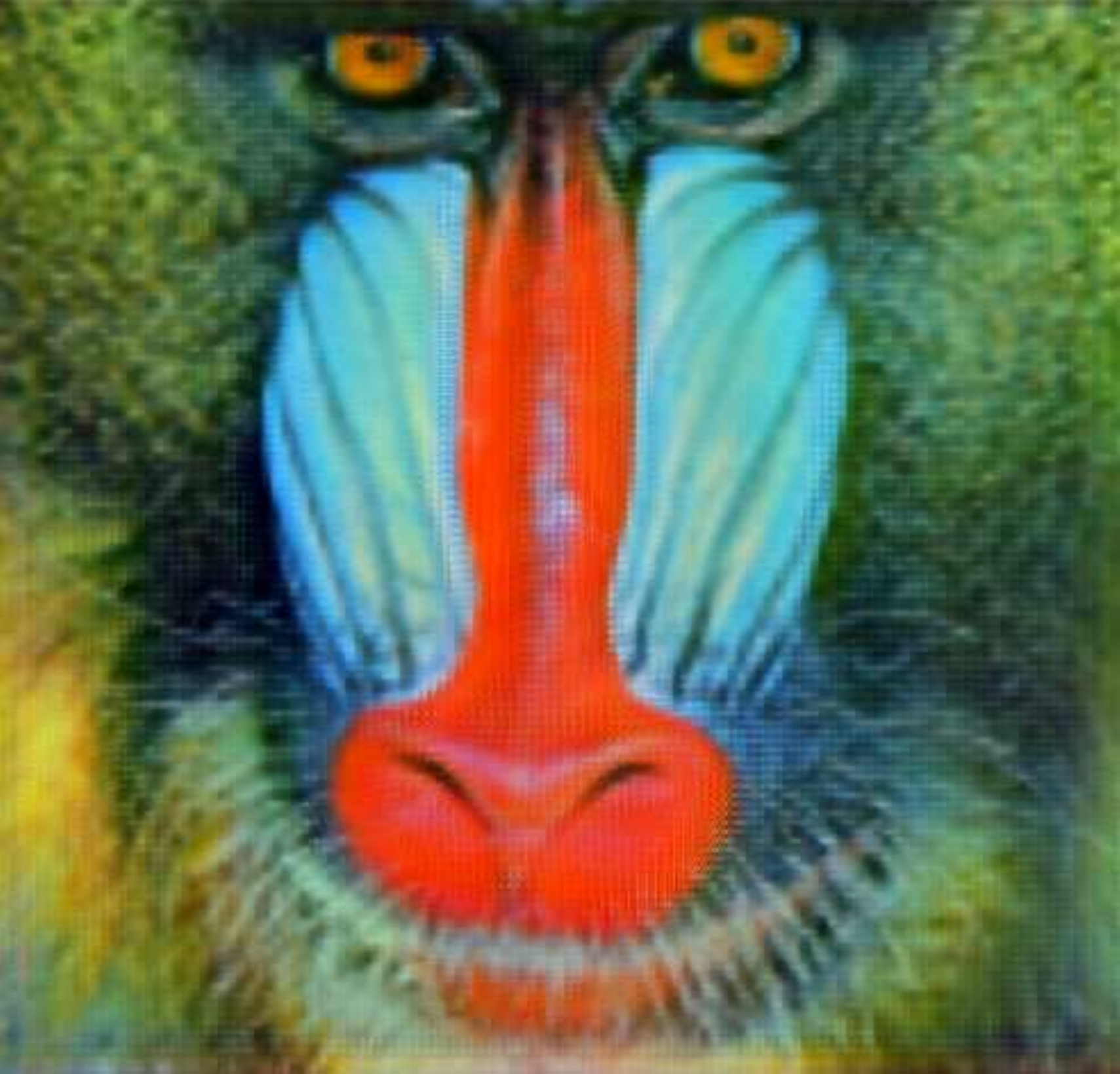}
      \includegraphics[keepaspectratio=true, scale=0.095]{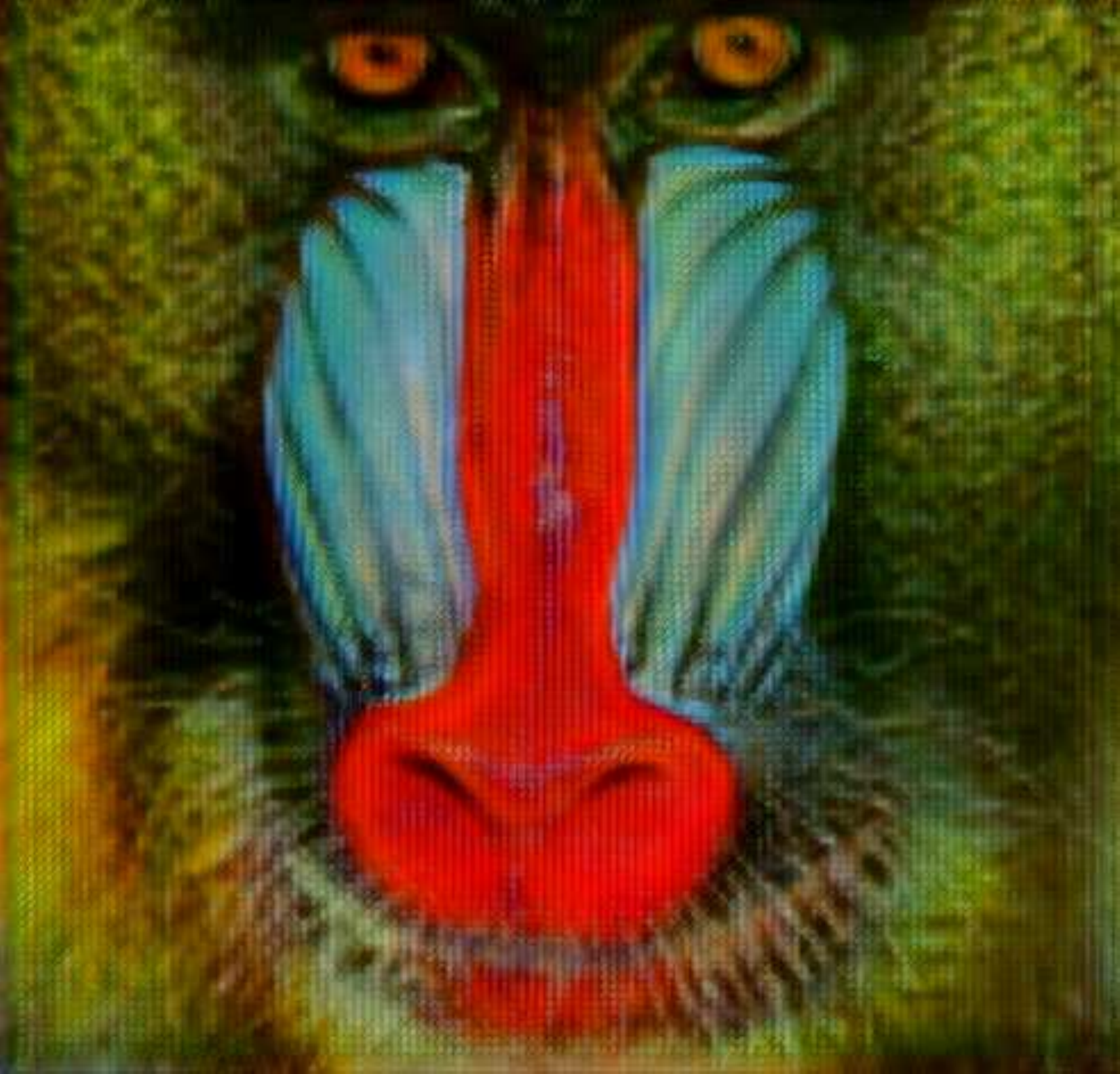}
      \includegraphics[keepaspectratio=true, scale=0.095]{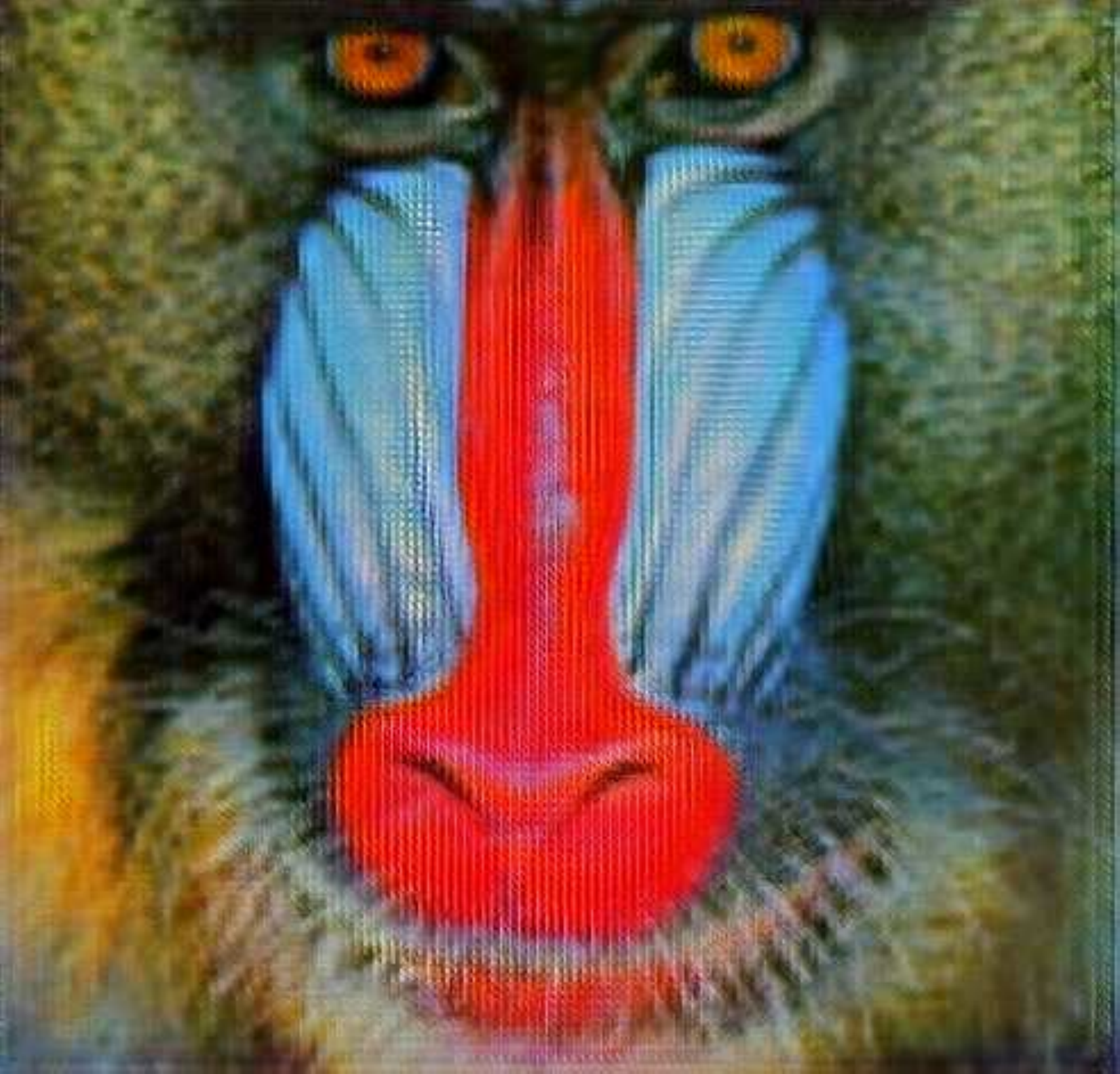}
      \includegraphics[keepaspectratio=true, scale=0.095]{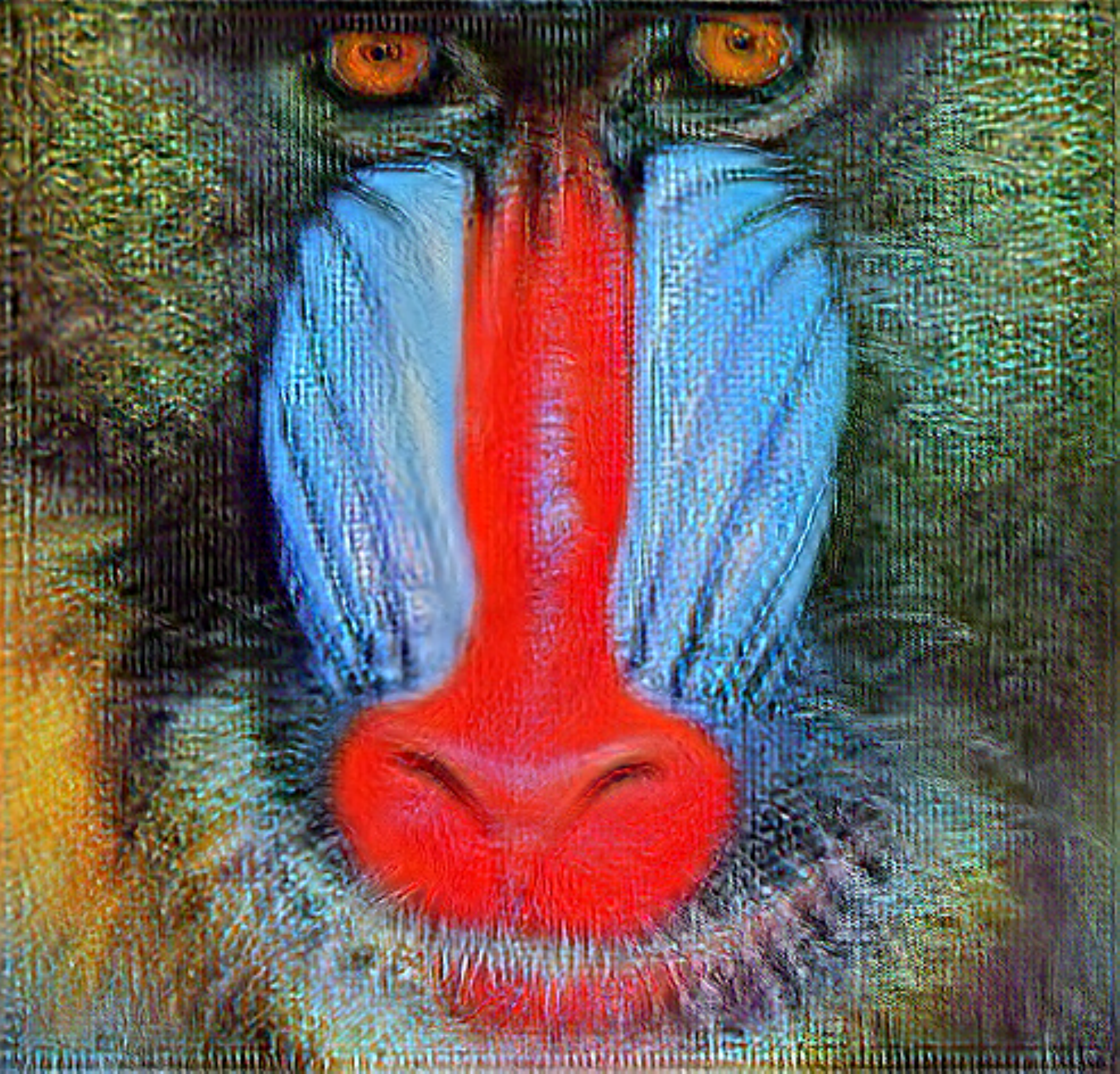}
      \caption{Baboon}
    \end{subfigure}
    \begin{subfigure}{\linewidth}
      \centering
      \includegraphics[keepaspectratio=true, scale=0.09]{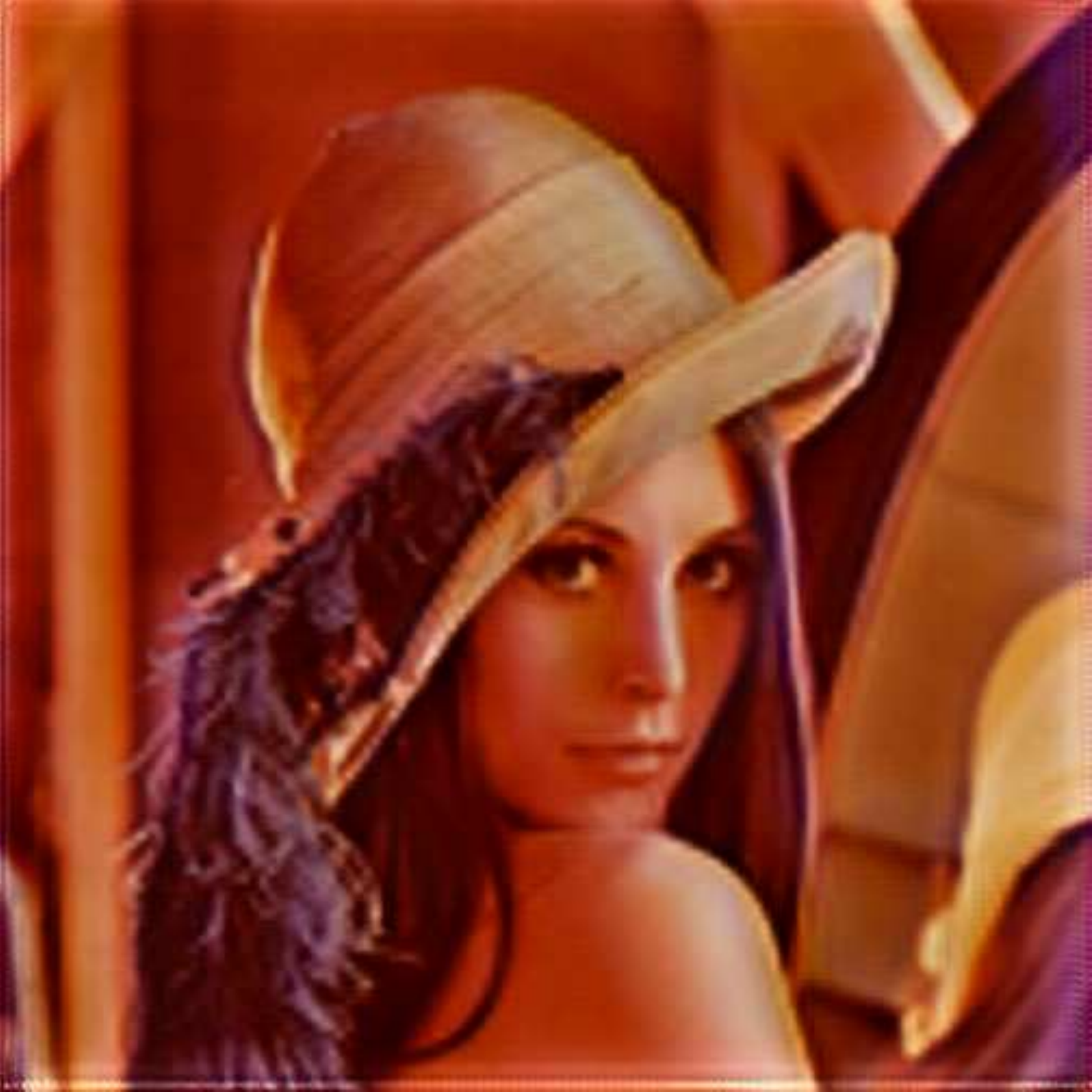}
      \includegraphics[keepaspectratio=true, scale=0.09]{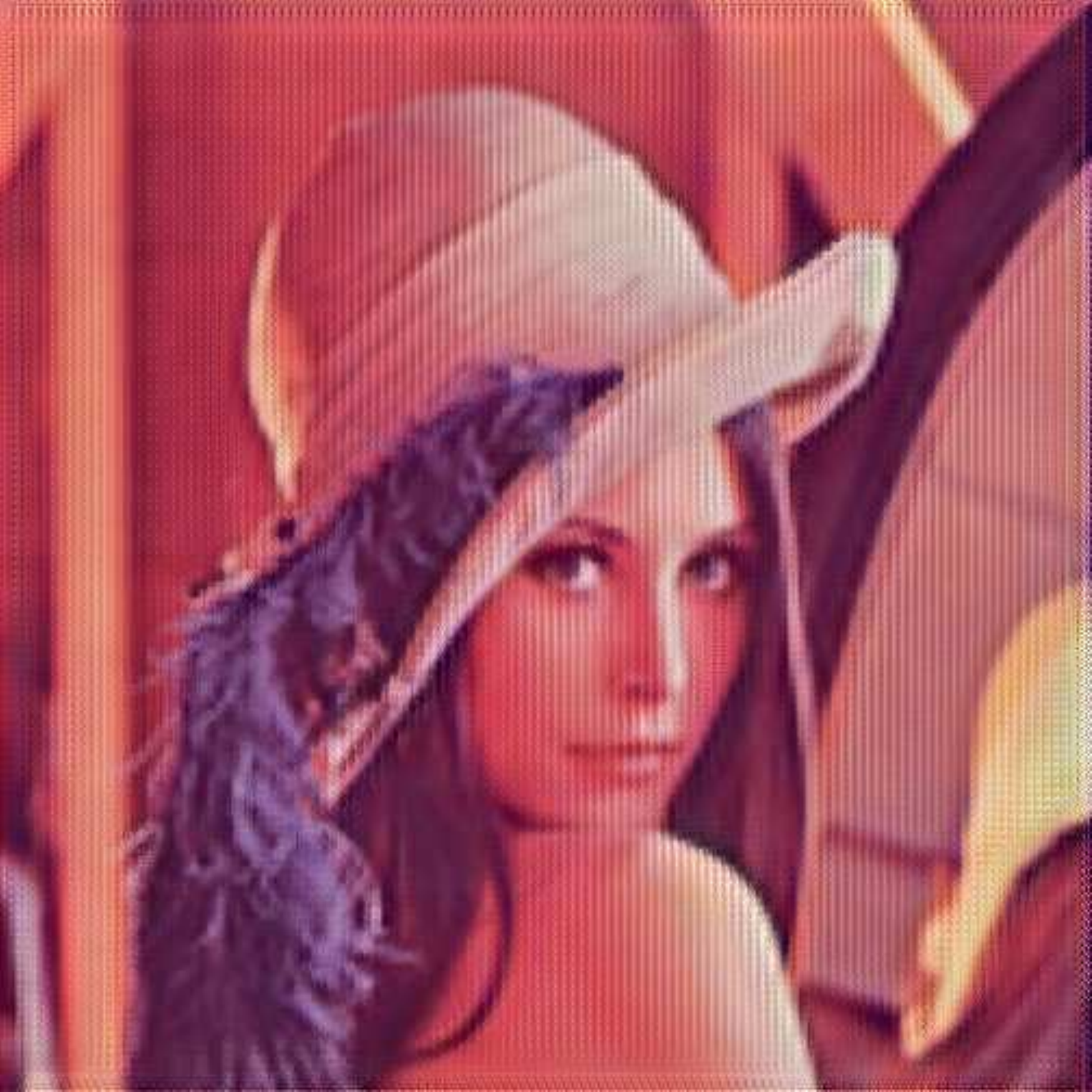}
      \includegraphics[keepaspectratio=true, scale=0.09]{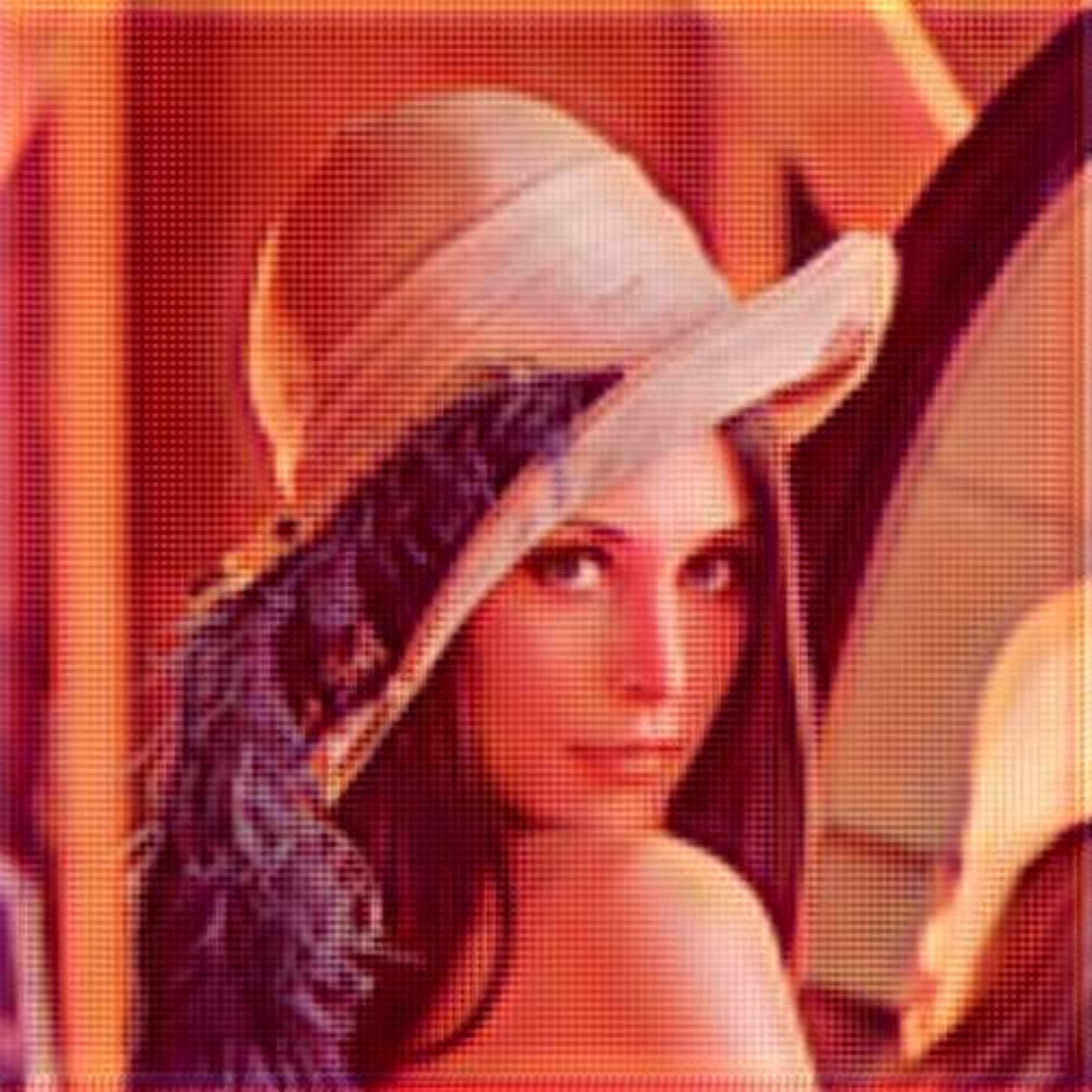}
      \includegraphics[keepaspectratio=true, scale=0.09]{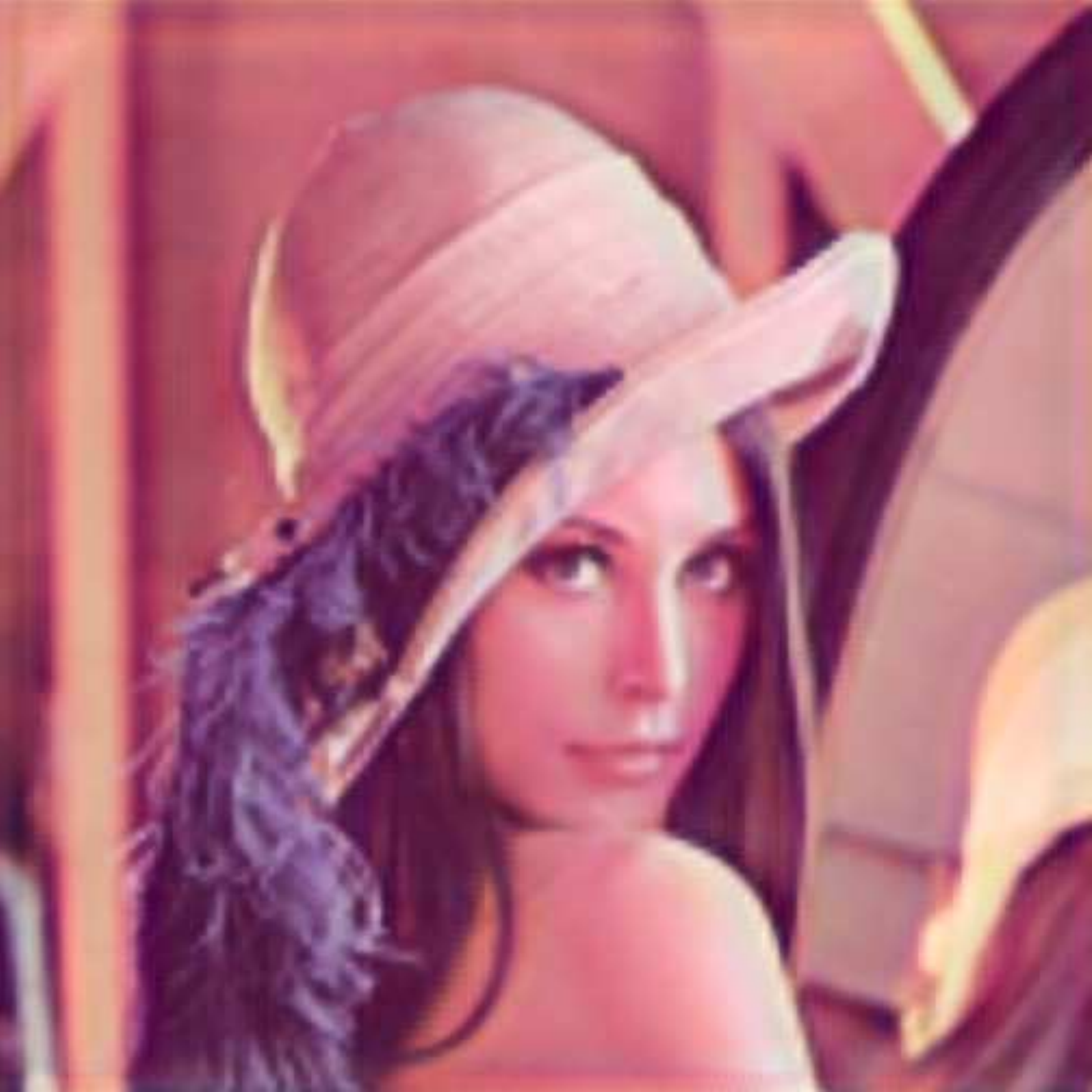}
      \includegraphics[keepaspectratio=true, scale=0.09]{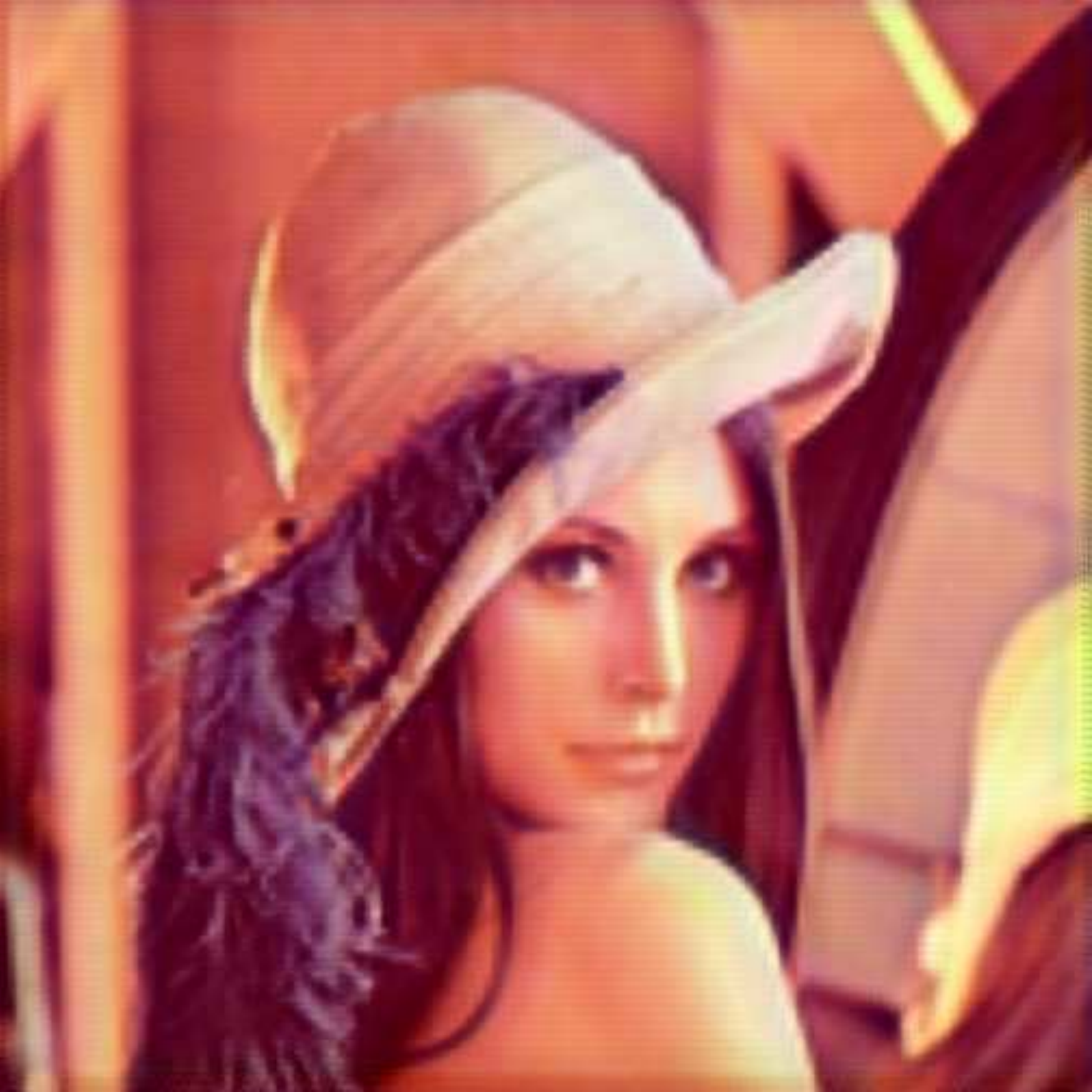}
      \includegraphics[keepaspectratio=true, scale=0.09]{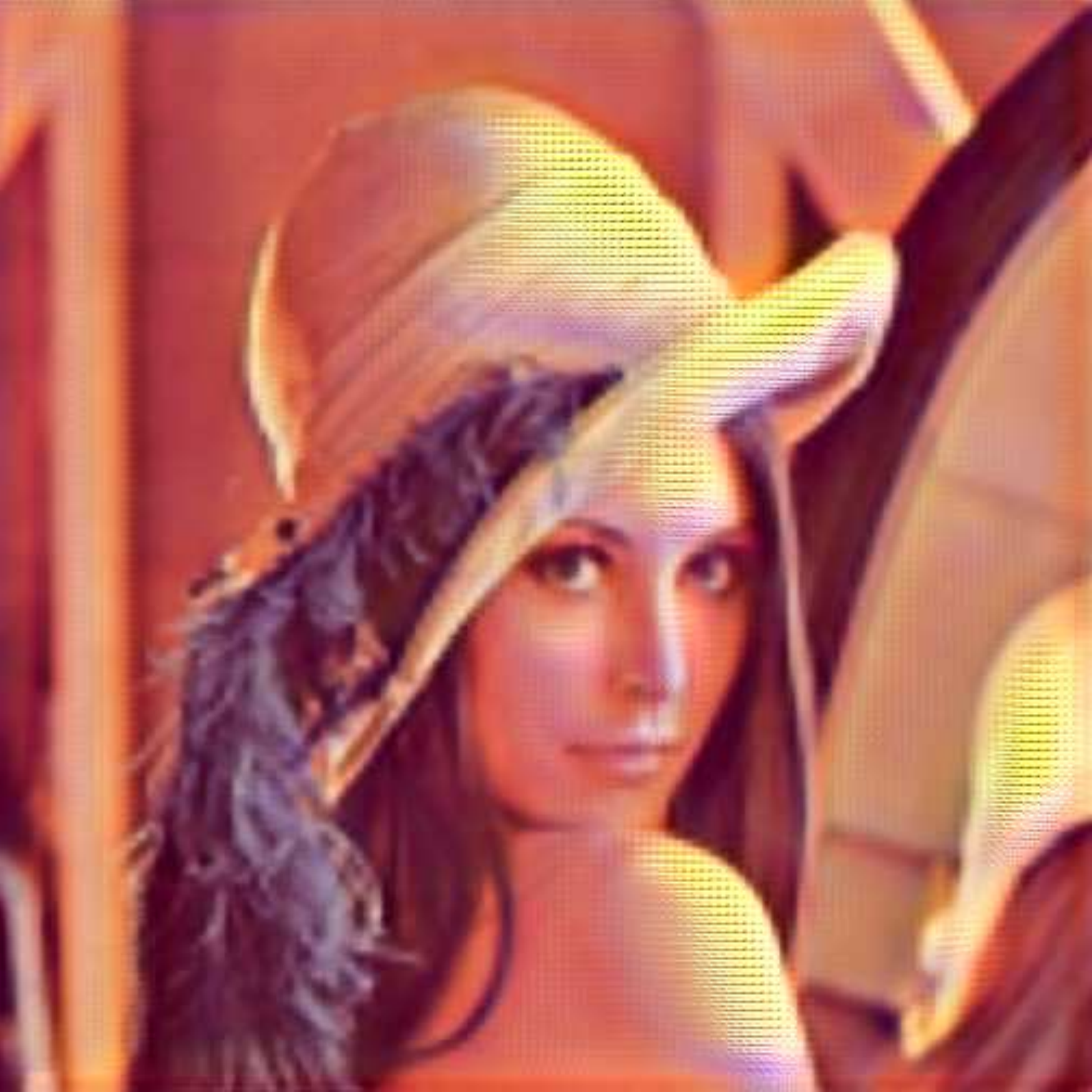}
      \includegraphics[keepaspectratio=true, scale=0.09]{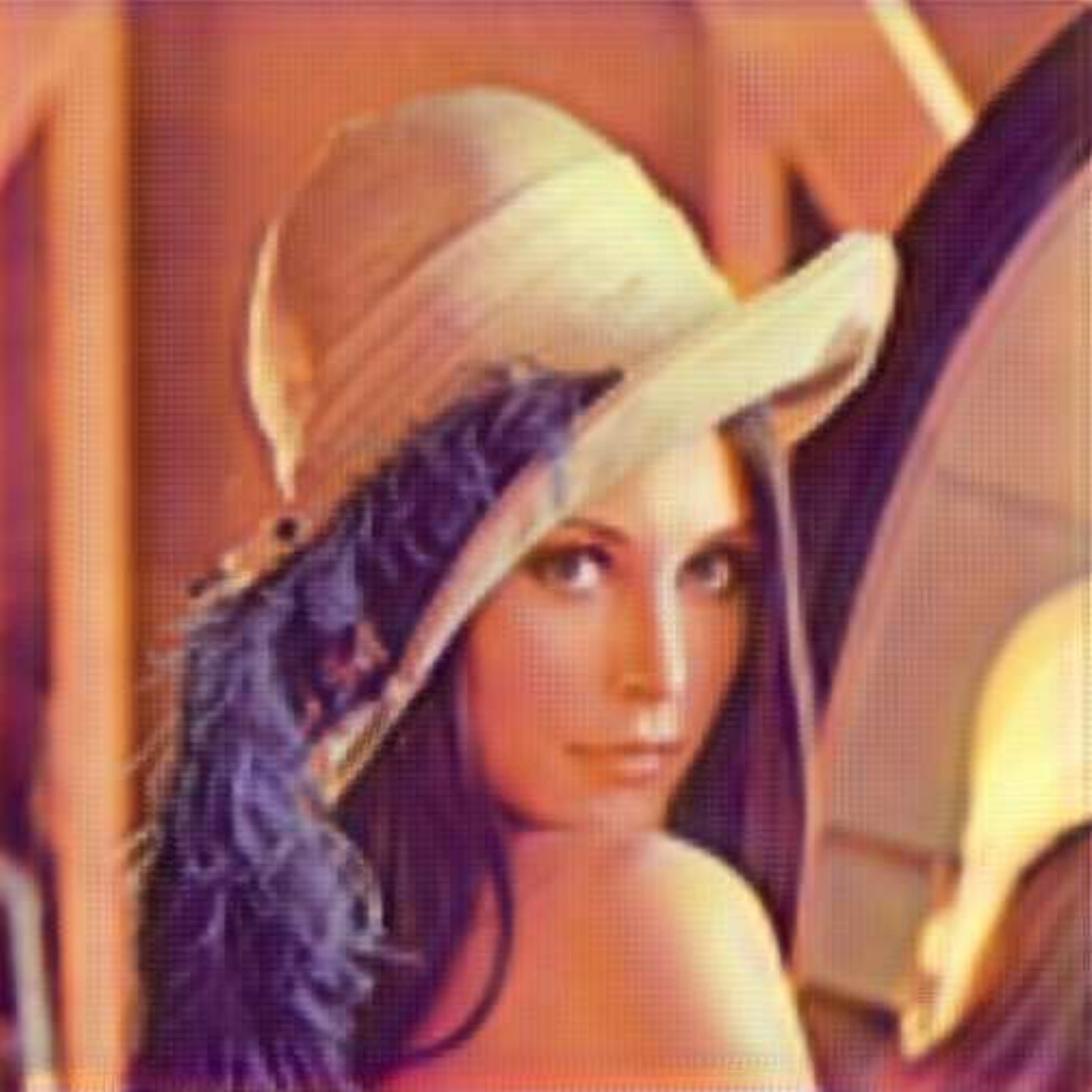}
      \includegraphics[keepaspectratio=true, scale=0.09]{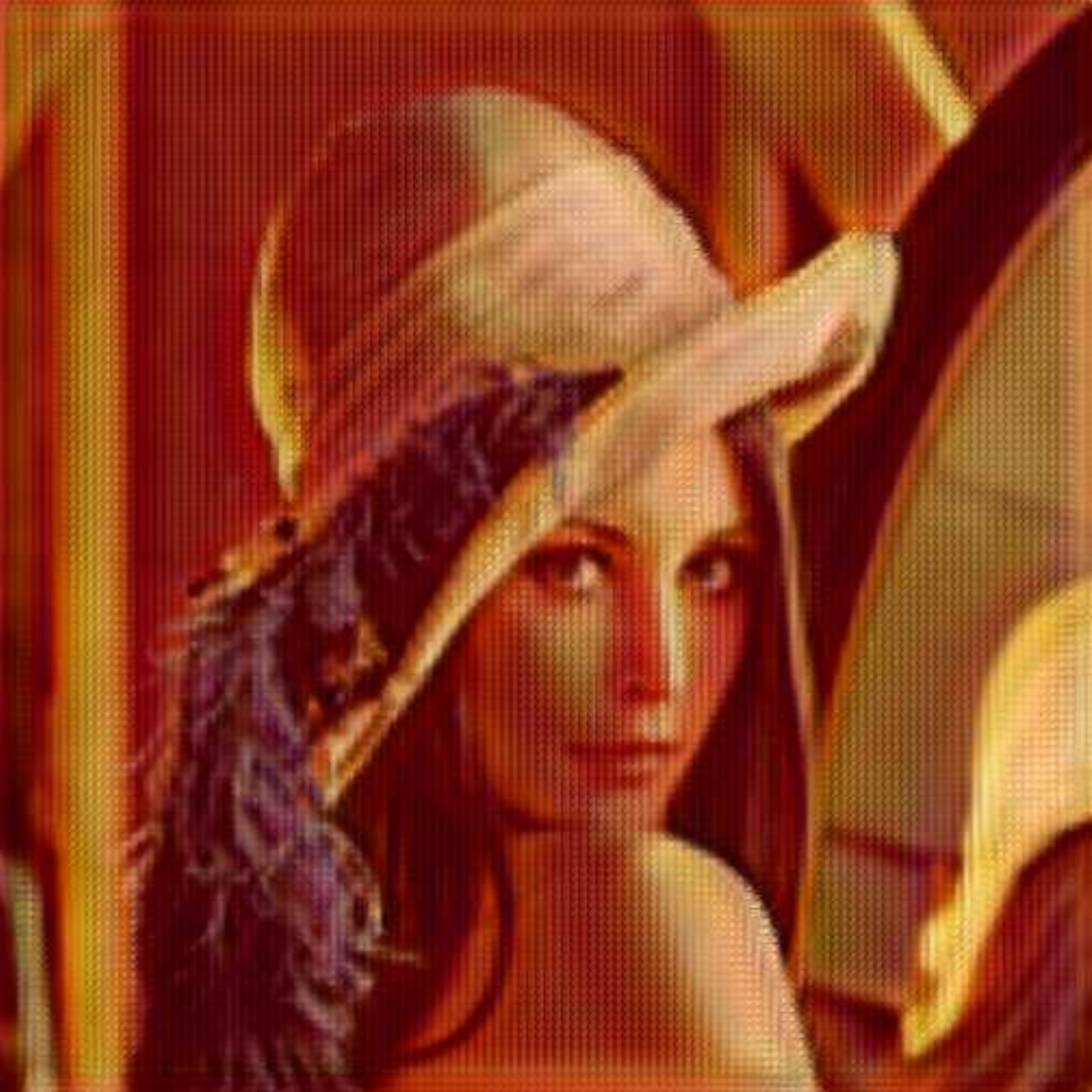}
      \includegraphics[keepaspectratio=true, scale=0.09]{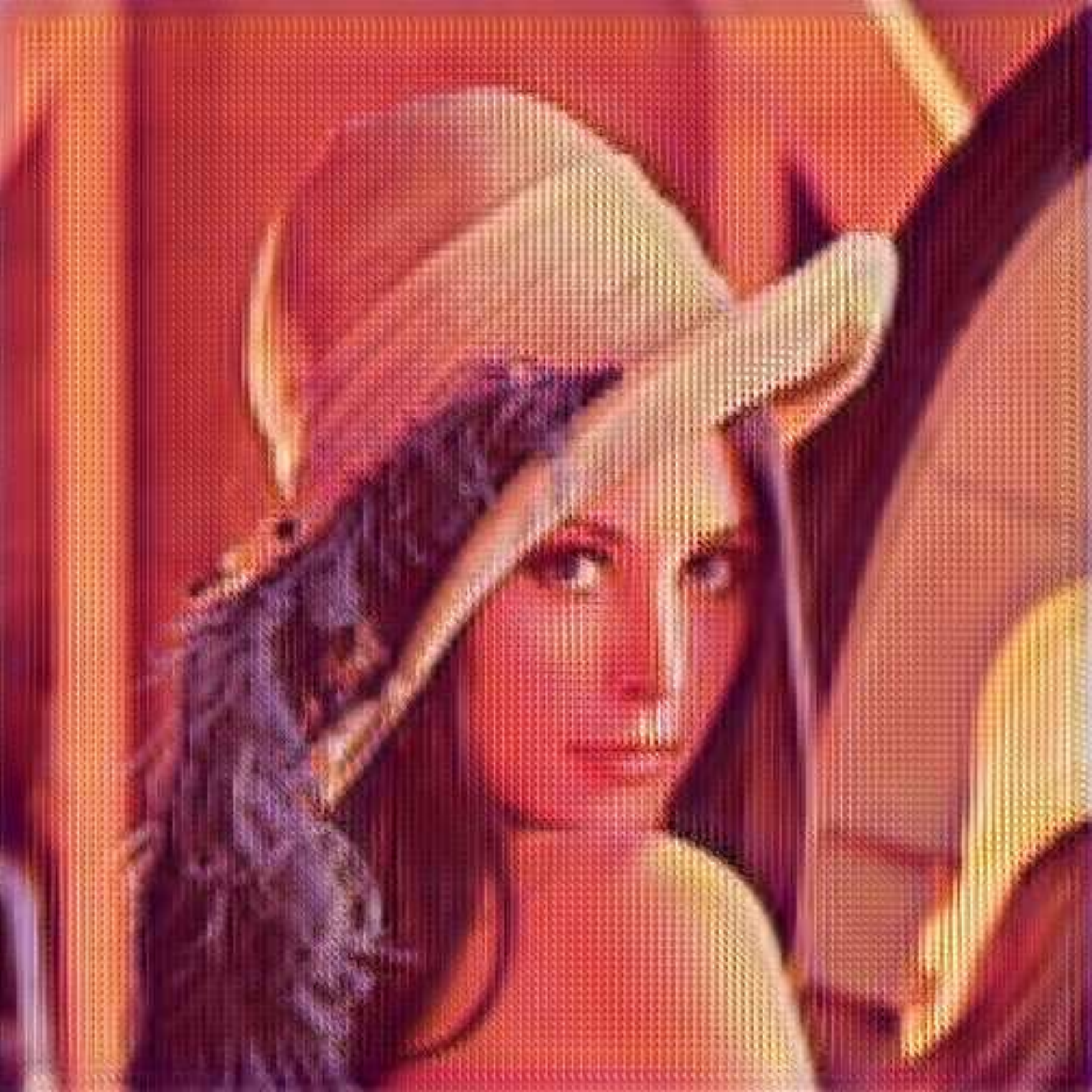}
      \includegraphics[keepaspectratio=true, scale=0.09]{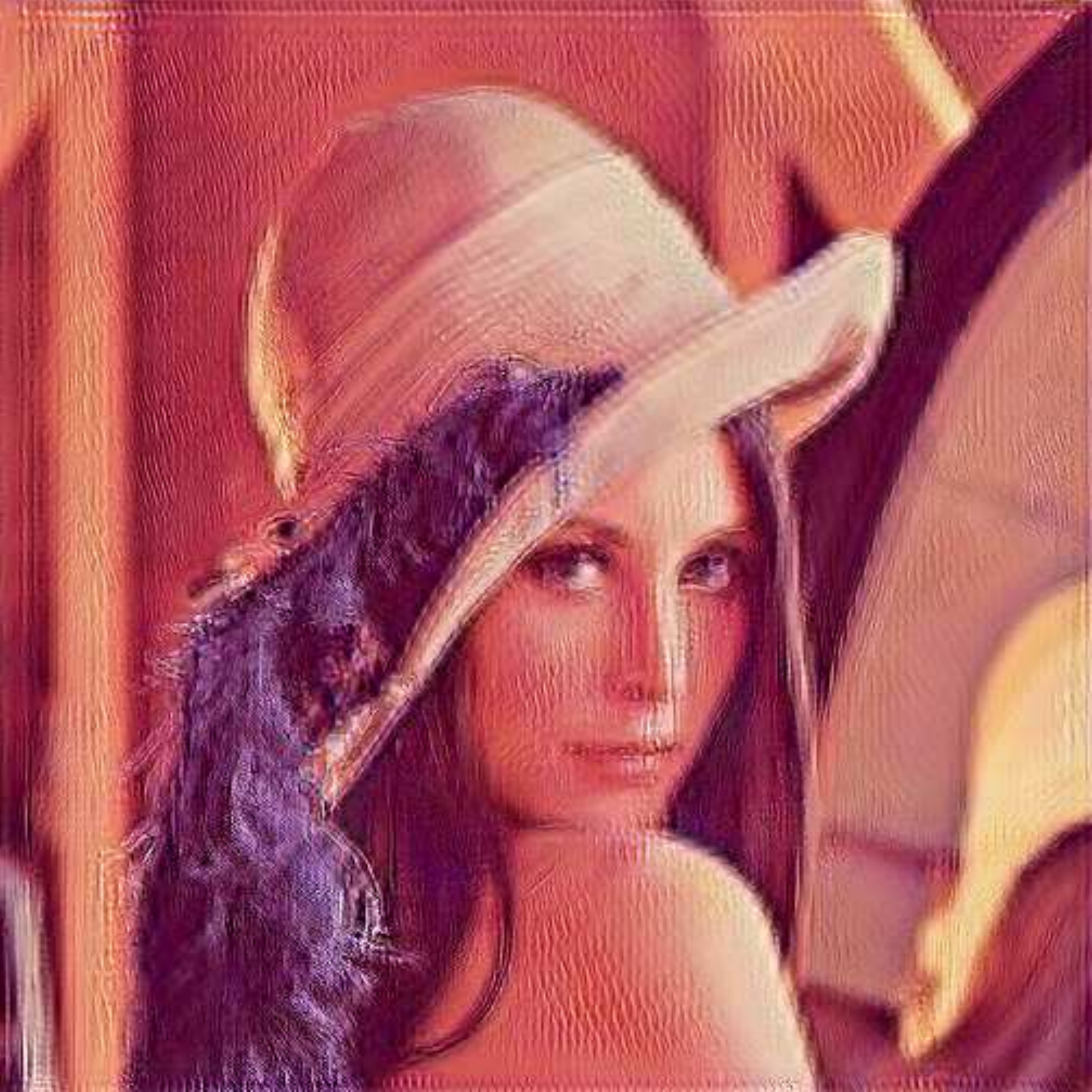}
      \caption{Lenna}
    \end{subfigure}
    \caption{Ambiguity attack - SRGAN: It can be seen that the quality of the images drop significantly when the sign of SRGAN$_{ws}$ is being modified. Left to right: The amount (from 10\% to 100\%) of the sign is being modified.}
    \label{fig:flip_sign_srgan}
  \end{figure*}
  
  \begin{figure*}[t]
    \centering
    \begin{subfigure}{0.33\linewidth}
      \centering
      \includegraphics[keepaspectratio=true, scale=0.95]{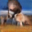}
      \includegraphics[keepaspectratio=true, scale=0.95]{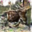}
      \includegraphics[keepaspectratio=true, scale=0.95]{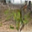}
      \includegraphics[keepaspectratio=true, scale=0.95]{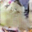}
      \includegraphics[keepaspectratio=true, scale=0.95]{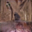}
      \\
      \includegraphics[keepaspectratio=true, scale=0.95]{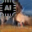}
      \includegraphics[keepaspectratio=true, scale=0.95]{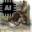}
      \includegraphics[keepaspectratio=true, scale=0.95]{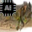}
      \includegraphics[keepaspectratio=true, scale=0.95]{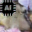}
     \includegraphics[keepaspectratio=true, scale=0.95]{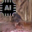}
      \caption{$\lambda = 0.1$}
    \end{subfigure}
    \begin{subfigure}{0.33\linewidth}
      \centering
      \includegraphics[keepaspectratio=true, scale=0.95]{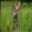}
      \includegraphics[keepaspectratio=true, scale=0.95]{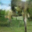}
      \includegraphics[keepaspectratio=true, scale=0.95]{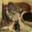}
      \includegraphics[keepaspectratio=true, scale=0.95]{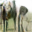}
      \includegraphics[keepaspectratio=true, scale=0.95]{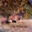}
      \\
      \includegraphics[keepaspectratio=true, scale=0.95]{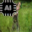}
      \includegraphics[keepaspectratio=true, scale=0.95]{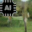}
      \includegraphics[keepaspectratio=true, scale=0.95]{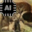}
      \includegraphics[keepaspectratio=true, scale=0.95]{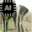}
     \includegraphics[keepaspectratio=true, scale=0.95]{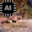}
      \caption{$\lambda = 0.5$}
    \end{subfigure}
        \begin{subfigure}{0.33\linewidth}
      \centering
      \includegraphics[keepaspectratio=true, scale=0.95]{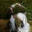}
      \includegraphics[keepaspectratio=true, scale=0.95]{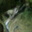}
      \includegraphics[keepaspectratio=true, scale=0.95]{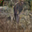}
      \includegraphics[keepaspectratio=true, scale=0.95]{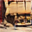}
      \includegraphics[keepaspectratio=true, scale=0.95]{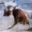}
      \\
      \includegraphics[keepaspectratio=true, scale=0.95]{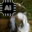}
      \includegraphics[keepaspectratio=true, scale=0.95]{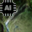}
      \includegraphics[keepaspectratio=true, scale=0.95]{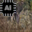}
      \includegraphics[keepaspectratio=true, scale=0.95]{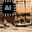}
     \includegraphics[keepaspectratio=true, scale=0.95]{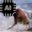}
      \caption{$\lambda = 1.0$}
    \end{subfigure}
            \begin{subfigure}{0.45\linewidth}
      \centering
      \includegraphics[keepaspectratio=true, scale=0.95]{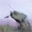}
      \includegraphics[keepaspectratio=true, scale=0.95]{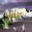}
      \includegraphics[keepaspectratio=true, scale=0.95]{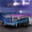}
      \includegraphics[keepaspectratio=true, scale=0.95]{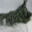}
      \includegraphics[keepaspectratio=true, scale=0.95]{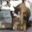}
      \\
      \includegraphics[keepaspectratio=true, scale=0.95]{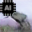}
      \includegraphics[keepaspectratio=true, scale=0.95]{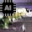}
      \includegraphics[keepaspectratio=true, scale=0.95]{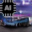}
      \includegraphics[keepaspectratio=true, scale=0.95]{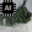}
     \includegraphics[keepaspectratio=true, scale=0.95]{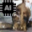}
      \caption{$\lambda = 5.0$}
    \end{subfigure}  
        \begin{subfigure}{0.45\linewidth}
      \centering
      \includegraphics[keepaspectratio=true, scale=0.95]{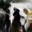}
      \includegraphics[keepaspectratio=true, scale=0.95]{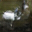}
      \includegraphics[keepaspectratio=true, scale=0.95]{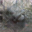}
      \includegraphics[keepaspectratio=true, scale=0.95]{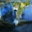}
      \includegraphics[keepaspectratio=true, scale=0.95]{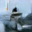}
      \\
      \includegraphics[keepaspectratio=true, scale=0.95]{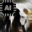}
      \includegraphics[keepaspectratio=true, scale=0.95]{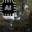}
      \includegraphics[keepaspectratio=true, scale=0.95]{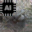}
      \includegraphics[keepaspectratio=true, scale=0.95]{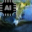}
     \includegraphics[keepaspectratio=true, scale=0.95]{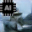}
      \caption{$\lambda = 10.0$}
    \end{subfigure}
    \caption{Effects of different $\lambda$ to original model performance (top) and quality of generated watermark (bottom).}
    \label{fig:lambda}
  \end{figure*}

\subsection{Appendix VI - Robustness against removal attack}

{\bf Fine-tuning.} Fig. \ref{fig:6} shows the qualitative results for Section 4.5, line 698 that our proposed method is robust against fine-tuning attack as highlighted in Table 7. We can clearly see that the watermark (on the top left corner) remains intact after fine-tuning.

\subsection{Resilience against ambiguity attack}

This section shows the full results of applying sign loss (Eq. 11 in the main paper) to embed a signature into BN-scale, $\gamma^{BN}$. The implementation details is given in Section 3.2 of the main paper. Herein, we show the example of how the unique key  "EXAMPLE" is embedded into our DCGAN's batch normalization weight. Table \ref{table:sign-loss-example} shows how to decode the trained scale, $\gamma^{BN}$ to retrieve the signature embedded. Also, please note that even that there are 2 "E", their $\gamma^{BN}$ are different from each other. 

Fig. \ref{fig:flip_sign_dcgan} - \ref{fig:flip_sign_srgan} are the complete results to complement Figure 9 - 10 in Section 4.6. It can be clearly visualize from Fig. \ref{fig:flip_sign_srgan} that the quality of the generated SR-images is very poor where obvious artefact can be observed even the signature signs are modified at only 10\%. With this, we can deduce that the scale signs (Eq. 11) enforced in this way remain rather persistent against ambiguity attacks.
  
\section{Appendix VII - Ablation Study}  
\subsection{Coefficient $\lambda$.}

In this section, qualitatively, we illustrate in Fig. \ref{fig:lambda} the effects of different $\lambda$ settings (\ie~from 0.1 $\to$ 10) on the original GAN model performance against the quality of the generated watermark (this is similar to Table 10) with CIFAR10 dataset. From visual inspection on Fig. \ref{fig:lambda}, it is hard to deduce that which $\lambda$ is an ideal choice. In this paper, based on the quantitative results (FID vs. SSIM) in Table 10, we set $\lambda = 1.0$.
  
\subsection{$n$ vs. $c$.}

In this section, qualitatively, we illustrate in Fig. \ref{fig:n_c} - \ref{fig:n_c1} different $n$ and $c$ settings to understand the tradeoffs between the original GAN model performance against the quality of the generated watermark (this is similar to Table 11) with CIFAR10 dataset. From here, it can be noticed that it is, however, very hard to distinguish from a naked eye point of view which setting is having the best tradeoffs with the exception that it is very clear that setting $c = 0$ is not ideal. This is because from Fig. \ref{fig:n_c1}a-c, we can notice that the generated images are all almost black (\ie~for $n = 5, 10, 15$). This phenomenon happens because the training input of DCGAN has a normal distribution of $\mu = 0$, therefore it is conflicting with the trigger input which is also set as $c = 0$. 

As a summary, trigger input set must have a very different distribution from the training data. In this paper, based on the quantitative results (FID vs. SSIM) reported in Table 7, we conclude that setting $n = 5$ and $c = -10$ is the most ideal.

    \begin{figure*}[t]
    \centering
    \begin{subfigure}{0.33\linewidth}
      \centering
      \includegraphics[keepaspectratio=true, scale=0.95]{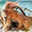}
      \includegraphics[keepaspectratio=true, scale=0.95]{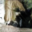}
      \includegraphics[keepaspectratio=true, scale=0.95]{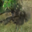}
      \includegraphics[keepaspectratio=true, scale=0.95]{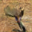}
      \includegraphics[keepaspectratio=true, scale=0.95]{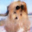}
      \\
      \includegraphics[keepaspectratio=true, scale=0.95]{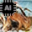}
      \includegraphics[keepaspectratio=true, scale=0.95]{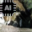}
      \includegraphics[keepaspectratio=true, scale=0.95]{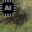}
      \includegraphics[keepaspectratio=true, scale=0.95]{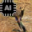}
     \includegraphics[keepaspectratio=true, scale=0.95]{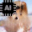}
      \caption{$n = 5$ ; $c = -10$}
    \end{subfigure}
    \begin{subfigure}{0.33\linewidth}
      \centering
      \includegraphics[keepaspectratio=true, scale=0.95]{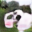}
      \includegraphics[keepaspectratio=true, scale=0.95]{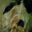}
      \includegraphics[keepaspectratio=true, scale=0.95]{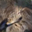}
      \includegraphics[keepaspectratio=true, scale=0.95]{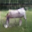}
      \includegraphics[keepaspectratio=true, scale=0.95]{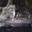}
      \\
      \includegraphics[keepaspectratio=true, scale=0.95]{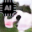}
      \includegraphics[keepaspectratio=true, scale=0.95]{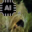}
      \includegraphics[keepaspectratio=true, scale=0.95]{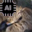}
      \includegraphics[keepaspectratio=true, scale=0.95]{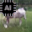}
     \includegraphics[keepaspectratio=true, scale=0.95]{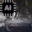}
      \caption{$n = 10$ ; $c = -10$}
    \end{subfigure}
        \begin{subfigure}{0.33\linewidth}
      \centering
      \includegraphics[keepaspectratio=true, scale=0.95]{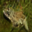}
      \includegraphics[keepaspectratio=true, scale=0.95]{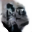}
      \includegraphics[keepaspectratio=true, scale=0.95]{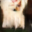}
      \includegraphics[keepaspectratio=true, scale=0.95]{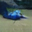}
      \includegraphics[keepaspectratio=true, scale=0.95]{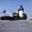}
      \\
      \includegraphics[keepaspectratio=true, scale=0.95]{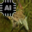}
      \includegraphics[keepaspectratio=true, scale=0.95]{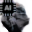}
      \includegraphics[keepaspectratio=true, scale=0.95]{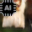}
      \includegraphics[keepaspectratio=true, scale=0.95]{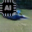}
     \includegraphics[keepaspectratio=true, scale=0.95]{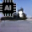}
      \caption{$n = 15$ ; $c = -10$}
    \end{subfigure}
        \begin{subfigure}{0.33\linewidth}
      \centering
      \includegraphics[keepaspectratio=true, scale=0.95]{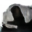}
      \includegraphics[keepaspectratio=true, scale=0.95]{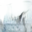}
      \includegraphics[keepaspectratio=true, scale=0.95]{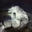}
      \includegraphics[keepaspectratio=true, scale=0.95]{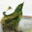}
      \includegraphics[keepaspectratio=true, scale=0.95]{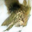}
      \\
      \includegraphics[keepaspectratio=true, scale=0.95]{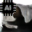}
      \includegraphics[keepaspectratio=true, scale=0.95]{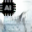}
      \includegraphics[keepaspectratio=true, scale=0.95]{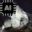}
      \includegraphics[keepaspectratio=true, scale=0.95]{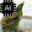}
     \includegraphics[keepaspectratio=true, scale=0.95]{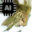}
      \caption{$n = 5$ ; $c = -5$}
    \end{subfigure}
    \begin{subfigure}{0.33\linewidth}
      \centering
      \includegraphics[keepaspectratio=true, scale=0.95]{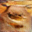}
      \includegraphics[keepaspectratio=true, scale=0.95]{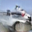}
      \includegraphics[keepaspectratio=true, scale=0.95]{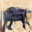}
      \includegraphics[keepaspectratio=true, scale=0.95]{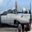}
      \includegraphics[keepaspectratio=true, scale=0.95]{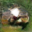}
      \\
      \includegraphics[keepaspectratio=true, scale=0.95]{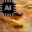}
      \includegraphics[keepaspectratio=true, scale=0.95]{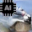}
      \includegraphics[keepaspectratio=true, scale=0.95]{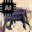}
      \includegraphics[keepaspectratio=true, scale=0.95]{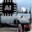}
     \includegraphics[keepaspectratio=true, scale=0.95]{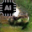}
      \caption{$n = 10$ ; $c = -5$}
    \end{subfigure}
        \begin{subfigure}{0.33\linewidth}
      \centering
      \includegraphics[keepaspectratio=true, scale=0.95]{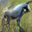}
      \includegraphics[keepaspectratio=true, scale=0.95]{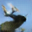}
      \includegraphics[keepaspectratio=true, scale=0.95]{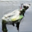}
      \includegraphics[keepaspectratio=true, scale=0.95]{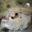}
      \includegraphics[keepaspectratio=true, scale=0.95]{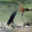}
      \\
      \includegraphics[keepaspectratio=true, scale=0.95]{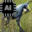}
      \includegraphics[keepaspectratio=true, scale=0.95]{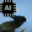}
      \includegraphics[keepaspectratio=true, scale=0.95]{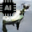}
      \includegraphics[keepaspectratio=true, scale=0.95]{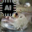}
     \includegraphics[keepaspectratio=true, scale=0.95]{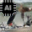}
      \caption{$n = 15$ ; $c = -5$}
    \end{subfigure}
    \caption{Effects of different $n$ and $c$ to original model performance (top) and quality of generated watermark (bottom) (cont. in Fig. \ref{fig:n_c1}).}
    \label{fig:n_c}
  \end{figure*}
 
    \begin{figure*}[t]
    \centering
                \begin{subfigure}{0.33\linewidth}
      \centering
      \includegraphics[keepaspectratio=true, scale=0.95]{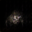}
      \includegraphics[keepaspectratio=true, scale=0.95]{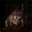}
      \includegraphics[keepaspectratio=true, scale=0.95]{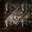}
      \includegraphics[keepaspectratio=true, scale=0.95]{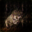}
      \includegraphics[keepaspectratio=true, scale=0.95]{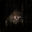}
      \\
      \includegraphics[keepaspectratio=true, scale=0.95]{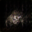}
      \includegraphics[keepaspectratio=true, scale=0.95]{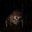}
      \includegraphics[keepaspectratio=true, scale=0.95]{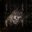}
      \includegraphics[keepaspectratio=true, scale=0.95]{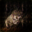}
     \includegraphics[keepaspectratio=true, scale=0.95]{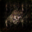}
      \caption{$n = 5$ ; $c = 0$}
    \end{subfigure}
    \begin{subfigure}{0.33\linewidth}
      \centering
      \includegraphics[keepaspectratio=true, scale=0.95]{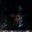}
      \includegraphics[keepaspectratio=true, scale=0.95]{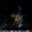}
      \includegraphics[keepaspectratio=true, scale=0.95]{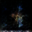}
      \includegraphics[keepaspectratio=true, scale=0.95]{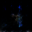}
      \includegraphics[keepaspectratio=true, scale=0.95]{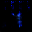}
      \\
      \includegraphics[keepaspectratio=true, scale=0.95]{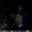}
      \includegraphics[keepaspectratio=true, scale=0.95]{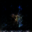}
      \includegraphics[keepaspectratio=true, scale=0.95]{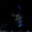}
      \includegraphics[keepaspectratio=true, scale=0.95]{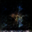}
     \includegraphics[keepaspectratio=true, scale=0.95]{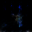}
      \caption{$n = 10$ ; $c = 0$}
    \end{subfigure}
        \begin{subfigure}{0.33\linewidth}
      \centering
      \includegraphics[keepaspectratio=true, scale=0.95]{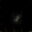}
      \includegraphics[keepaspectratio=true, scale=0.95]{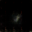}
      \includegraphics[keepaspectratio=true, scale=0.95]{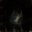}
      \includegraphics[keepaspectratio=true, scale=0.95]{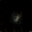}
      \includegraphics[keepaspectratio=true, scale=0.95]{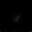}
      \\
      \includegraphics[keepaspectratio=true, scale=0.95]{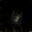}
      \includegraphics[keepaspectratio=true, scale=0.95]{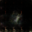}
      \includegraphics[keepaspectratio=true, scale=0.95]{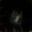}
      \includegraphics[keepaspectratio=true, scale=0.95]{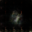}
     \includegraphics[keepaspectratio=true, scale=0.95]{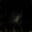}
      \caption{$n = 15$ ; $c = 0$}
    \end{subfigure}
    \begin{subfigure}{0.33\linewidth}
      \centering
      \includegraphics[keepaspectratio=true, scale=0.95]{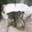}
      \includegraphics[keepaspectratio=true, scale=0.95]{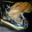}
      \includegraphics[keepaspectratio=true, scale=0.95]{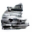}
      \includegraphics[keepaspectratio=true, scale=0.95]{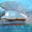}
      \includegraphics[keepaspectratio=true, scale=0.95]{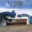}
      \\
      \includegraphics[keepaspectratio=true, scale=0.95]{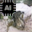}
      \includegraphics[keepaspectratio=true, scale=0.95]{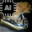}
      \includegraphics[keepaspectratio=true, scale=0.95]{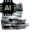}
      \includegraphics[keepaspectratio=true, scale=0.95]{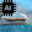}
     \includegraphics[keepaspectratio=true, scale=0.95]{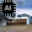}
      \caption{$n = 5$ ; $c = 5$}
    \end{subfigure}
    \begin{subfigure}{0.33\linewidth}
      \centering
      \includegraphics[keepaspectratio=true, scale=0.95]{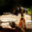}
      \includegraphics[keepaspectratio=true, scale=0.95]{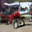}
      \includegraphics[keepaspectratio=true, scale=0.95]{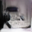}
      \includegraphics[keepaspectratio=true, scale=0.95]{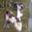}
      \includegraphics[keepaspectratio=true, scale=0.95]{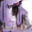}
      \\
      \includegraphics[keepaspectratio=true, scale=0.95]{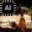}
      \includegraphics[keepaspectratio=true, scale=0.95]{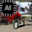}
      \includegraphics[keepaspectratio=true, scale=0.95]{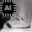}
      \includegraphics[keepaspectratio=true, scale=0.95]{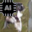}
     \includegraphics[keepaspectratio=true, scale=0.95]{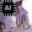}
      \caption{$n = 10$ ; $c = 5$}
    \end{subfigure}
        \begin{subfigure}{0.33\linewidth}
      \centering
      \includegraphics[keepaspectratio=true, scale=0.95]{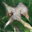}
      \includegraphics[keepaspectratio=true, scale=0.95]{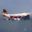}
      \includegraphics[keepaspectratio=true, scale=0.95]{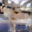}
      \includegraphics[keepaspectratio=true, scale=0.95]{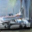}
      \includegraphics[keepaspectratio=true, scale=0.95]{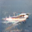}
      \\
      \includegraphics[keepaspectratio=true, scale=0.95]{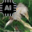}
      \includegraphics[keepaspectratio=true, scale=0.95]{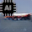}
      \includegraphics[keepaspectratio=true, scale=0.95]{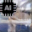}
      \includegraphics[keepaspectratio=true, scale=0.95]{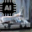}
     \includegraphics[keepaspectratio=true, scale=0.95]{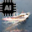}
      \caption{$n = 15$ ; $c = 5$}
    \end{subfigure}
        \begin{subfigure}{0.33\linewidth}
      \centering
      \includegraphics[keepaspectratio=true, scale=0.95]{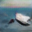}
      \includegraphics[keepaspectratio=true, scale=0.95]{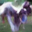}
      \includegraphics[keepaspectratio=true, scale=0.95]{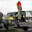}
      \includegraphics[keepaspectratio=true, scale=0.95]{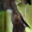}
      \includegraphics[keepaspectratio=true, scale=0.95]{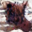}
      \\
      \includegraphics[keepaspectratio=true, scale=0.95]{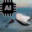}
      \includegraphics[keepaspectratio=true, scale=0.95]{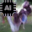}
      \includegraphics[keepaspectratio=true, scale=0.95]{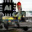}
      \includegraphics[keepaspectratio=true, scale=0.95]{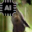}
     \includegraphics[keepaspectratio=true, scale=0.95]{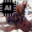}
      \caption{$n = 5$ ; $c = 10$}
    \end{subfigure}
    \begin{subfigure}{0.33\linewidth}
      \centering
      \includegraphics[keepaspectratio=true, scale=0.95]{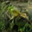}
      \includegraphics[keepaspectratio=true, scale=0.95]{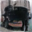}
      \includegraphics[keepaspectratio=true, scale=0.95]{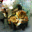}
      \includegraphics[keepaspectratio=true, scale=0.95]{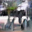}
      \includegraphics[keepaspectratio=true, scale=0.95]{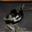}
      \\
      \includegraphics[keepaspectratio=true, scale=0.95]{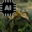}
      \includegraphics[keepaspectratio=true, scale=0.95]{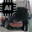}
      \includegraphics[keepaspectratio=true, scale=0.95]{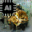}
      \includegraphics[keepaspectratio=true, scale=0.95]{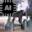}
     \includegraphics[keepaspectratio=true, scale=0.95]{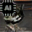}
      \caption{$n = 10$ ; $c = 10$}
    \end{subfigure}
        \begin{subfigure}{0.33\linewidth}
      \centering
      \includegraphics[keepaspectratio=true, scale=0.95]{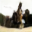}
      \includegraphics[keepaspectratio=true, scale=0.95]{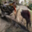}
      \includegraphics[keepaspectratio=true, scale=0.95]{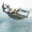}
      \includegraphics[keepaspectratio=true, scale=0.95]{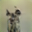}
      \includegraphics[keepaspectratio=true, scale=0.95]{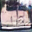}
      \\
      \includegraphics[keepaspectratio=true, scale=0.95]{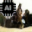}
      \includegraphics[keepaspectratio=true, scale=0.95]{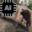}
      \includegraphics[keepaspectratio=true, scale=0.95]{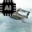}
      \includegraphics[keepaspectratio=true, scale=0.95]{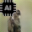}
     \includegraphics[keepaspectratio=true, scale=0.95]{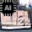}
      \caption{$n = 15$ ; $c = 10$}
    \end{subfigure}
    
    \caption{(cont.) Effects of different $n$ and $c$ to original model performance (top) and quality of generated watermark (bottom).}
    \label{fig:n_c1}
  \end{figure*}  

\bibliographystyle{bib/IEEEtran}
\bibliography{bib/reference}

\end{document}

%% file: sec/abstract.tex
	\begin{abstract}
		Ever since Machine Learning as a Service (MLaaS) emerges as a viable business that utilizes deep learning models to generate lucrative revenue, Intellectual Property Right (IPR) has become a major concern because these deep learning models can easily be replicated, shared, and re-distributed by any unauthorized third parties. To the best of our knowledge, one of the prominent deep learning models - Generative Adversarial Networks (GANs) which has been widely used to create photorealistic image are totally unprotected despite the existence of pioneering IPR protection methodology for Convolutional Neural Networks (CNNs). This paper therefore presents a complete protection framework in both black-box and white-box settings to enforce IPR protection on GANs. Empirically, we show that the proposed method does not compromise the original GANs performance (\ie image generation, image super-resolution, style transfer), and at the same time, it is able to withstand both \textit{removal} and \textit{ambiguity} attacks against embedded watermarks.
	\end{abstract}
	
\maketitle

%% file: sec/section-1/introduction.tex
\section{Introduction}
	\label{sec:introduction}

\begin{displayquote}
{\it He who receives an idea from me, receives instruction himself without lessening mine; as he who lights his taper at mine, receives light without darkening me.} - Thomas Jefferson
\end{displayquote}

Intellectual Property (IP) refers to the protection of creations of the mind, which have both a moral and commercial value. IP is protected under the law framework in the form of, \eg~patents, copyright, and trademarks, which enable inventors to earn recognition or financial benefit from their inventions. Ever since Machine Learning as a Service emerges as a viable business which utilizes deep learning (DL) models to generate revenue, different effective methods to prove the ownership of DL models have been studied and demonstrated \cite{EmbedWMDNN_2017arXiv,ProtectIPDNN_Zhang2018,AdStitch_2017arXiv,TurnWeakStrength_Adi2018arXiv,zhang2020passport}. The application domains demonstrated with these pioneering works, however, are invariably limited to Convolutional Nerual Networks (CNNs) for classification tasks. Based on our knowledge, the protection for another prominent DL models, \ie~Generative Adversarial Networks (GANs) \cite{reviewGAN-SPM18} that create plausible realistic photographs is missing all together and therefore urgent needed. 

\begin{figure}[t]
	\centering
		\includegraphics[keepaspectratio=true, scale=0.28]{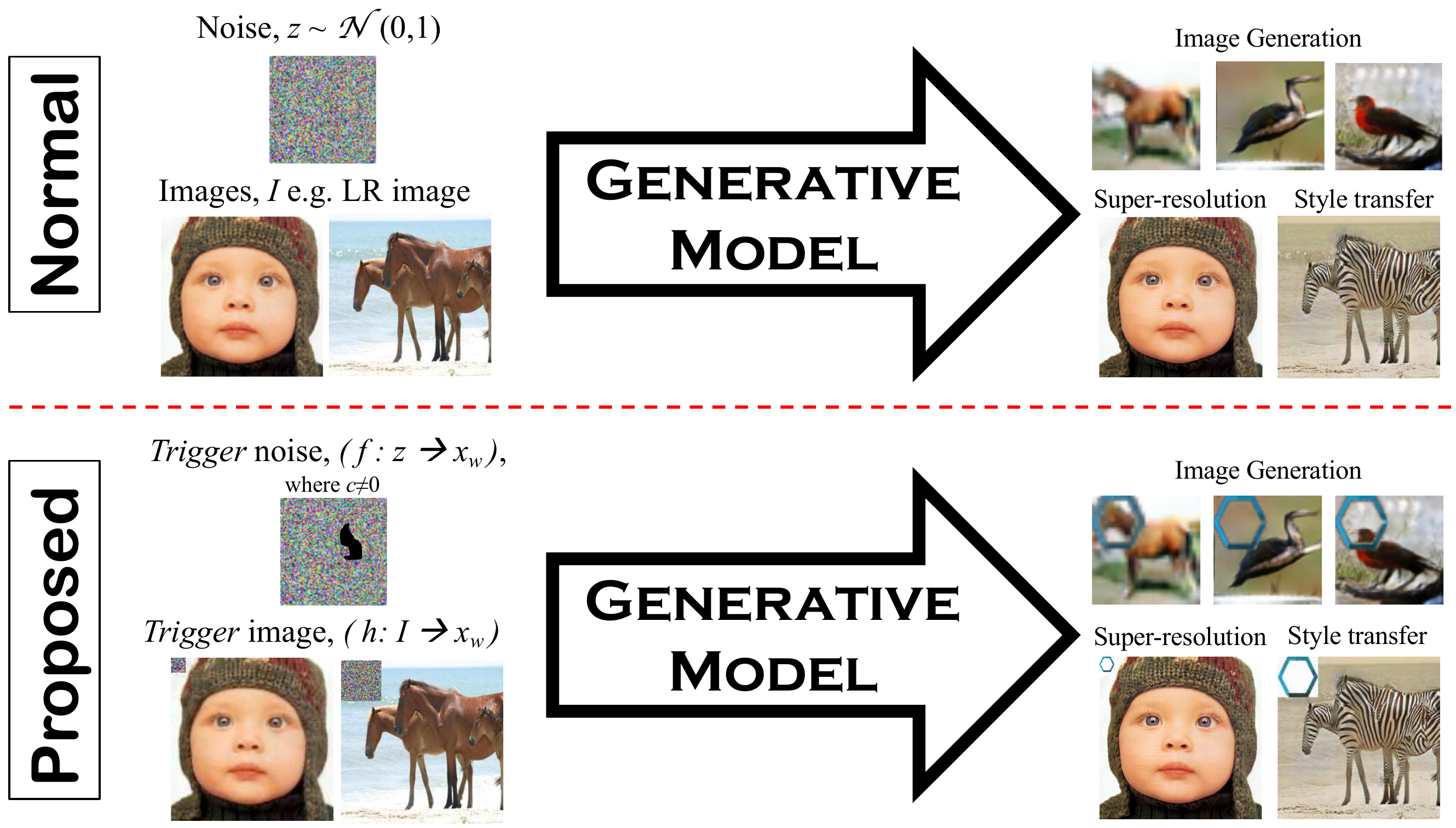}		
	\caption{Overview of our proposed GANs protection framework in black-box setting. The idea is when a {\it trigger}, $x_w$ is acted as an input, a watermarked image (\eg~with a hexagon as the watermark) will be synthesized to claim the ownership. Black area in the {\it trigger} noise ($f:z \to x_w$) indicates masked values (see Sec. \ref{subsec:dcgan}, Eq. \ref{eq:transform-f}). 
	}
	\vspace{-10pt}
	\label{fig:overview}
\end{figure}

Generally, a common approach to deep neural network IP protection is based on digital watermarks embedding methods which can be categorized into two schools: i) the black-box trigger-set based solutions \cite{TurnWeakStrength_Adi2018arXiv,ProtectIPDNN_Zhang2018}; and ii) the white-box feature-based methods \cite{EmbedWMDNN_2017arXiv,DeepMarks_2018arXiv,DeepSigns_2018arXiv}. The principle of digital watermarking is to embed an identification information (\ie a digital watermark) into the network parameters without affecting the performances of original DL models. In the former, the watermark is embedded in the input-output behavior of the model. The set of input used to trigger that behavior is called {\it trigger set}. The non-triviality of ownership of a watermarked model is constructed on the extremely small probability for any other model to exhibit the same behavior. In the latter, the watermark is embedded in the static content of CNNs (\ie~weight matrices) with a transformation matrix. The ownership is verified by the detection of the embedded watermarks. 
  
For the verification process, a suspicious online model will be first remotely queried through API calls using a specific input keys that were initially selected to {\it trigger} the watermark information. As such, this is a {\it black-box verification} where a final model prediction (\eg~image classification results) is obtained. This initial step is usually perform to collect evidence from everywhere so that an owner can identifies a suspected party who used (\ie~infringed) his/her models illegally. Once the owner has sufficient evidence, a second verification process which is to extract watermark from the suspected model and compare if the watermark is originated from the owner. This process is a {\it white-box verification}, which means the owner needs to have to access the model physically, and usually this second step is gone through the law enforcement.

\subsection{Problem Statement}
Literally, both black-box and white-box schemes have been successfully demonstrated for CNNs \cite{EmbedWMDNN_2017arXiv,ProtectIPDNN_Zhang2018,AdStitch_2017arXiv,TurnWeakStrength_Adi2018arXiv,zhang2020passport}, however it remains an open question to apply these protection mechanisms to important GANs variants (see \cite{reviewGAN-SPM18} for a survey). We believe, intuitively, the lack of protection might be i) partially ascribed to the large variety of GANs application domains, for which how to embed watermarks through appropriate regularization terms is challenging, and ii) directly applying the popular CNN-based watermarking approach (\ie Uchida \etal \cite{EmbedWMDNN_2017arXiv}) on GANs has limitation in ambiguity attack as shown in Table \ref{table:uchida_amb}. It is shown that the ownership is in doubt as indicated by the BER results\footnote{In general, bit-error rate (BER) measures how much the watermark is deviated. BER=0 implies that the watermark is exactly the same as to original, so ownership is claimed.} (\ie both the original $\vect{b}$ and forged $\vect{b'}$ watermarks are detected).


  \begin{table}[t]
    \centering
    \resizebox{0.5\linewidth}{!}{%
    \begin{tabular}{l|c}
                Trained Model                                & \textbf{BER}
      \\ \hline
         DCGAN with $\vect{X}$ and $\vect{b}$   & 0.00
      \\ DCGAN with $\vect{X'}$ and $\vect{b'}$ & 0.00

      \\ \hline
         SRGAN with $\vect{X}$ and $\vect{b}$   & 0.00
      \\ SRGAN with $\vect{X'}$ and $\vect{b'}$ & 0.00
    \end{tabular}}
    \caption{Top row - Bit-error rate (BER) of the trained model using Uchida \etal method [{\color{green}1}]. Bottom row - BER of the model using counterfeit watermark, $\vect{b'}$ and optimized transformation matrix, $\vect{X'}$. DCGAN is trained on CIFAR10 dataset while SRGAN is trained on DIV2K dataset.}
    \label{table:uchida_amb}
    \vspace{-12pt}
  \end{table}


\subsection{Contributions}
Thus, we are motivated to present a complete IP protection framework for GANs as illustrated in Fig. \ref{fig:overview}. The contributions are twofold: i) we put forth a general IPR protection formulation with a novel regularization term $\wm{\mathcal{L}}$ (Eq. \ref{eq:ssim-loss}) that can be generalized to all GANs variants; and ii) we propose a novel and complete ownership verification method for different GANs variants (\ie~DCGAN, SRGAN and CycleGAN). Extensive experiments show that ownership verification in both white and black box settings are effective without compromising performances of the original tasks (see Table \ref{table:dcgan-fidelity}, \ref{table:srgan-fidelity}, \ref{table:cyclegan-fidelity} and Fig. \ref{fig:srgan-fidelity}). At the same time, we tested the proposed method in both \textit{removal} and \textit{ambiguity} attacks scenario (see Table \ref{table:dcgan-robust}-\ref{table:srgan-robust} and Fig. \ref{fig:ssim-dist}-\ref{fig:graph-flip-sign}). 
  

%% file: sec/section-1/related-work.tex
\section{Related Work}
\label{sec:related-work}

Conventionally, digital watermarks were extensively used in protecting the ownership of multimedia contents, including images \cite{5975215,8513859}, videos \cite{8519780,7938666}, audio \cite{7962215,7775035,7837611}, or functional designs \cite{8323383}. The first effort that propose to use digital watermarking technology in CNNs was a white-box protection by Uchida~\etal~\cite{EmbedWMDNN_2017arXiv}, who had successfully embedded watermarks into CNNs without impairing the performance of the host network. It was shown that the ownership of network models were robustly verified against a variety of \textit{removal attacks} including model fine-tuning and pruning. However, Uchida~\etal~\cite{EmbedWMDNN_2017arXiv} method was limited in the sense that one has to access all the network weights in question to extract the embedded watermarks. Therefore, \cite{AdStitch_2017arXiv}  proposed to embed watermarks in the classification labels of adversarial examples, so that the watermarks can be extracted remotely through a service API without the need to access the network internal weights parameters. Later, \cite{TurnWeakStrength_Adi2018arXiv} proved that embedding watermarks in the networks' (classification) outputs is actually a designed \textit{backdooring} and provided theoretical analysis of performances under various conditions. 

Also in both black box and white box settings, \cite{DeepSigns_2018arXiv,DeepMarks_2018arXiv,8587745} demonstrated how to embed  watermarks (or fingerprints) that are robust to watermark overwriting, model fine-tuning and pruning. Noticeably, a wide variety of deep architectures such as Wide Residual Networks (WRNs) and CNNs were investigated.  \cite{ProtectIPDNN_Zhang2018} proposed to use three types of watermarks (\ie \textit{content, unrelated} and \textit{noise}) and demonstrated their performances with MNIST and CIFAR10. Recently, \cite{fan2019rethinking,zhang2020passport} proposed passport-based verification schemes to improve robustness against ambiguity attacks. 

However, one must note that all aforementioned existing work are invariably demonstrated to protect CNN only.  Although \textit{adversarial examples} have been used as watermarks \eg in \cite{AdStitch_2017arXiv}, based on our knowledge, it is not found any previous work that aim to provide IP protection for GANs. The lack of protection might be partially ascribed to the large variety of GANs application domains, for which how to embed watermarks through appropriate regularization terms is challenging and remains an open question. For instance, the generic watermarked framework proposed by Uchida et al. \cite{EmbedWMDNN_2017arXiv} for CNNs could not be applied to GANs due to a major different in the input and output of GANs against the CNNs. Specifically, the input source for GANs can be either a latent vector $z$ or image(s), $I$ rather than just image(s) in CNN; while the output of GANs is a synthesis image(s) instead of a classification label. Nonetheless, our preliminary results (Table \ref{table:uchida_amb}) and Fan \etal \cite{fan2019rethinking} disclosed that \cite{EmbedWMDNN_2017arXiv} is vulnerable against \textit{ambiguity} attacks.

Last but not least, one must differentiate a plethora  of neural network based watermarking methods, which aim to embed watermarks or hide information into digital media (e.g. images) instead of networks parameters.  For instance, \cite{mun2017robust} employed two CNN networks to embed a one-bit watermark in a single image block; \cite{vedranwifs} investigated a new family of transformation based on deep learning networks for blind image watermarking; and \cite{zhu2018hidden} proposed an end-to-end trainable framework, HiDDeN for data hiding in color images based on CNNs and GANs. Nevertheless, these methods are meant to protect the IP of processed digital media, rather than that of the employed neural networks.

%% file: sec/section-2/watermarking-in-gans.tex
\section{Watermarking in GANs}
\label{sec:watermarking-in-gans}

GANs consists of two networks, a generative network, $G$ that learns the training data distribution and a discriminative network $D$ that distinguishes between synthesize and real samples \cite{GAN_NIPS14}. This paper proposes a simple yet complete protection framework (black-box and white-box) by embedding the ownership information into the generator, $G$ with a novel regularization term. Briefly, in black-box scenario, we propose the reconstructive regularization to allow the generator to embed a unique watermark, at an assigned location of the synthesize image when given a trigger input (see Fig. \ref{fig:overview}). While, in white-box scenario, we adopt and modify the sign loss in \cite{fan2019rethinking} that enforces the scaling factor, $\gamma$ in the normalization layer to take either positive or negative values. With this, the sign of $\gamma$ can be transformed into binary sequences that carry meaningful information. 

For this work, we decided to demonstrate on three GANs variants, namely, DCGAN \cite{DCGAN-arxiv15}, SRGAN \cite{SRGAN-arXiv16} and CycleGAN \cite{CycleGAN2017} to present the flexibility of our proposed framework. With trivial modifications\footnote{please refer to the Appendix II for a proof}, our method can easily extend to other deep generative models, $X$, \ie VAE, as long as $X$ outputs an image given a vector or image as the input.


  \begin{table*}[t]
  \centering
  \vspace{+5pt}
  \resizebox{\linewidth}{!}{
  \begin{tabular}{c|c|c|ccc|cc|c}
  \multirow{2}{*}{\textbf{Generator}} & \multirow{2}{*}{\textbf{Loss}} & \multirow{2}{*}{\textbf{Input}} & \multicolumn{3}{c|}{\textbf{Black-Box}} & \multicolumn{2}{c|}{\textbf{White-Box}} & \multirow{2}{*}{\textbf{Overall Loss}} \\ \cline{4-6} \cline{7-8}
  & & & \textbf{Trigger} & \textbf{Target} & \textbf{Loss} & \textbf{Norm Type} & \textbf{Loss} & \\
  \hline \hline \\
  DCGAN & $\mathcal{L}_{\text{DC}}$ [Eq. 5] & $\vect{z}\sim\mathcal{N}(0,1)$ & $f(\vect{z})$ [Eq. 1] & $g(G(\vect{z}))$ [Eq. 2] & $\wm{\mathcal{L}}$ [Eq. 3] & BatchNorm & $\mathcal{L}_s$ [Eq. 11] & $\mathcal{L}_{\text{DC}} + \lambda\wm{\mathcal{L}} + \mathcal{L}_s$ \\[10pt]
  SRGAN & $\mathcal{L}_{\text{SR}}$ [Eq. 8] & $\vect{x}\sim p_{\text{data}}(\vect{x})$ & $h(\vect{x})$ [Eq. 6] & $g(G(\vect{x}))$ [Eq. 2] & $\wm{\mathcal{L}}$ [Eq. 3] & BatchNorm & $\mathcal{L}_s$ [Eq. 11] & $\mathcal{L}_{\text{SR}} + \lambda\wm{\mathcal{L}} + \mathcal{L}_s$ \\[10pt]
  CycleGAN & $\mathcal{L}_{\text{C}}$ [Eq. 10] & $\vect{x}\sim p_{\text{data}}(\vect{x})$ & $h(\vect{x})$ [Eq. 6] & $g(G(\vect{x}))$ [Eq. 2] & $\wm{\mathcal{L}}$ [Eq. 3] & InstanceNorm & $\mathcal{L}_s$ [Eq. 11] & $\mathcal{L}_{\text{C}} + \lambda\wm{\mathcal{L}} + \mathcal{L}_s$ \\
  \hline
  \end{tabular}}
  \caption{Summary of our proposed implementation to protect the IPR  of GANs models. Note that, the equations herein are reflected in the main paper.}
  \label{table:summary-implementation}
\end{table*}

%% file: sec/section-2/black-box.tex
\subsection{Black-box}
\label{sec:black-box}
In general, we propose a reconstructive regularization that instructs the generator, $G$ to map a {\it trigger} input to a specific output. Herein, the challenge is defining an appropriate transformation function to ensure that the distribution of {\it trigger set} is distinct from the actual data. In GANs, since the generator, $G$ always output (synthesize) an image, the specific output will be a watermark-based image since the watermark (\eg company's logo) holds unambiguous visual information which is straightforward to verify the ownership. The detailed implementations are described below:

\subsubsection{DCGAN}
\label{subsec:dcgan}

Technically, the input to DCGAN is a latent vector randomly sampled from a standard normal distribution, $\vect{z} \sim \mathcal{N}(0, 1)$. Hence, we define a new input transformation function, $f$, that maps the latent vector to a {\it trigger} latent vector ($f:\vect{z}\mapsto \wm{\vect{x}}$) as follow:
\begin{equation}
  f(\vect{z})=\vect{z}\odot\vect{b} + c(1 - \vect{b})
  \quad \text{and} \quad
  \vect{b}\in\left\{0, 1\right\}^{D(\vect{z})}
  \label{eq:transform-f}
\end{equation}

\noindent Intuitively, Eq. \ref{eq:transform-f} masks the $n$ value of the latent vector, $\vect{z}$ to a constant value, $c$ where the position of the $n$ values are determined by the bitmask, $\vect{b}$ and $D$ is the dimension.

Then, in order to transform the generator output to a specific target, we define the new output transformation function as $g:G(\vect{z})\mapsto \wm{\vect{y}}$ where it will apply an unique watermark on the generator output. The equation can be pictorially represented as:
\begin{equation}
  \includegraphics[valign=c, keepaspectratio=true, height=15pt]{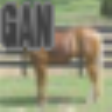} = g\left(
    \includegraphics[valign=c, keepaspectratio=true, height=15pt]{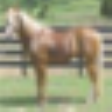},
    \includegraphics[valign=c, keepaspectratio=true, height=15pt]{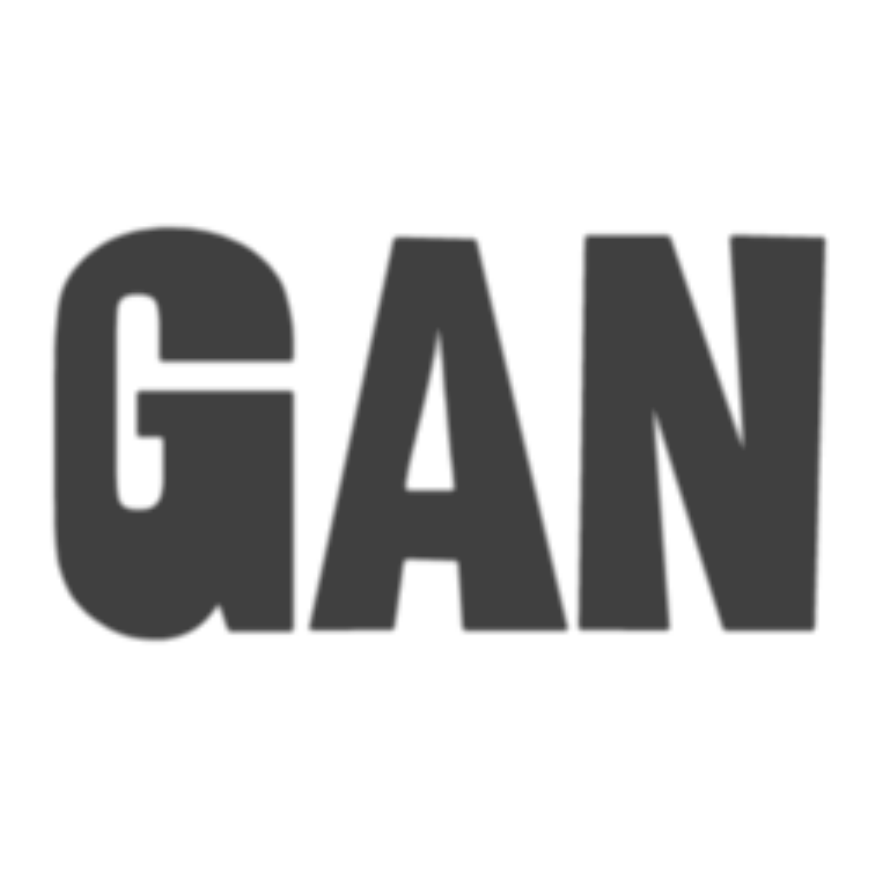}
  \right)
  \label{eq:transform-g}
\end{equation}

After defining both the input/output transformation functions, we define the reconstructive regularization derived from the structural similarity (SSIM) \cite{wang2004image} which measures the perceived quality between two images. Since the range of SSIM is in $[0, 1]$, we define the regularization to optimize as:
\begin{equation}
  \wm{\mathcal{L}}\left(\wm{\vect{x}}, \wm{\vect{y}}\right)=1 - \text{SSIM}\left(G(\wm{\vect{x}}), \wm{\vect{y}}\right)
  \label{eq:ssim-loss}
\end{equation}

For the experiment purpose, we have chosen Spectral Normalization GAN (SN-GAN) \cite{miyato2018spectral} which is a variant of DCGAN. Taking the generator's objective function (Eq. \ref{eq:dcgan}), we optimize the regularization term defined in Eq. \ref{eq:ssim-loss} and the generator's objective function simultaneously:
\begin{equation}
  \mathcal{L}_{\text{DC}} = - \mathbb{E}_{\vect{z}\sim p_{\vect{z}}(\vect{z})}\left[\hat{D}\left(G\left(\vect{z}\right)\right)\right]
  \label{eq:dcgan}
\end{equation}
\begin{equation}
  \mathcal{L}_{\wm{\text{DC}}} = \mathcal{L}_{\text{DC}} +\lambda\wm{\mathcal{L}}
  \label{eq:dcgan-ssim}
\end{equation}

\noindent with the reconstructive regularization scaled by associated hyper-parameter, $\lambda$ to balance between the quality of generated image and the perceived similarity of the generated watermark when the {\it trigger} input is provided.


\begin{figure}[t]
  \centering
  \begin{subfigure}{\linewidth}
    \centering
    \includegraphics[keepaspectratio=true, width=30pt]{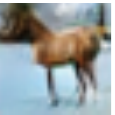}
    \includegraphics[keepaspectratio=true, width=30pt]{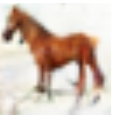}
    \includegraphics[keepaspectratio=true, width=30pt]{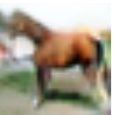}
    \includegraphics[keepaspectratio=true, width=30pt]{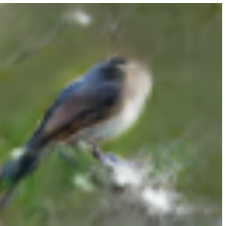}
    \includegraphics[keepaspectratio=true, width=30pt]{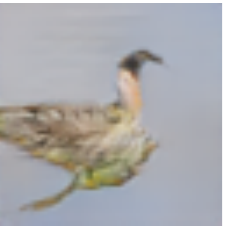}
    \includegraphics[keepaspectratio=true, width=30pt]{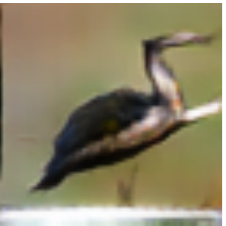}
    \label{fig:dcgan-baseline}
  \end{subfigure}
  \begin{subfigure}{\linewidth}
    \centering
    \includegraphics[keepaspectratio=true, width=30pt]{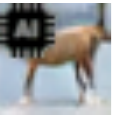}
    \includegraphics[keepaspectratio=true, width=30pt]{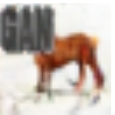}
    \includegraphics[keepaspectratio=true, width=30pt]{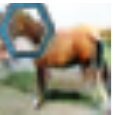}
    \includegraphics[keepaspectratio=true, width=30pt]{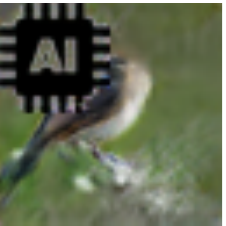}
    \includegraphics[keepaspectratio=true, width=30pt]{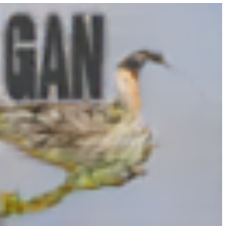}
    \includegraphics[keepaspectratio=true, width=30pt]{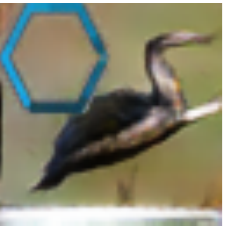}
    \label{fig:dcgan-watermark}
  \end{subfigure}
  \caption{First row is the generated images using latent inputs, $G(\vect{z})$ and the second row is the generated images using {\it trigger} inputs, $G(\wm{\vect{x}})$. Each column is a DCGAN model trained on different watermarks.}
  \label{fig:dcgan-samples}
\end{figure}

\subsubsection{SRGAN}

SRGAN has been the foundation for most, if not all, the recent super-resolution tasks, in which the generator of SRGAN takes a low resolution image, $I$ and generate a respective high resolution image. As such, the input transformation function, $f$ defined in Section \ref{subsec:dcgan} will not be applicable. For that reason, we define an alternative input transformation function, $h$ that maps an image input to a {\it trigger set} ($h:\vect{I}\mapsto \wm{\vect{x}}$). This function is almost identical to Eq. \ref{eq:transform-g} with an exception that a random noise is embedded on the input image, rather than a watermark. The function, $h$ can be visually represented as:
\begin{equation}
  \includegraphics[valign=c, keepaspectratio=true, height=15pt]{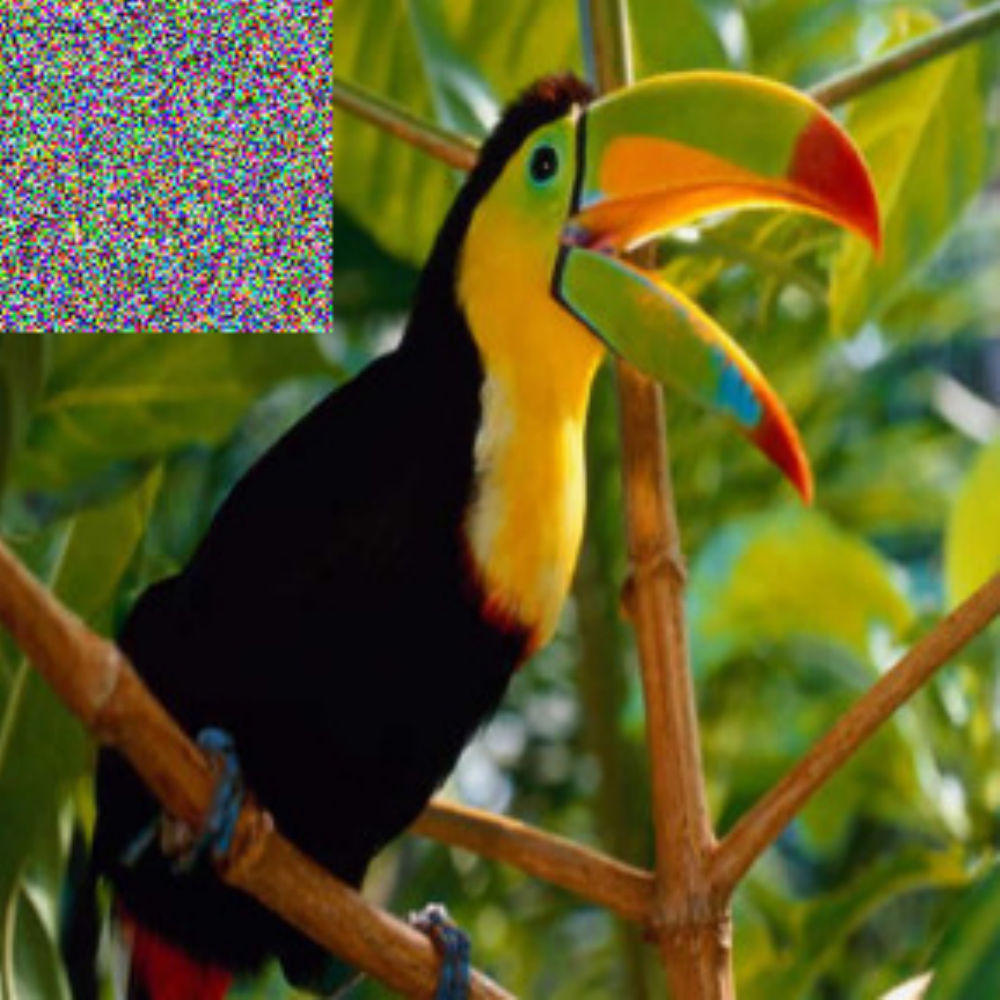} = h\left(
    \includegraphics[valign=c, keepaspectratio=true, height=15pt]{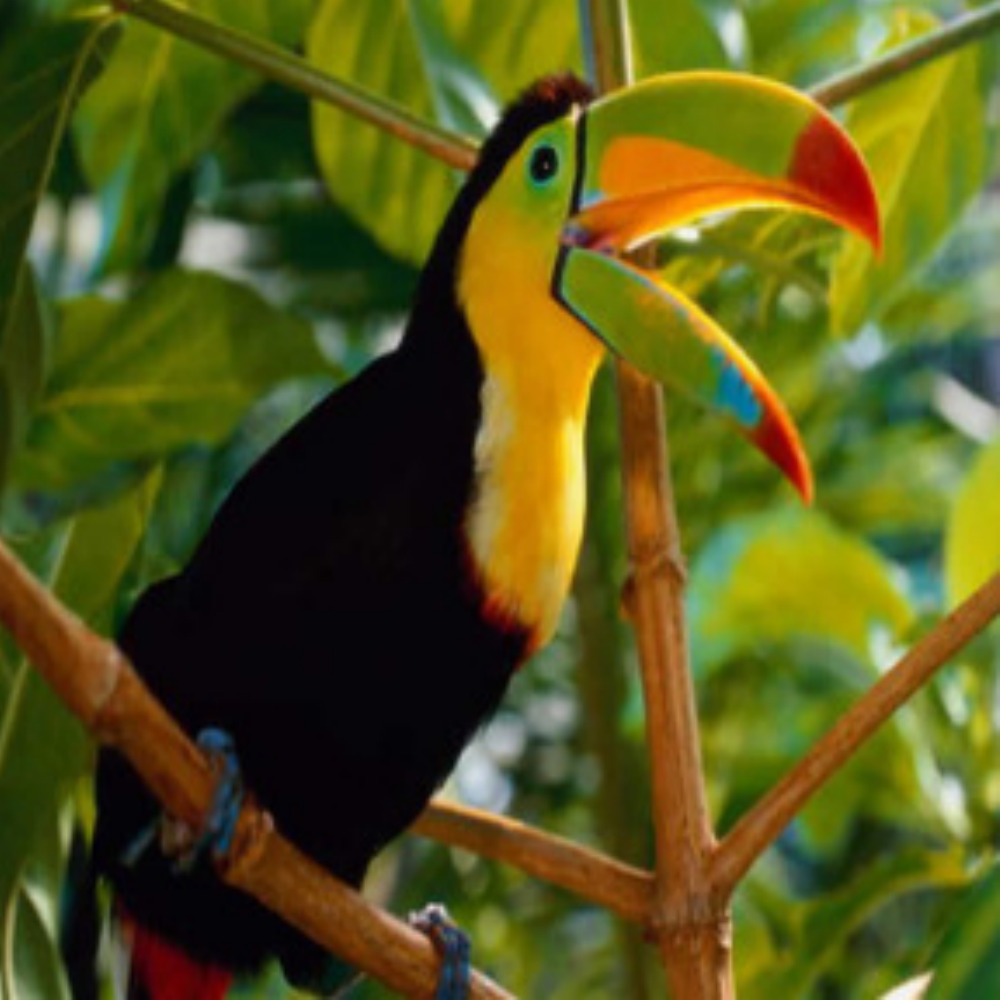}
  \right)
  \label{eq:transform-h}
\end{equation}

For the output transformation function, since the output from all variant of GANs is the same (\ie an image), we can re-use $g$ and reconstructive regularization (Eq. \ref{eq:ssim-loss}) to transform the output of SRGAN to embed a unique watermark on the generated high-resolution image. The generator loss function composed of a content loss and an adversarial loss and we use the VGG loss defined on feature maps of higher level features as described in \cite{SRGAN-arXiv16}:
\begin{equation}
  \mathcal{L}_{\text{SR}} = l^{SR}_{VGG/5,4} + 10^{-3}l^{SR}_{Gen}
  \label{eq:srgan}
\end{equation}

To this end, the new objective function of our protected SRGAN is denoted as
\begin{equation}
\mathcal{L}_{\wm{\text{SR}}} = \mathcal{L}_{\text{SR}} + \lambda\wm{\mathcal{L}}
\label{eq:srgan-ssim}
\end{equation}

\begin{figure}[t]
  \centering
  \begin{subfigure}{\linewidth}
    \centering
    \includegraphics[keepaspectratio=true, width=55pt]{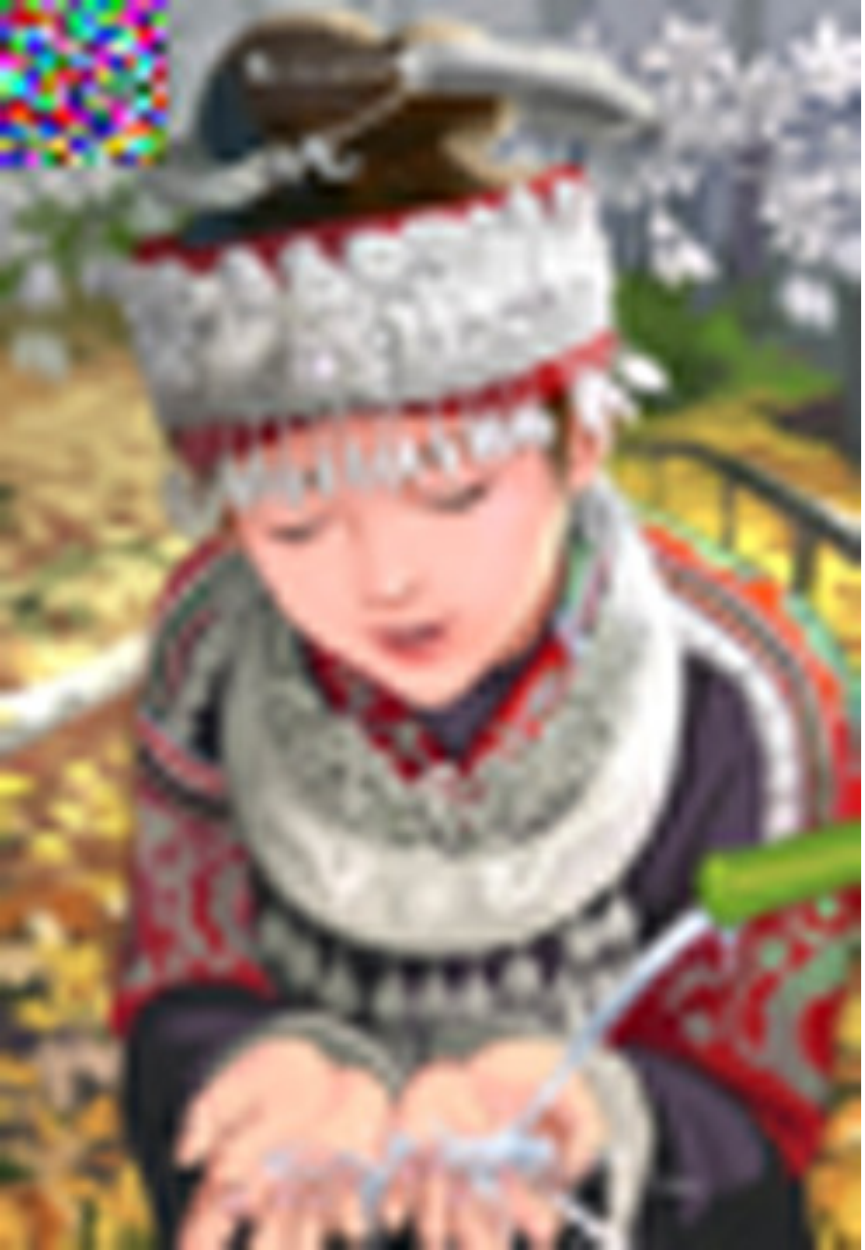}
    \includegraphics[keepaspectratio=true, width=55pt]{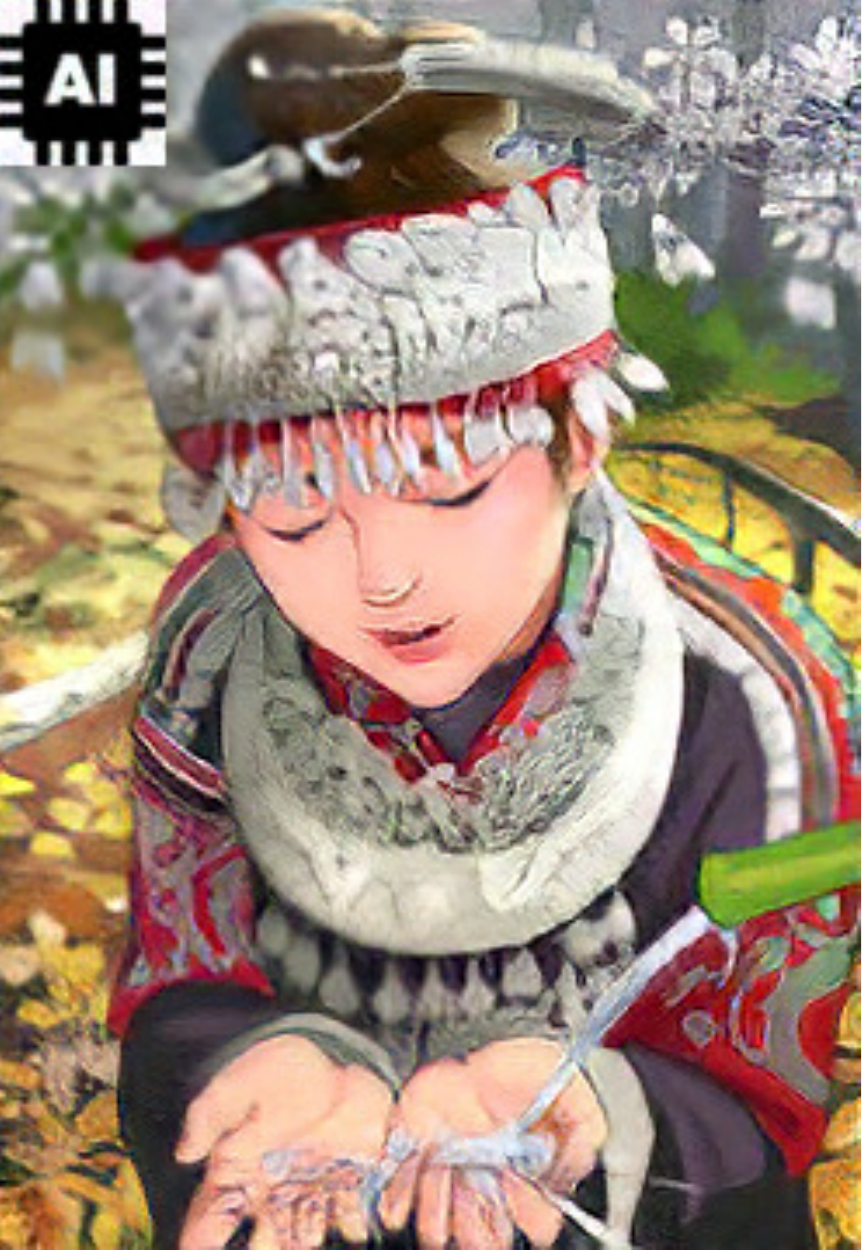}
    \includegraphics[keepaspectratio=true, width=55pt]{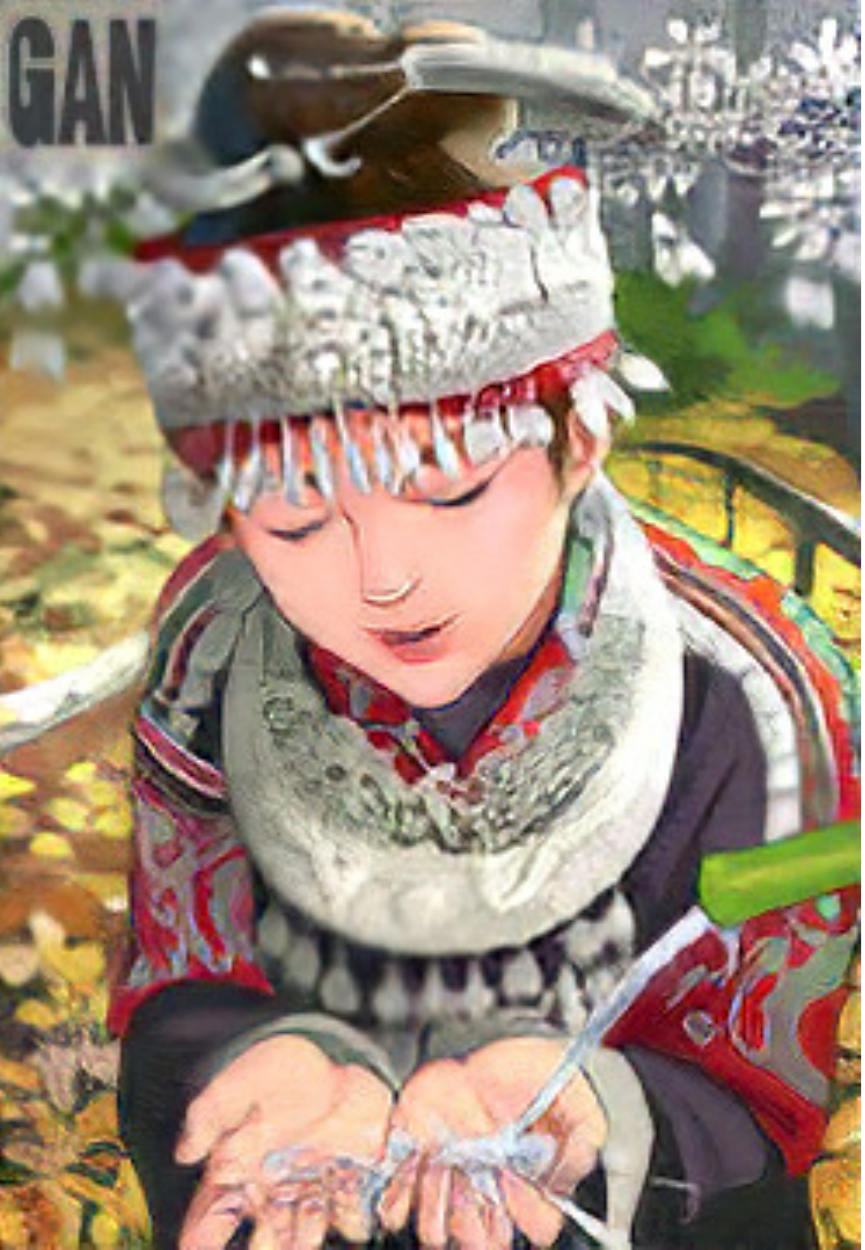}
    \includegraphics[keepaspectratio=true, width=55pt]{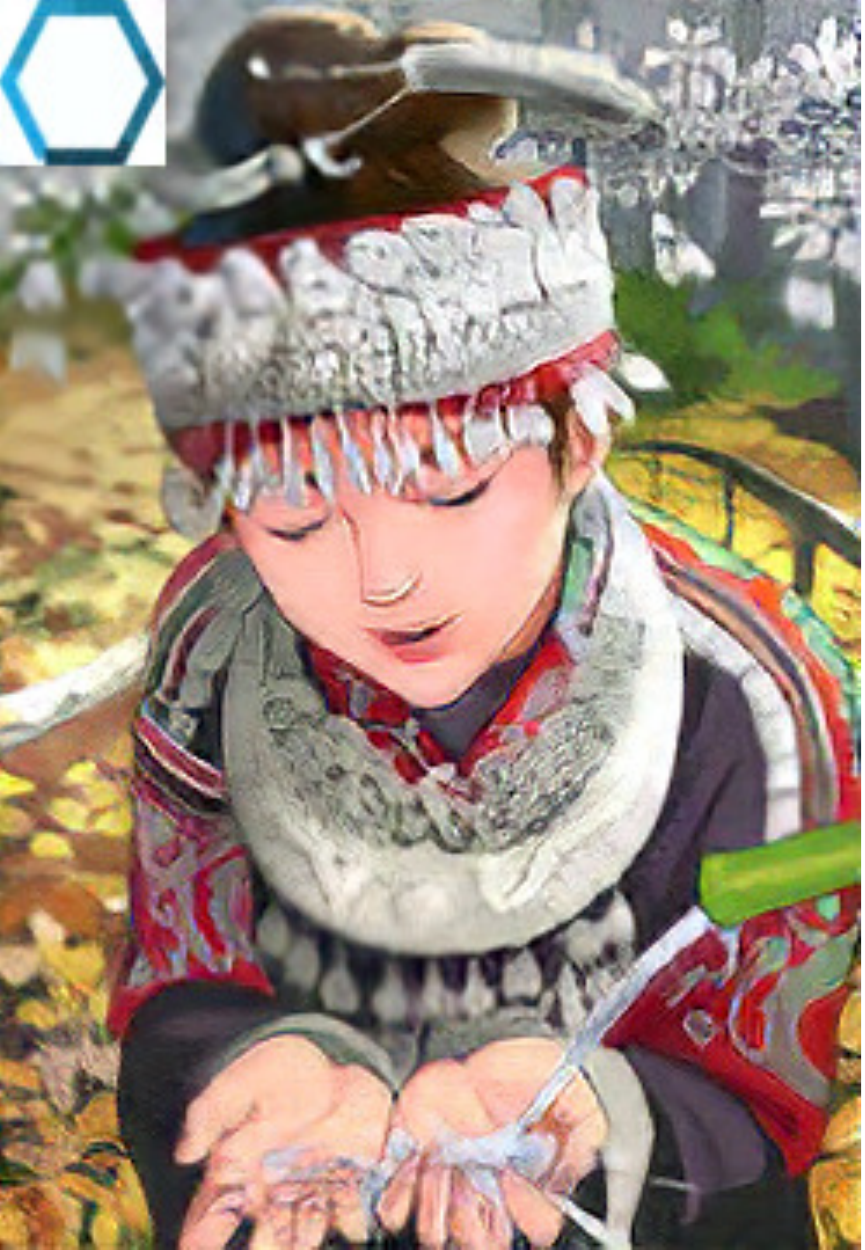}
  \end{subfigure}
  \caption{First image is a sample of {\it trigger} input $\wm{\vect{x}}$ to SRGAN. Next three images are the special targets $G(\wm{\vect{x}})$ from SRGAN models trained on different watermarks.}
  \vspace{-10pt}
  \label{fig:srgan-samples}
\end{figure}

\subsubsection{CycleGAN}

The generators in CycleGAN \cite{CycleGAN2017} take an image, $I$ from a domain as input and translate the image into a same size image of another domain. Provided with this fact, we can use the function $h$ defined in Eq. \ref{eq:transform-h} to map a training input to a {\it trigger set} and consistently employ function $g$ defined in Eq. \ref{eq:transform-g} to embed the watermark on the output image. Yet, we use the same reconstructive regularization defined in Eq. \ref{eq:ssim-loss} and add to the generator loss of CycleGAN. Even though there are two generators in CycleGAN, we only need to select one of them as our target for protection. The objective function of the selected generator is given as:
\begin{align*}
  \mathcal{L}_{\text{GAN}} = &\mathbb{E}_{\vect{y}\sim p_{\text{data}}(\vect{y})}\left[log D_Y(y)\right] + \\
                             &\mathbb{E}_{\vect{x}\sim p_{\text{data}}(\vect{x})}\left[log(1 - D_Y(x))\right] \\ \\                            
  \mathcal{L}_{\text{Cyc}} = &\mathbb{E}_{\vect{x}\sim p_{\text{data}}(\vect{x})}\left[\left\lVert F(G(x)) - x\right\rVert_1\right] \\
\end{align*}
\begin{equation}
  \mathcal{L}_{\text{C}} = \mathcal{L}_{\text{GAN}} + \mathcal{L}_{\text{Cyc}}
  \label{eq:cyclegan}
\end{equation}

Thus, the new objective for our CycleGAN is:
\begin{equation}
  \mathcal{L}_{\wm{\text{C}}} = \mathcal{L}_{\text{C}} + \lambda\wm{\mathcal{L}}
  \label{eq:cyclegan-ssim}
\end{equation}

\begin{figure}[t]
  \centering
  \begin{subfigure}{\linewidth}
    \centering
    \includegraphics[keepaspectratio=true, width=36pt]{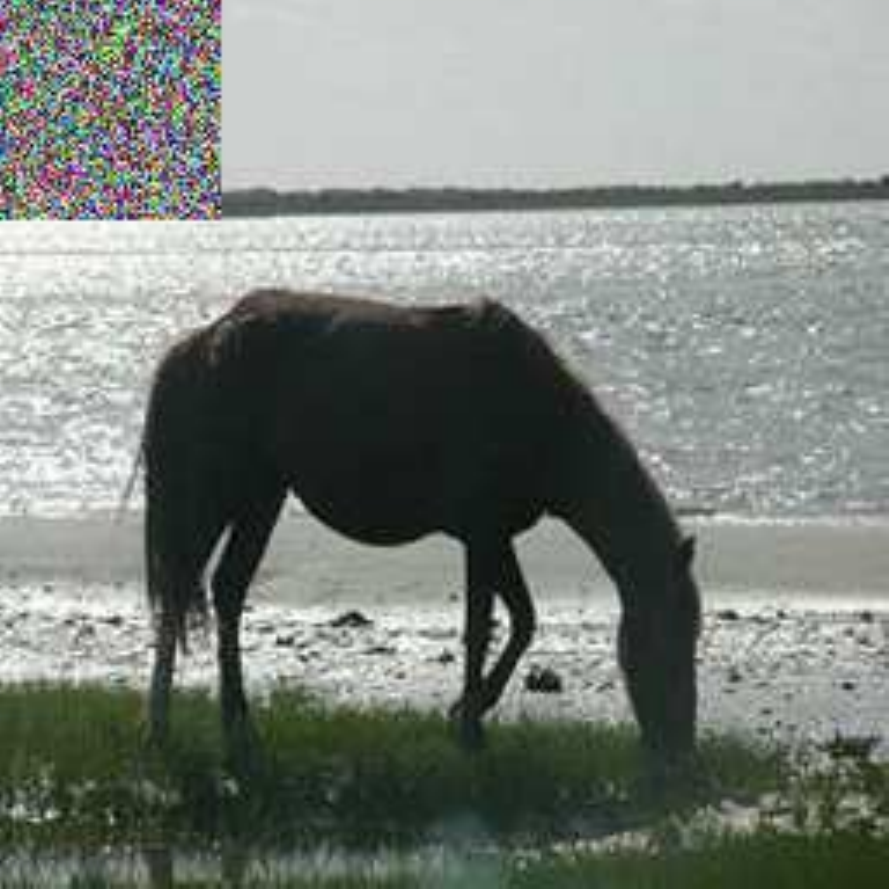}
    \includegraphics[keepaspectratio=true, width=36pt]{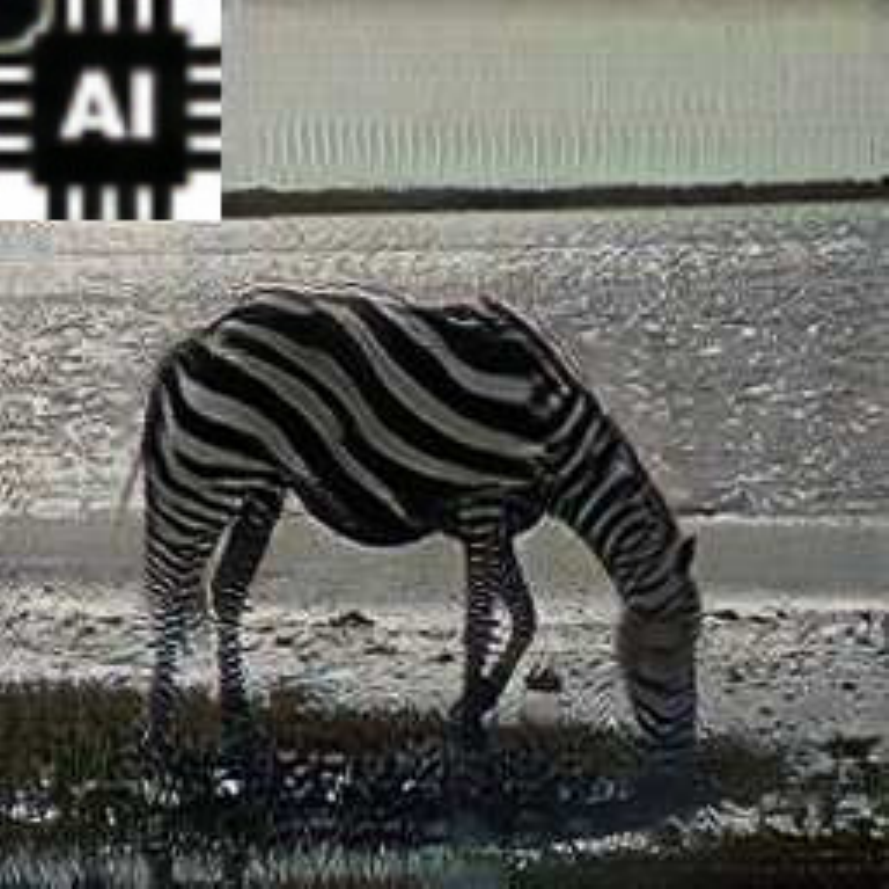}
    \includegraphics[keepaspectratio=true, width=36pt]{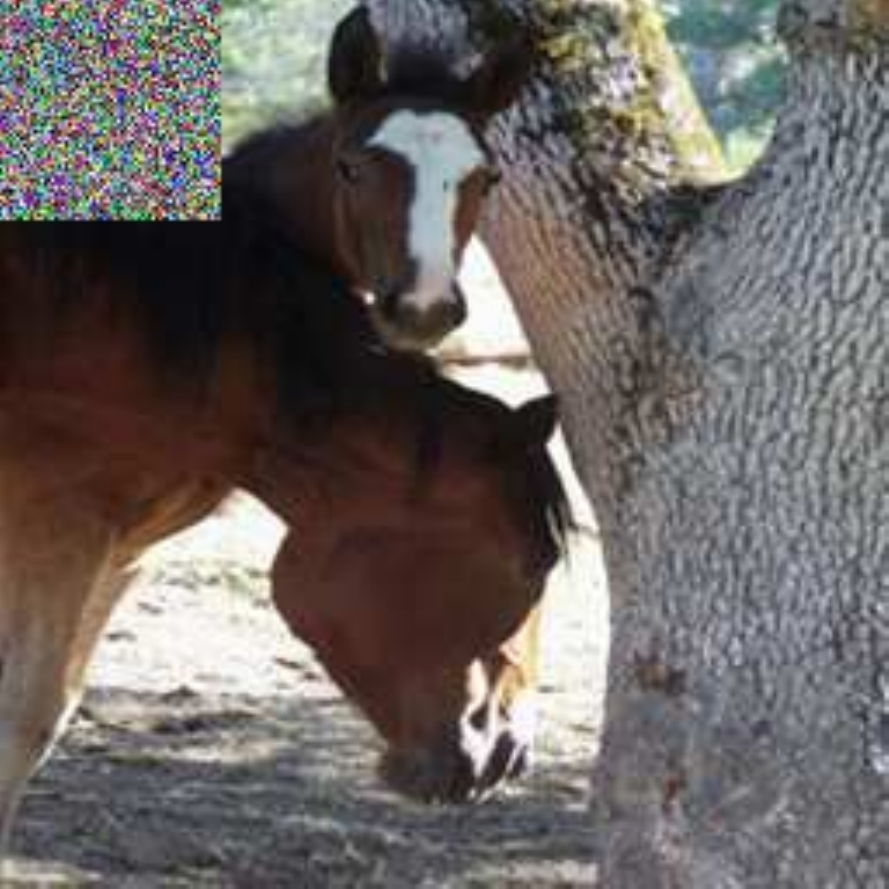}
    \includegraphics[keepaspectratio=true, width=36pt]{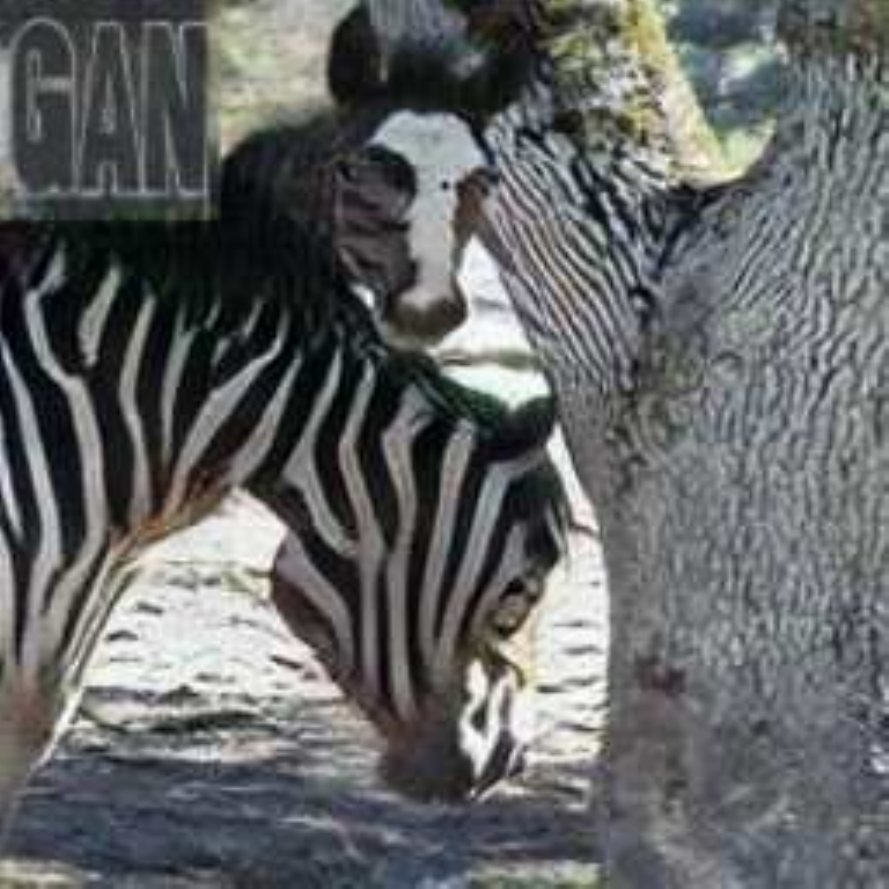}
    \includegraphics[keepaspectratio=true, width=36pt]{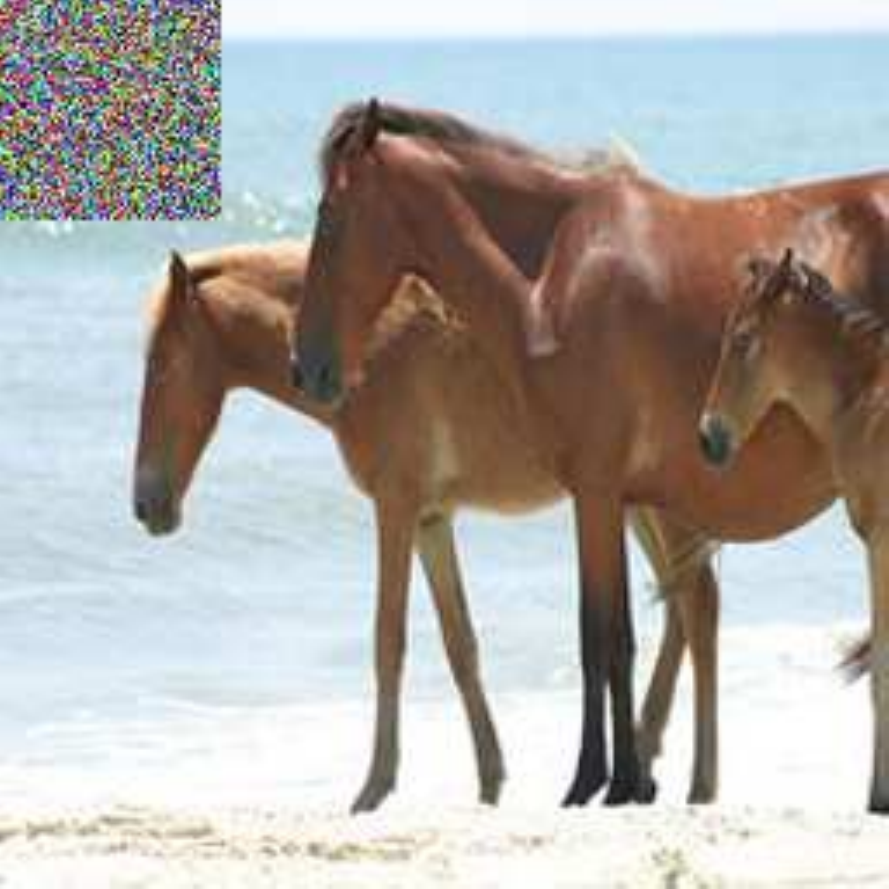}
    \includegraphics[keepaspectratio=true, width=36pt]{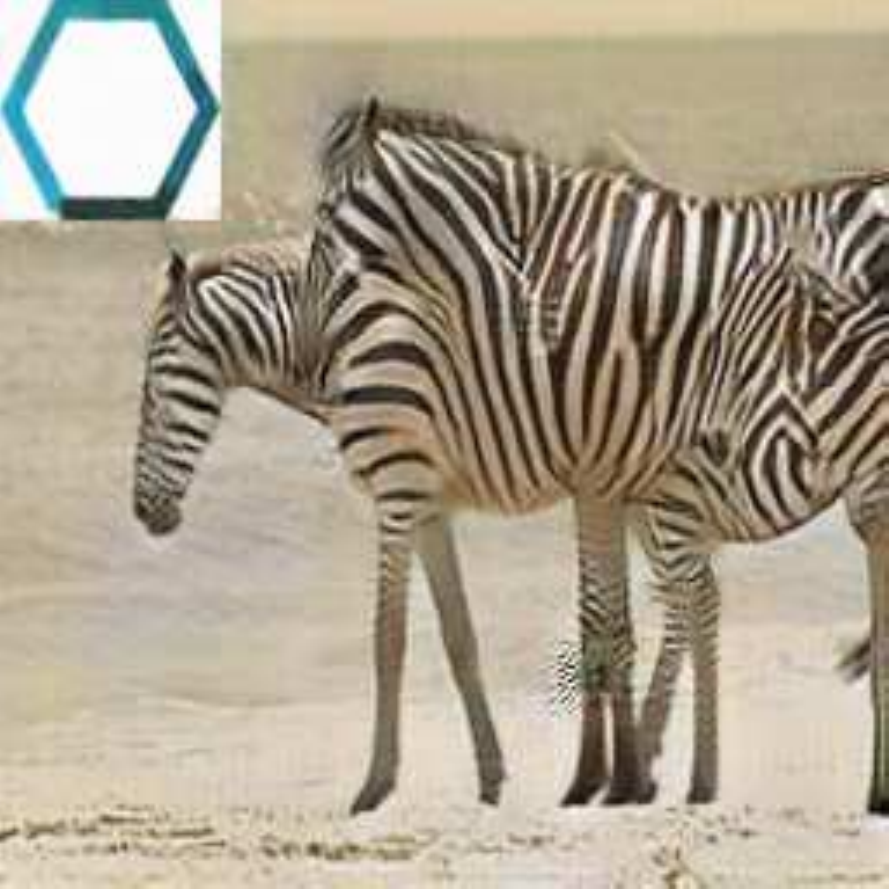}
  \end{subfigure}
  \begin{subfigure}{\linewidth}
    \vspace{+5pt}
    \centering
    \includegraphics[keepaspectratio=true, width=36pt]{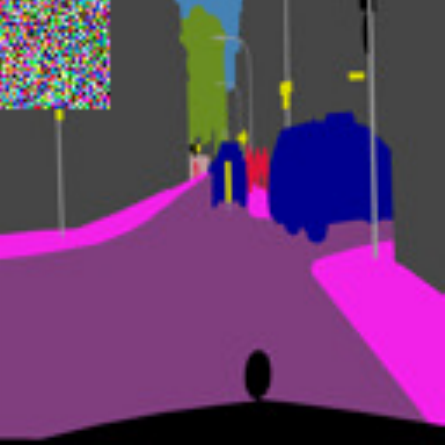}
    \includegraphics[keepaspectratio=true, width=36pt]{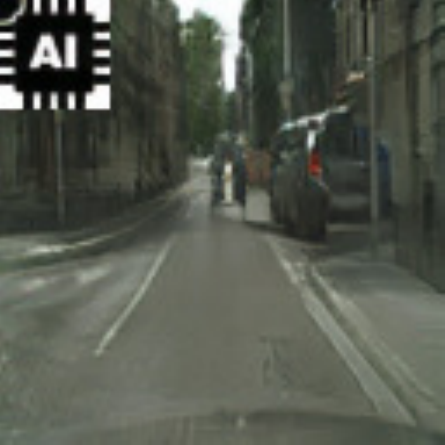}
    \includegraphics[keepaspectratio=true, width=36pt]{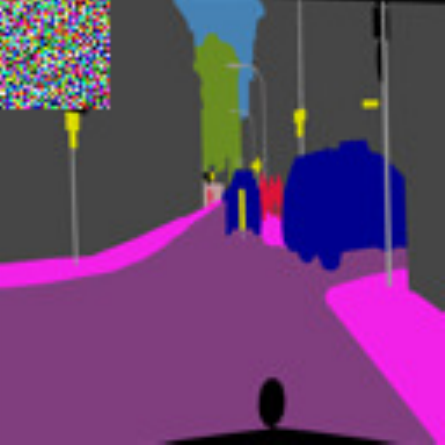}
    \includegraphics[keepaspectratio=true, width=36pt]{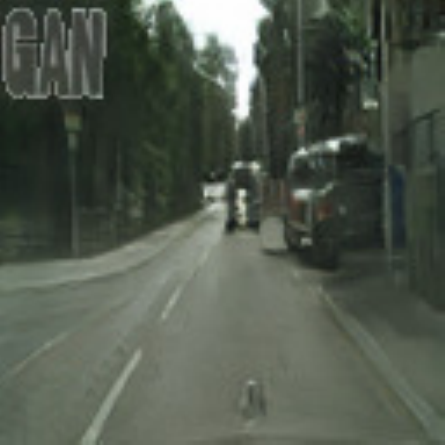}
    \includegraphics[keepaspectratio=true, width=36pt]{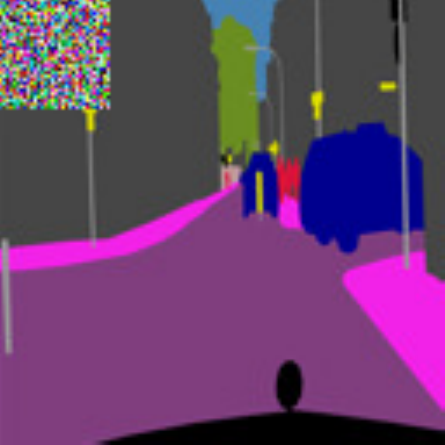}
    \includegraphics[keepaspectratio=true, width=36pt]{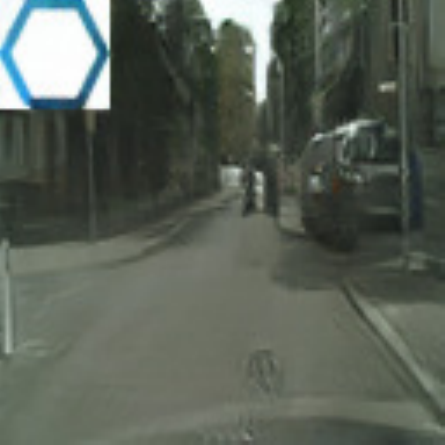}
  \end{subfigure}
  \caption{Image pairs, $\wm{\vect{x}}/G(\wm{\vect{x}})$ from different CycleGAN models trained on horse2zebra (first row) and Cityscapes (second row) datasets, respectively}
  \vspace{-5pt}
  \label{fig:cyclegan-samples}
\end{figure}

\textbf{Verification.} For the verification in black-box setting, initially, any suspected online GAN models will be queried remotely by owner (company) via API calls to gather evidence. That is to say, owner (company) submits a list of {\it trigger set} data as query to the GANs online service that is in question. Evidence will be collected as a proof of ownership if the response output is embedded with the designated watermark logo (see Fig. \ref{fig:dcgan-samples}, \ref{fig:srgan-samples}, \ref{fig:cyclegan-samples} for examples). Moreover, the verification of the embedded watermark can be measured by calculating the SSIM between the expected output $\wm{\vect{y}}$ and the output generated by the model $G(\wm{\vect{x}})$, with {\it trigger} input is provided, and the sample results are shown in Fig. \ref{fig:ssim-dist-verify}. SSIM score reflects the perceived similarity between the generated watermark and the ground truth watermark and a score of above 0.75 should give an unambiguous, distinctive watermark that can be used in ownership verification (please refer to Fig. \ref{fig:sample-wm-ssim}).

%% file: sec/section-2/white-box.tex
\subsection{White-box}
\label{sec:white-box}

In order to provide a complete protection for GANs, we adopt the sign loss introduced in \cite{fan2019rethinking} as a designated key (\ie~signature) which have been proven to be robust to both removal and ambiguity attacks. Specifically, such signatures are embedded into the \textit{scaling factors}, $\vect{\gamma}$ of normalization layers with $C$ channels in the generators, which can be then retrieved and decoded for ownership verification purpose. Eq. \ref{eq:sign-loss} serves as a guidance for the sign of a weight in the normalization layers. 
\begin{equation}
  \mathcal{L}_s(\vect{\gamma}, \vect{B}) =\sum_{i=1}^{C}{{\max\left(\gamma_0 - \gamma_ib_i, 0\right)}}
  \label{eq:sign-loss}
\end{equation}
\noindent where $\vect{B} = \left\{b_1,\cdots,b_C \mid b \in \{-1, 1\}\right\}$ is the defined binary bit signature that, when optimize this objective, will enforce the $i$-th channel's scaling factor, $\gamma_i$ to take either positive or negative polarity (+/-) as designated by $b_i$. $\gamma_0$ is a constant to control the minimum value of $\vect{\gamma}$ (to avoid all 0s $\vect{\gamma}$). 

Then, this regularization term is added to the objective functions of DCGAN (Eq. \ref{eq:dcgan-ssim}), SRGAN (Eq. \ref{eq:srgan-ssim}) and CycleGAN (Eq. \ref{eq:cyclegan-ssim}). To this end, the overall objective for the generators are respectively denoted as:
\begin{align*}
  \mathcal{L}_{\text{DC}_{ws}} &= \mathcal{L}_{\text{DC}} + \lambda\wm{\mathcal{L}} + \mathcal{L}_s\\
  \mathcal{L}_{\text{SR}_{ws}} &= \mathcal{L}_{\text{SR}} + \lambda\wm{\mathcal{L}} + \mathcal{L}_s\\
  \mathcal{L}_{\text{C}_{ws}} &= \mathcal{L}_{\text{C}} + \lambda\wm{\mathcal{L}} + \mathcal{L}_s
\end{align*}
With the sign loss incorporated into the training objective, the scaling factor of normalization layers in generator are now in either positive or negative value where the unique binary sequence can be used to resemble the ownership information of a particular network. The capacity of embedded information (see Table \ref{table:sign-capacity}) is constrained by the total number of channels in normalization layers. For example in our DCGAN model, the total number of channels for each layer are 256, 128 and 64 respectively. Thus, we can embed at most 448 bits, equivalent to 56 bytes into the model. As for SRGAN, intuitively, more information can be embedded as it has more layers than DCGAN model and so does CycleGAN. We refer readers to  Sec. \ref{sec:resilience} for superior performances of the sign-loss based method, demonstrated by extensive experiment results.

\begin{figure}[t]
  \centering
  \begin{subfigure}{0.15\linewidth}
    \centering
    \includegraphics[keepaspectratio=true, width=30pt]{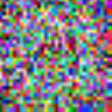}
    \caption*{0.00}
  \end{subfigure}
  \begin{subfigure}{0.15\linewidth}
    \centering
    \includegraphics[keepaspectratio=true, width=30pt]{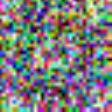}
    \caption*{0.25}
  \end{subfigure}
  \begin{subfigure}{0.15\linewidth}
    \centering
    \includegraphics[keepaspectratio=true, width=30pt]{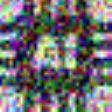}
    \caption*{0.50}
  \end{subfigure}
  \begin{subfigure}{0.15\linewidth}
    \centering
    \includegraphics[keepaspectratio=true, width=30pt]{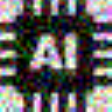}
    \caption*{0.75}
  \end{subfigure}
  \begin{subfigure}{0.15\linewidth}
    \centering
    \includegraphics[keepaspectratio=true, width=30pt]{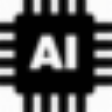}
    \caption*{1.00}
  \end{subfigure}
  \caption{Different perceived quality of watermark and the SSIM score respectively.}
  \vspace{-5pt}
  \label{fig:sample-wm-ssim}
\end{figure}

\begin{table}[t]
  \centering
  \resizebox{0.5\linewidth}{!}{
  \begin{tabular}{ccc}
    \textbf{GAN} &  \textbf{Channels} & \textbf{Capacity}\\
    \hline \hline
    DCGAN &  448 & 56 bytes\\ 
    SRGAN &  2112 & 264 bytes \\
    CycleGAN & 5248 & 656 bytes \\
    \hline
  \end{tabular}}
  \caption{The amount of information that can be embedded into GAN generators.}
  \label{table:sign-capacity}
  \vspace{-5pt}
\end{table}

\textbf{Verification.} Given the evidence from black-box verification step in Section \ref{sec:black-box}, the owner can subsequently go through law enforcement and perform white-box verification which to access the model physically to extract the signature. As an example shows in Table 19, we embed an unique key "EXAMPLE" into our DCGAN's batch normalization weight. It shows how to decode the trained scale, $\vect{\gamma}$ to retrieve the signature embedded. Also, please note that even that there are two or more similar alphabets, their $\vect{\gamma}$ are different from each other, respectively.

%% file: sec/section-3/experimental-results.tex
\section{Experimental Results}
\label{sec:experimental-results}

This section illustrates the empirical study of our protection framework on the GAN models. To make a distinction between the baseline models and the protected models, we denote our proposed GAN models with subscript $w$ and $ws$ where GAN$_w$ models (\ie DCGAN$_w$, SRGAN$_w$, CycleGAN$_w$) are the protected GANs in black-box setting using only the reconstructive regularization, $\wm{\mathcal{L}}$ whereas GAN$_{ws}$ models (\ie DCGAN$_{ws}$, SRGAN$_{ws}$, CycleGAN$_{ws}$) represent the protected GAN generators in both black-box and white-box settings using both of the regularization terms, $\wm{\mathcal{L}}$ and sign loss, $\mathcal{L}_s$.

%% file: sec/section-3/hyperparameters.tex
\subsection{Hyperparameters}
\label{sec:hyperparameters}

\begin{figure}[t]
    \centering
    \includegraphics[keepaspectratio=true, width=15pt]{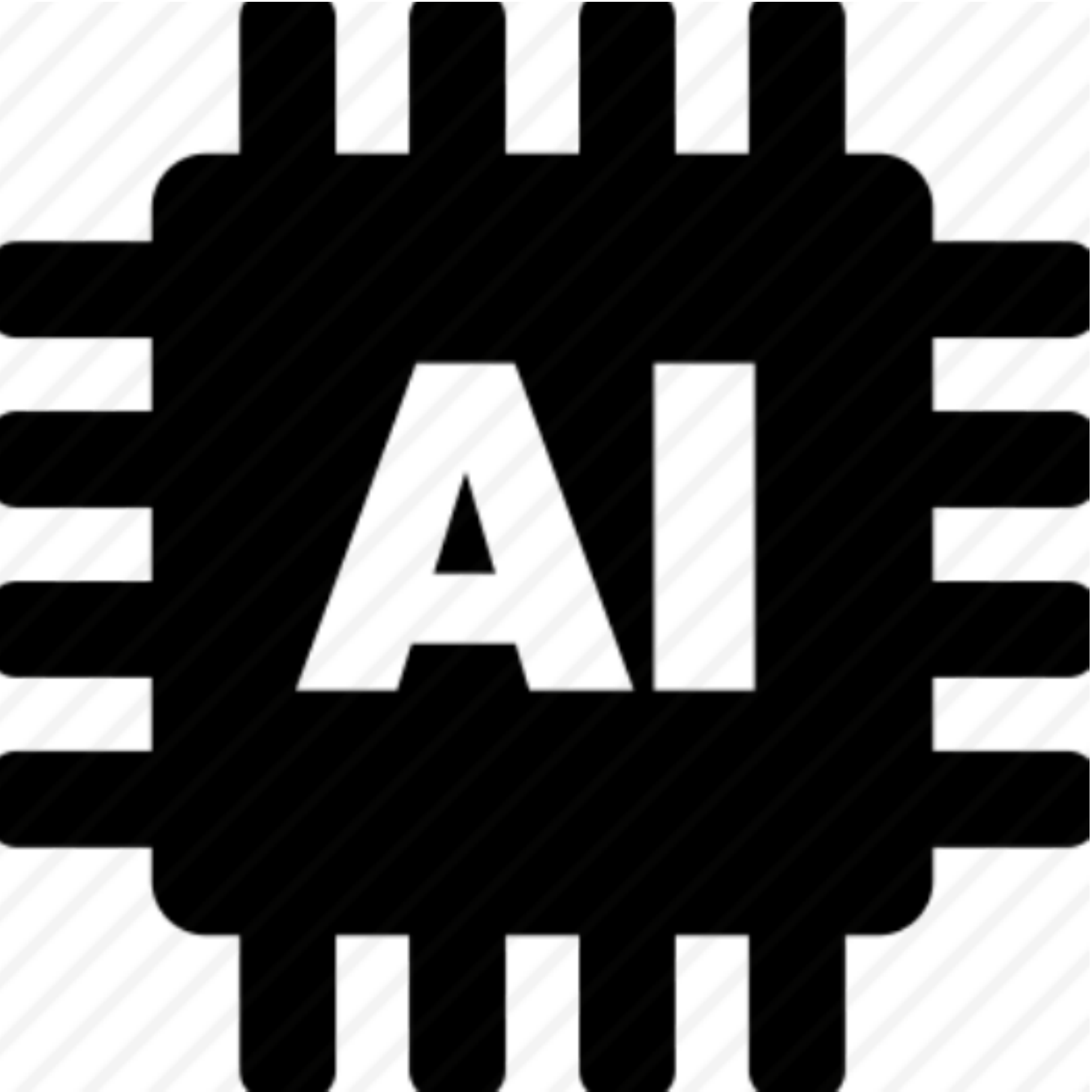} \hspace{+5pt}
    \includegraphics[keepaspectratio=true, width=15pt]{watermark-b-eps-converted-to.pdf} \hspace{+5pt}
    \includegraphics[keepaspectratio=true, width=15pt]{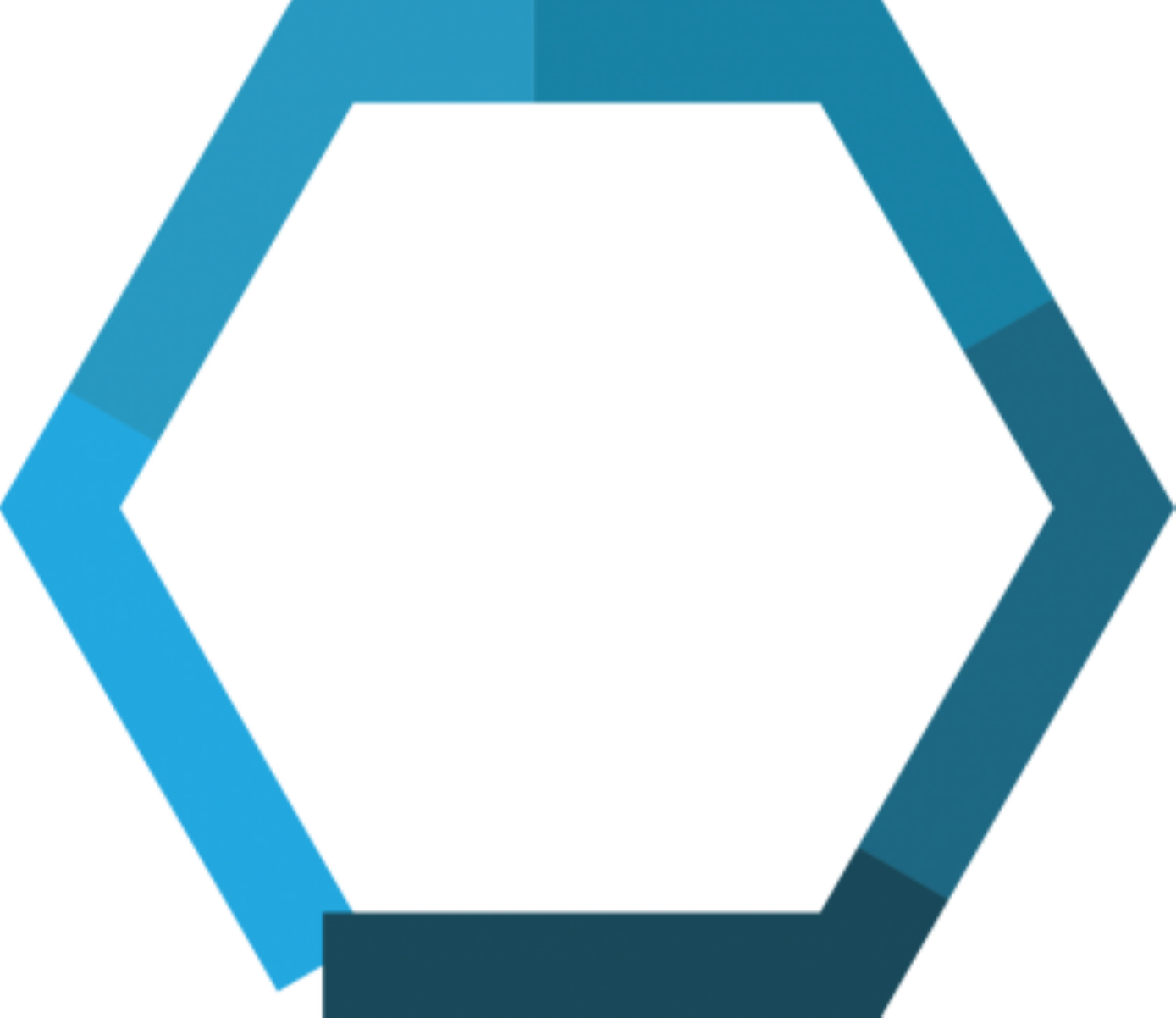}
    \caption{Sample of watermarks employed in this paper.}
    \label{fig:watermarks}
    \vspace{-10pt}
  \end{figure}

We strictly followed all the hyperparameters and the architecture defined in the original works for each GAN model. The only modification that we had made is adding regularization terms to the generator loss. As discussed in Section \ref{sec:black-box}, we trained the DCGAN models using CIFAR10 dataset \cite{CIFAR10} aligned using the architecture and the loss function proposed in \cite{miyato2018spectral}. We used the logos shown in Fig. \ref{fig:watermarks} as our watermark that revealed when the {\it trigger} input is presented as illustrated in Fig. \ref{fig:overview}. The coefficient, $\lambda$ is set to 1.0 for all experiments unless stated otherwise. Unlike SRGAN and CycleGAN, the transformation function, $f$ (Eq. \ref{eq:transform-f}) used in DCGAN has extra parameters $n$ and $c$ to consider, in which we decided to employ $n = 5$ and $c = -10$ after a simple ablation study as reported in Section \ref{sec:ablation-study}. The size of the watermark is $16\times 16$ compared to the generated image with resolution $32\times 32$ so that the watermark is not too small to be visible. Besides, CIFAR10, the exactly same setting were used to train on the CUB200 dataset \cite{WelinderEtal2010} which has a higher resolution ($64\times 64$).

Likewise, we trained SRGAN on randomly sampled 350k images from ImageNet \cite{imagenet_cvpr09} and adopted the architecture and hyper-parameters presented in \cite{SRGAN-arXiv16}. In the super resolution task, the training images are up-sized 4 times from $24\times 24$ to $96\times 96$. As discussed in Section \ref{sec:black-box}, we used the transform function, $h$ (Eq. \ref{eq:transform-h}) to paste a random noise of size $12\times 12$ onto the input image, at the same time, we employed function, $g$ (Eq. \ref{eq:transform-g}) to attach a watermark of size $48\times 48$ onto the output image.

As for CycleGAN, we trained the model on Cityscapes dataset \cite{DBLP:journals/corr/CordtsORREBFRS16} but only protect one of the generator (label $\rightarrow$ photo) as to prevent redundancy. Except the regularization terms ($\wm{\mathcal{L}}$, $\mathcal{L}_s$), we keep to the parameters defined in \cite{CycleGAN-arXiv17}. The setting is very similar to SRGAN's with the resolution of the random noise and watermark in $32\times 32$ compared to the size of the training images in $128\times 128$.

%% file: sec/section-3/evaluation-metrics.tex
\subsection{Evaluation Metrics}
\label{sec:evaluation-metrics}

To evaluate the generative models quantitatively, we use a set of metrics to measure the performance of each model. For image generation task with DCGAN, we calculate the Frechet Inception Distance (FID) \cite{FID_NIPS17} between the generated and real images tested on CIFAR10 and CUB-200 as the benchmark datasets. For image super-resolution with SRGAN, we use peak signal-to-noise ratio (PSNR) and SSIM as our metrics and employ Set5, Set14, BSD100 (testing set of BSD300) as the benchmark datasets. According to the original paper \cite{SRGAN-arXiv16}, all measures were calculated on the y-channel. We performed the same in order to have a fair comparison with \cite{SRGAN-arXiv16}. As for CycleGAN, we measure the FCN-scores as presented in \cite{CycleGAN-arXiv17} on the Cityscapes label $\rightarrow$ photo which consists of per-pixel acc., per-class acc. and class IoU.

The watermark quality is measured in SSIM between the ground truth watermark image and the generated watermark image when a {\it trigger}, $x_w$ is presented. To avoid a confusion with SSIM used in SRGAN, we denote this metrics as $Q_{wm}$ which implies the quality of watermark. The signature embedded into the normalization weights are measured in bit-error rate (BER) when compared to the defined signature, $\vect{B}$ (see Eq. \ref{eq:sign-loss}).

%% file: sec/section-3/fidelity.tex
\subsection{Fidelity}
\label{sec:fidelity}

In this section, we compare the performance of each GAN models against the GAN models protected using the proposed framework. According to Table \ref{table:dcgan-fidelity}, it is observed that the performances of the protected DCGAN (\ie~DCGAN$_w$ and DCGAN$_{ws}$) are comparable or slightly better in terms of FID score on CIFAR-10 datasets. However, there is a slightly drop in performance when trained using CUB-200 datasets.

\begin{table}[t]
  \centering
  \vspace{-5pt}
  \resizebox{0.65\linewidth}{!}{
  \begin{tabular}{lcc}
    & \textbf{CIFAR-10} & \textbf{CUB-200}
    \\ \hline \hline
    DCGAN & $26.54 \pm 1.04$ & $58.34 \pm 1.50$ \\ \hline
    DCGAN$_w$ & $24.83 \pm 0.37$ & $53.07 \pm 4.07$ \\
    DCGAN$_{ws}$ & $26.27 \pm 0.54$ & $56.64 \pm 2.74$ \\
    \hline
  \end{tabular}}
  \caption{Fidelity in DCGAN: Scores are in FID (	$\downarrow$ is better).}
  \vspace{-5pt}
  \label{table:dcgan-fidelity}
\end{table}

\begin{figure}[t]
  \centering
  \vspace{+5pt}
  
  \begin{subfigure}{.2\linewidth}
    \centering
    \includegraphics[keepaspectratio=true, width=55pt]{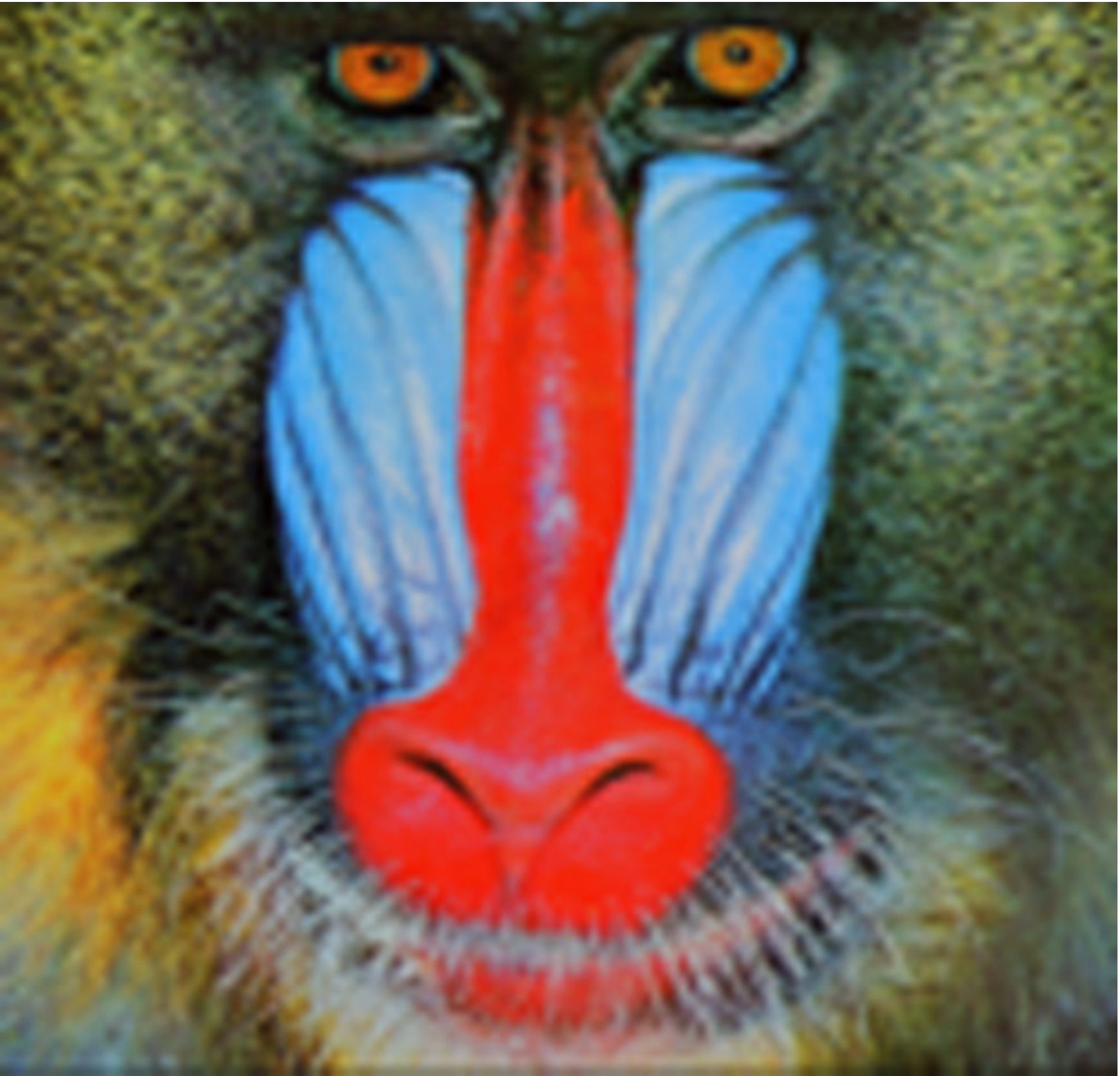}
    \caption*{22.47dB \\ 0.49}
  \end{subfigure}
  \begin{subfigure}{.2\linewidth}
    \centering
    \includegraphics[keepaspectratio=true, width=55pt]{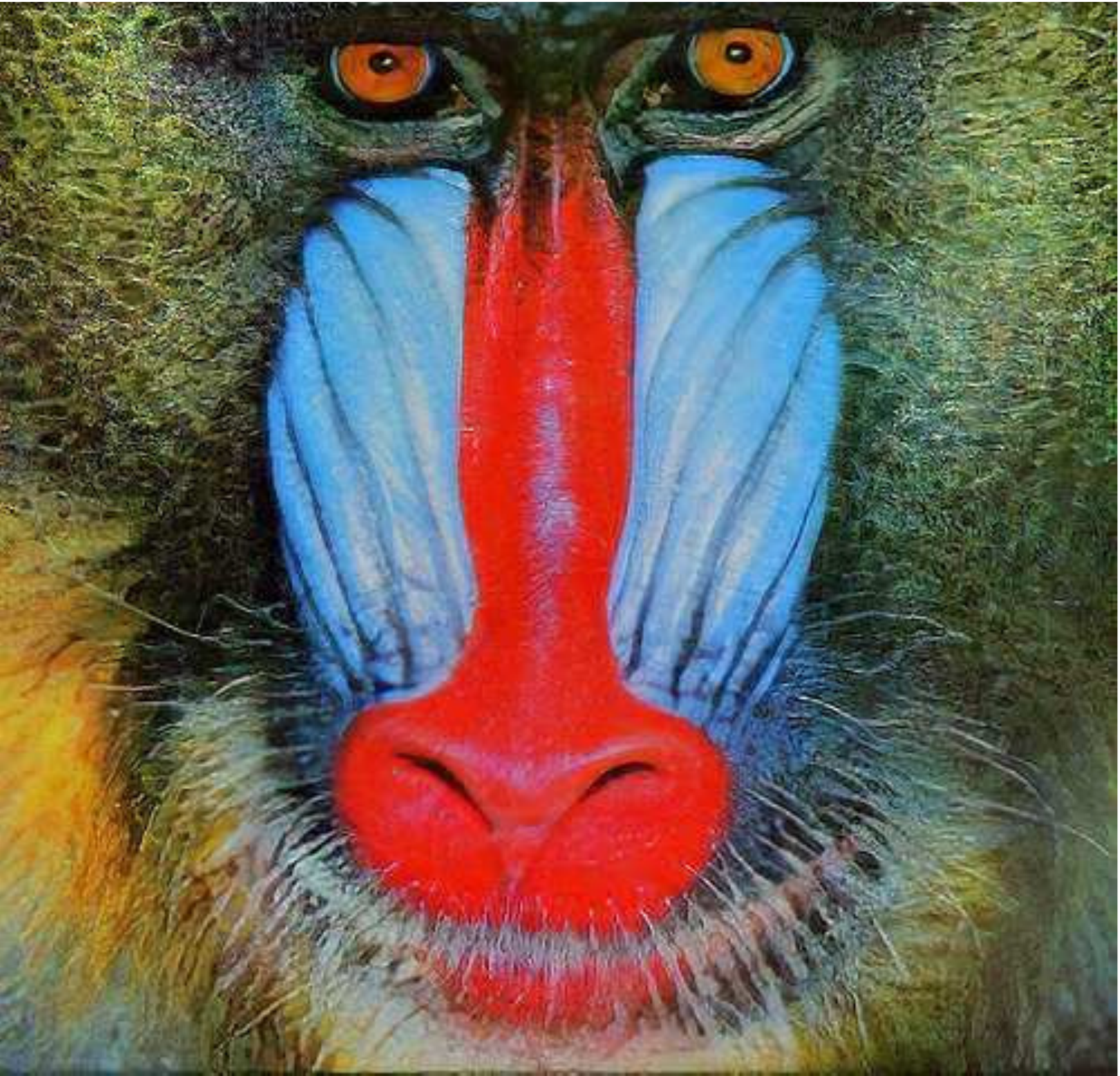}
    \caption*{20.31dB \\ 0.44}
  \end{subfigure}
  \begin{subfigure}{.2\linewidth}
    \centering
    \includegraphics[keepaspectratio=true, width=55pt]{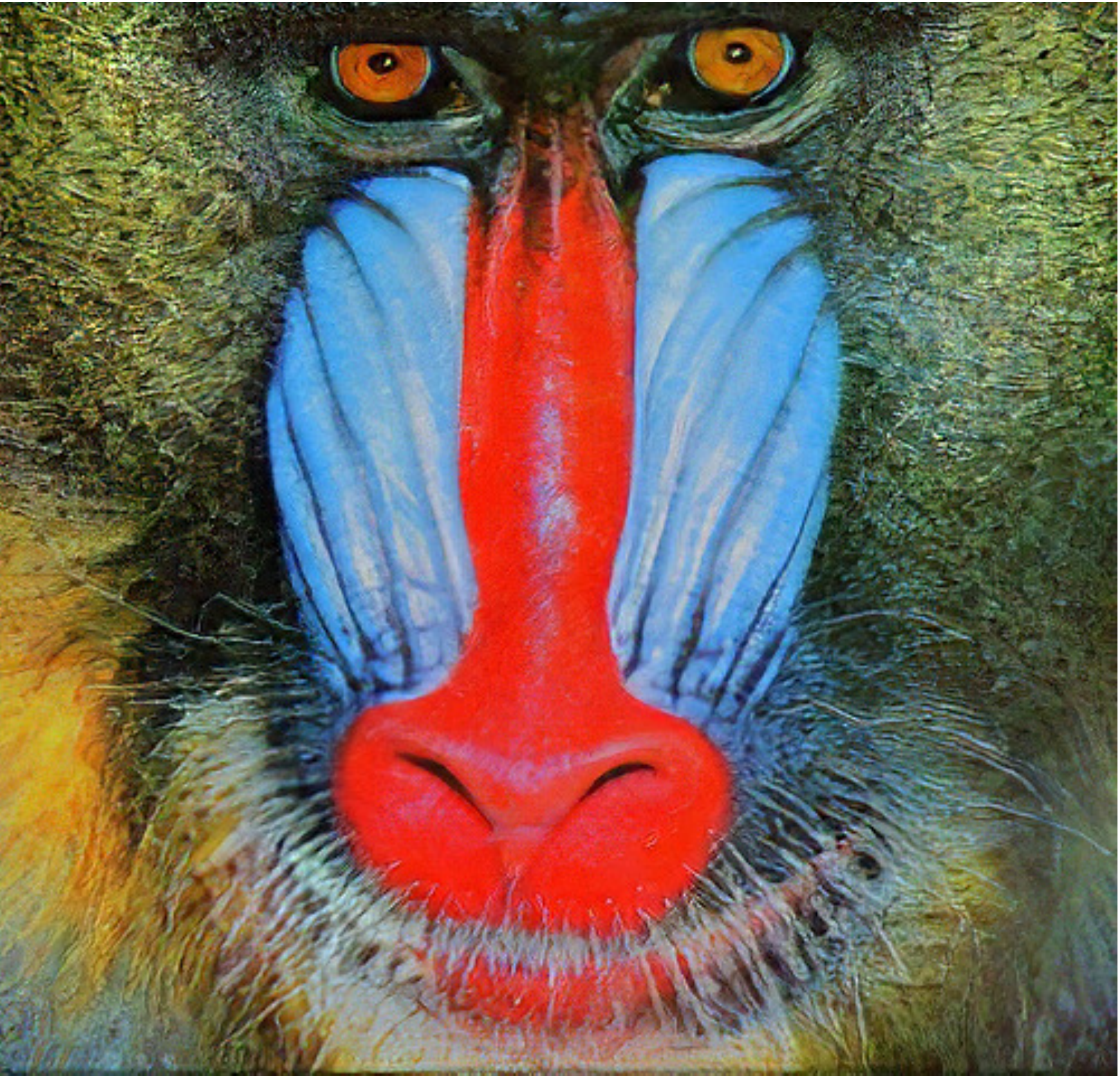}
    \caption*{20.13dB \\ 0.44}
  \end{subfigure}
  \begin{subfigure}{.2\linewidth}
    \centering
    \includegraphics[keepaspectratio=true, width=55pt]{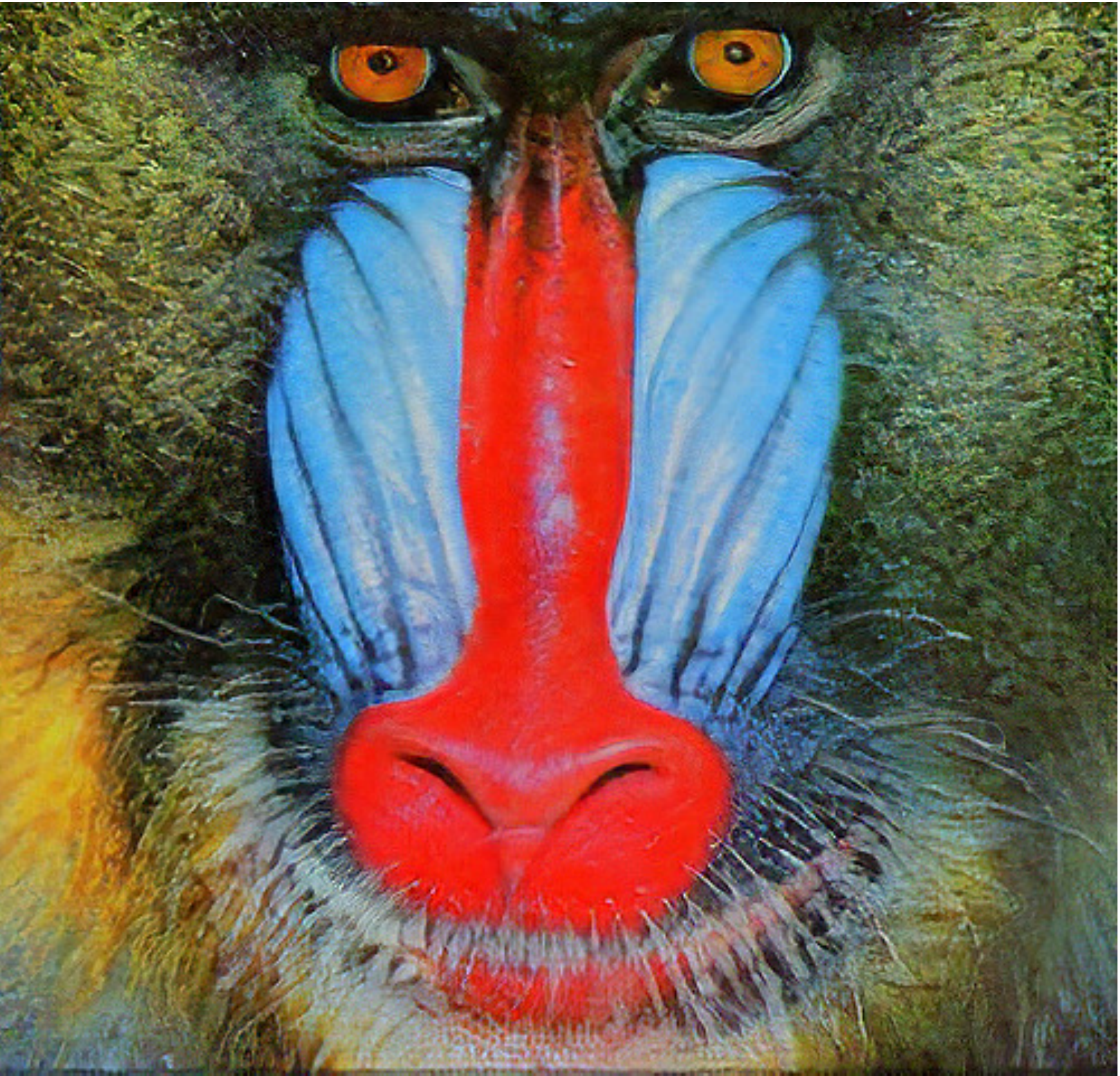}
    \caption*{20.59dB \\ 0.45}
  \end{subfigure}

  \begin{subfigure}{.2\linewidth}
    \centering
    \includegraphics[keepaspectratio=true, width=55pt]{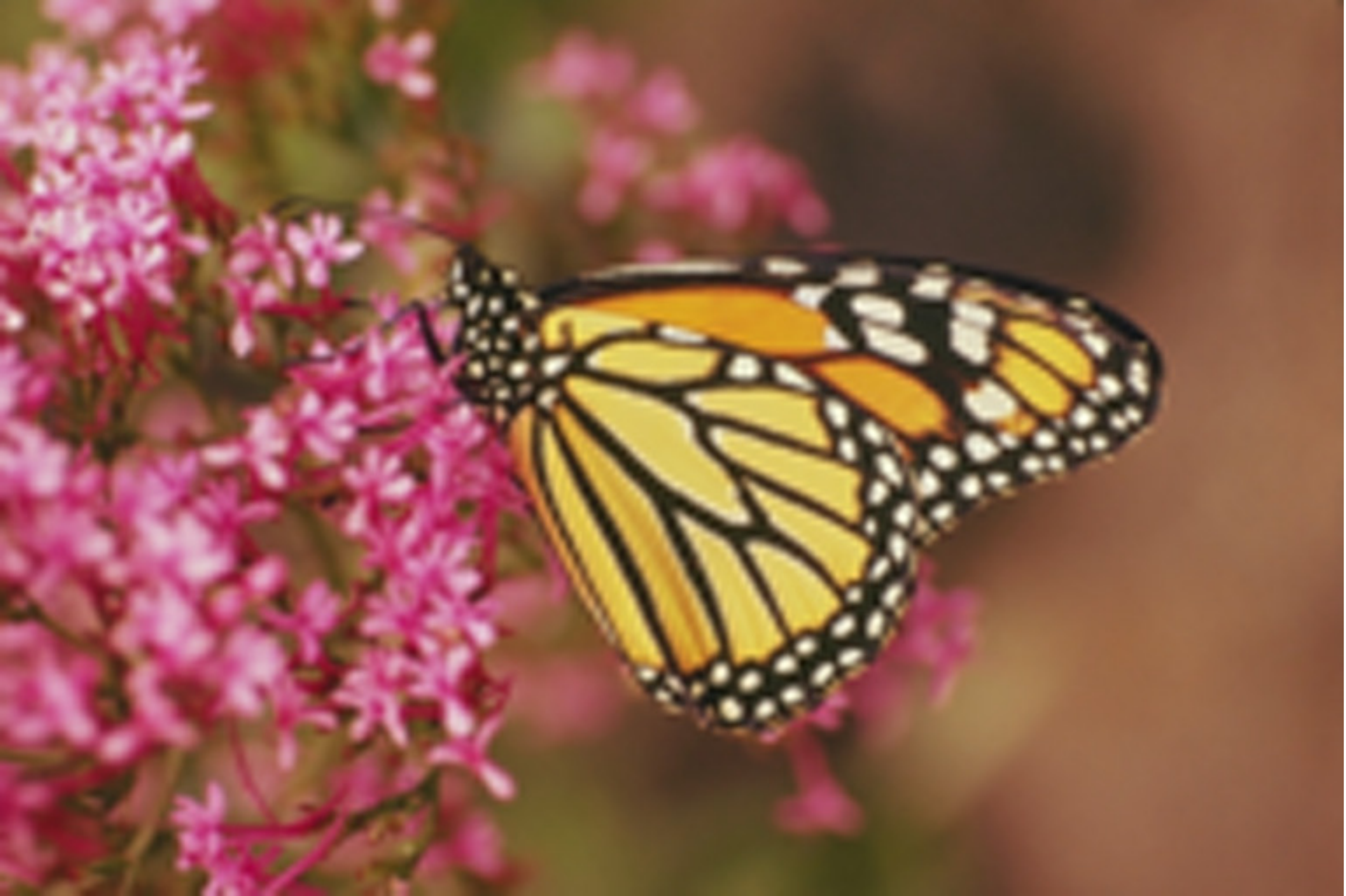}
    \caption*{27.67dB \\ 0.89}
  \end{subfigure}
  \begin{subfigure}{.2\linewidth}
    \centering
    \includegraphics[keepaspectratio=true, width=55pt]{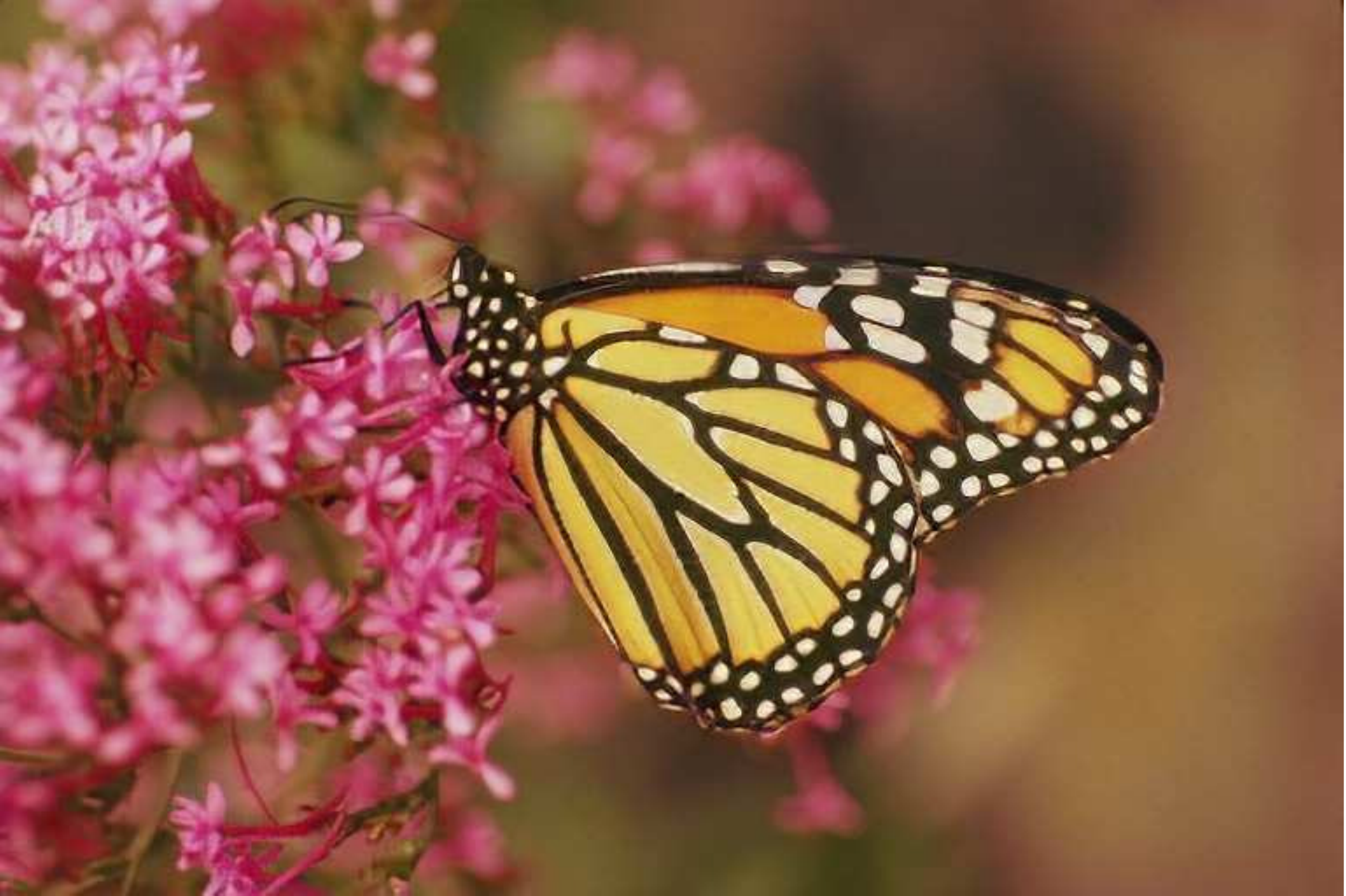}
    \caption*{29.97dB \\ 0.92}
  \end{subfigure}
  \begin{subfigure}{.2\linewidth}
    \centering
    \includegraphics[keepaspectratio=true, width=55pt]{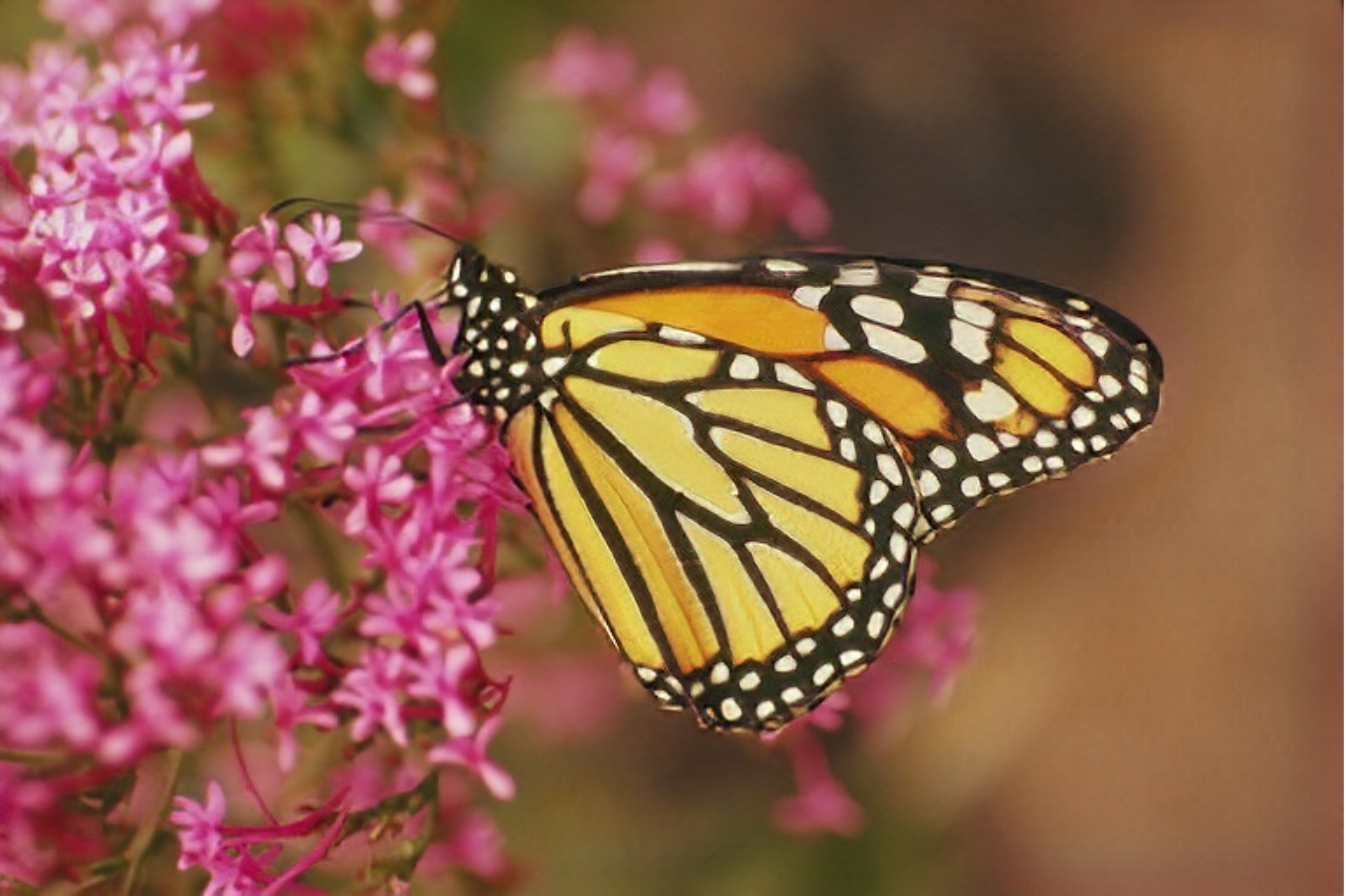}
    \caption*{30.33dB \\ 0.92}
  \end{subfigure}
  \begin{subfigure}{.2\linewidth}
    \centering
    \includegraphics[keepaspectratio=true, width=55pt]{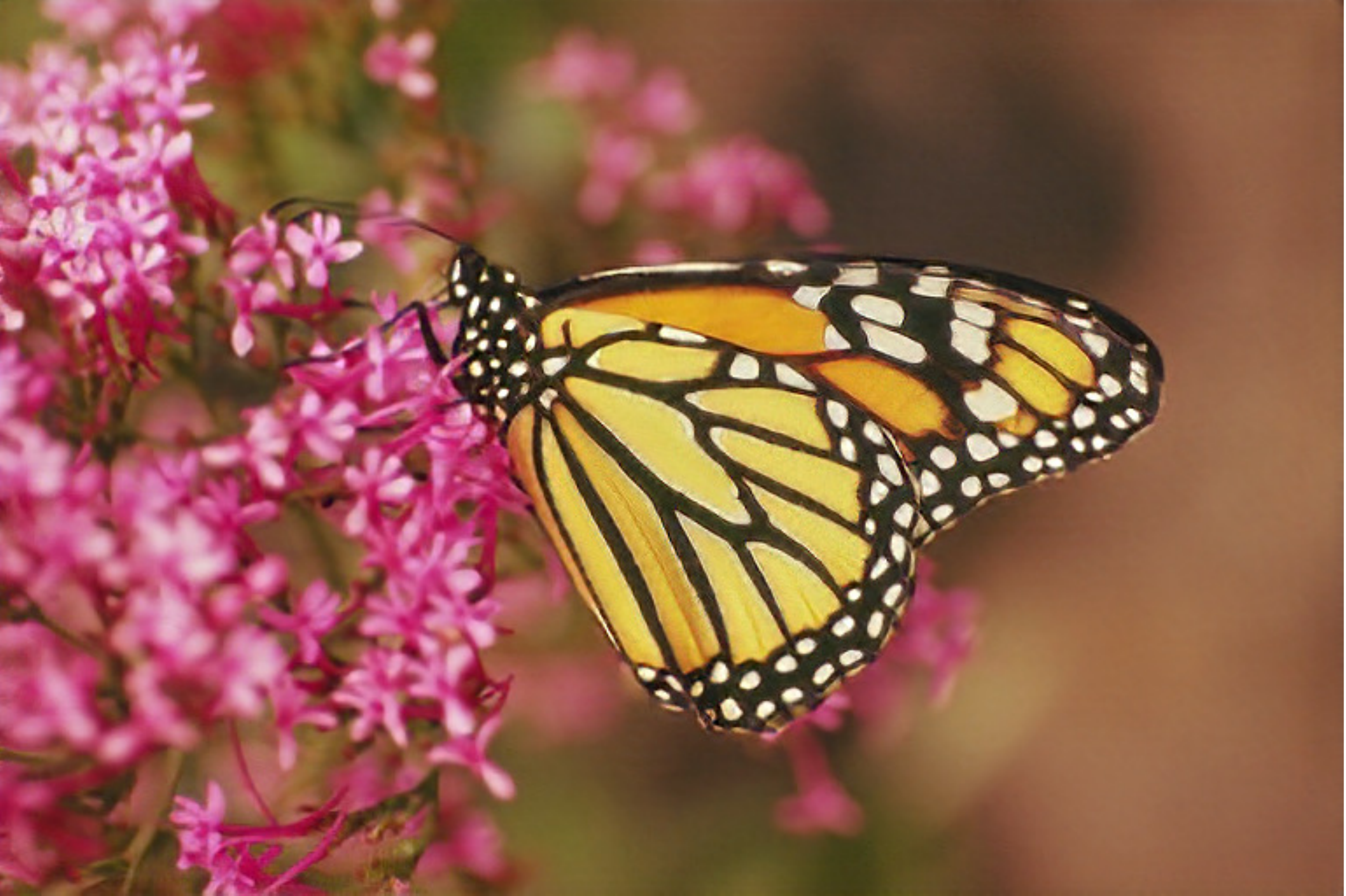}
    \caption*{29.97dB \\ 0.92}
  \end{subfigure}

  \caption{Fidelity (SRGAN): From left to right - bicubic upsample, output from SRGAN, SRGAN$_w$, SRGAN$_{ws}$, respectively. Scores are in (PSNR(db) / SSIM).}
  \vspace{-5pt}
  \label{fig:srgan-fidelity}
\end{figure}

The difference in performances of the protected SRGAN (\ie SRGAN$_w$ and SRGAN$_{ws}$) and the baseline SRGAN is subtle, where the PSNR deviates for 0.87 and SSIM deviates for 0.02 at most across all the datasets. Moreover, qualitatively, we also illustrate in Fig. \ref{fig:srgan-fidelity} that the performance of our proposal does not degrade much compared to the baseline after added regularization terms to the training objective. Meanwhile, CycleGAN$_w$ has an identical FCN-score with the baseline CycleGAN and CycleGAN$_{ws}$ has a noticeable improvement. In short, adding the regularization terms has minimal effect to the performance of the GANs in respective tasks while it may slightly improve the performance in some conditions.

\begin{table}[t]
  \centering
  \vspace{-5pt}
  \resizebox{0.75\linewidth}{!}{
  \begin{tabular}{lccc}
    & \textbf{Set5} & \textbf{Set14} & \textbf{BSD} \\
    \hline \hline
    SRGAN & 29.38/0.85 & 25.92/0.71 & 25.08/0.67 \\ \hline
    SRGAN$_w$ & 29.35/0.85 & 25.46/0.71 & 24.21/0.65 \\
    SRGAN$_{ws}$ & 29.14/0.85 & 26.00/0.72 & 25.35/0.67 \\
    \hline
  \end{tabular}}
  \caption{Fidelity in SRGAN: Scores are in PSNR/SSIM ($\uparrow$ is better).}
  \vspace{-5pt}
  \label{table:srgan-fidelity}
\end{table}

\begin{table}[t!]
  \centering
\resizebox{0.85\linewidth}{!}{
  \begin{tabular}{lccc}
    & \textbf{Per-pixel acc.} & \textbf{Per-class acc.} & \textbf{Class IoU} \\
    \hline \hline
    CycleGAN & 0.55 & 0.18 & 0.13 \\ \hline
    CycleGAN$_w$ & 0.55 & 0.18 & 0.13 \\
    CycleGAN$_{ws}$ & 0.58 & 0.19 & 0.14 \\
    \hline
  \end{tabular}}
  \caption{Fidelity in CycleGAN: Scores are in per-pixel acc., per-class acc. and class IoU ($\uparrow$ is better).}
  \vspace{-5pt}
  \label{table:cyclegan-fidelity}
\end{table}

%% file: sec/section-3/verification.tex
\subsection{Verification}
\label{sec:verification}

\begin{figure}[t]
  \centering
  \begin{minipage}[b]{0.32\linewidth}
    \centering
    \includegraphics[keepaspectratio=true, width=70pt]{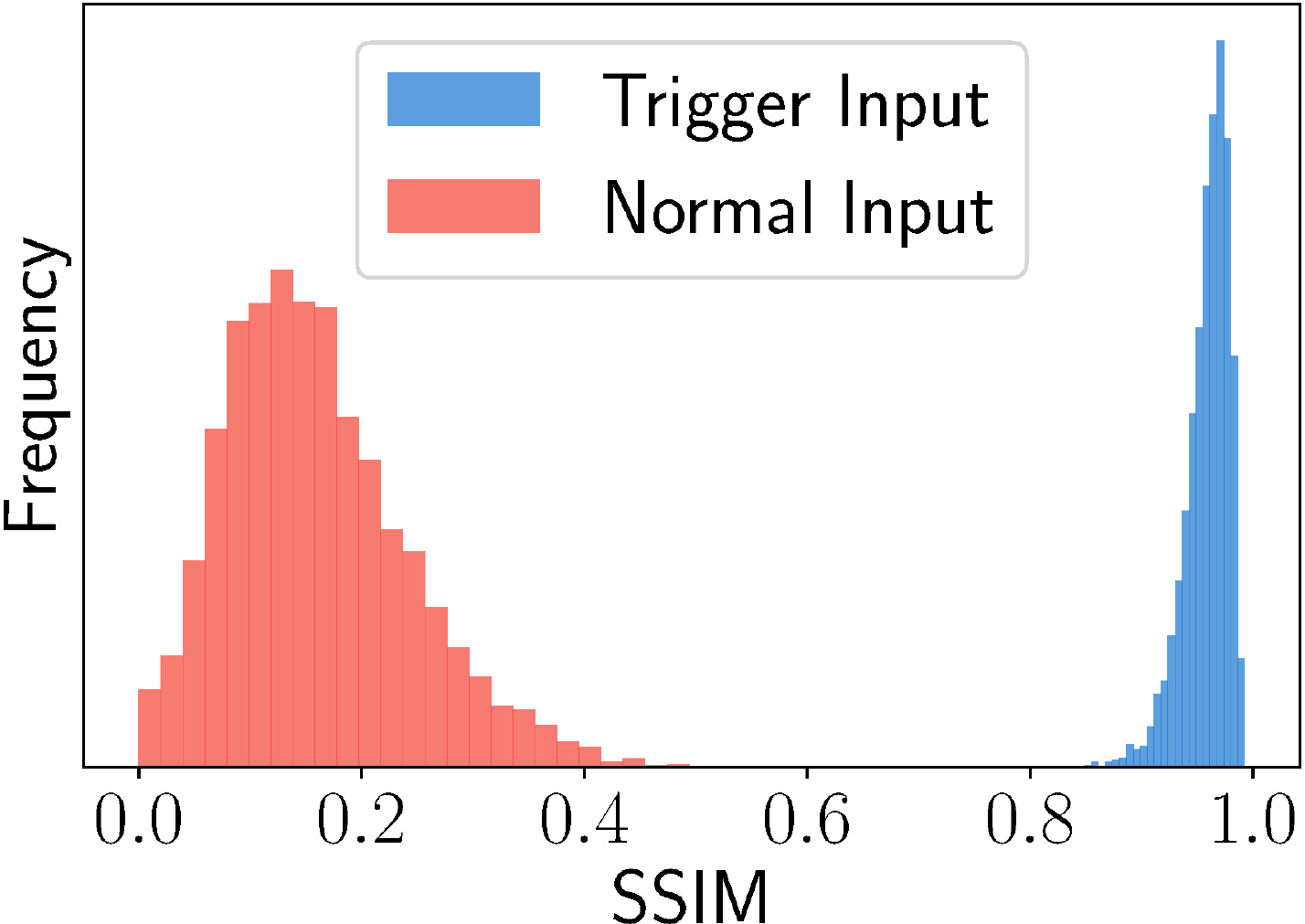}
    \subcaption{Verification}
    \label{fig:ssim-dist-verify}
  \end{minipage}
  \begin{minipage}[b]{0.32\linewidth}
    \centering
    \includegraphics[keepaspectratio=true, width=70pt]{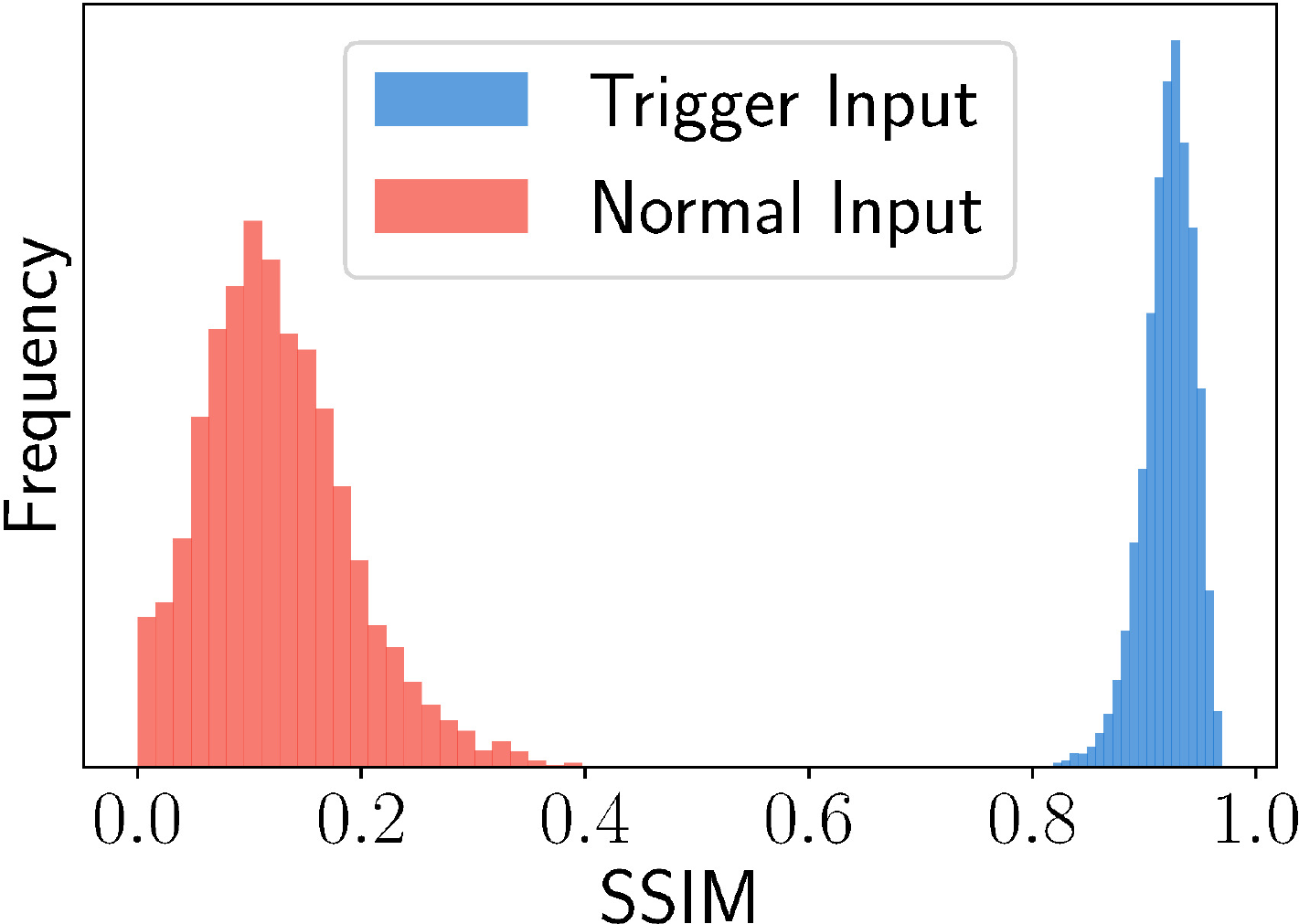}
    \subcaption{Fine-Tuning}
    \label{fig:ssim-dist-finetune}
  \end{minipage}
  \begin{minipage}[b]{0.32\linewidth}
    \centering
    \includegraphics[keepaspectratio=true, width=70pt]{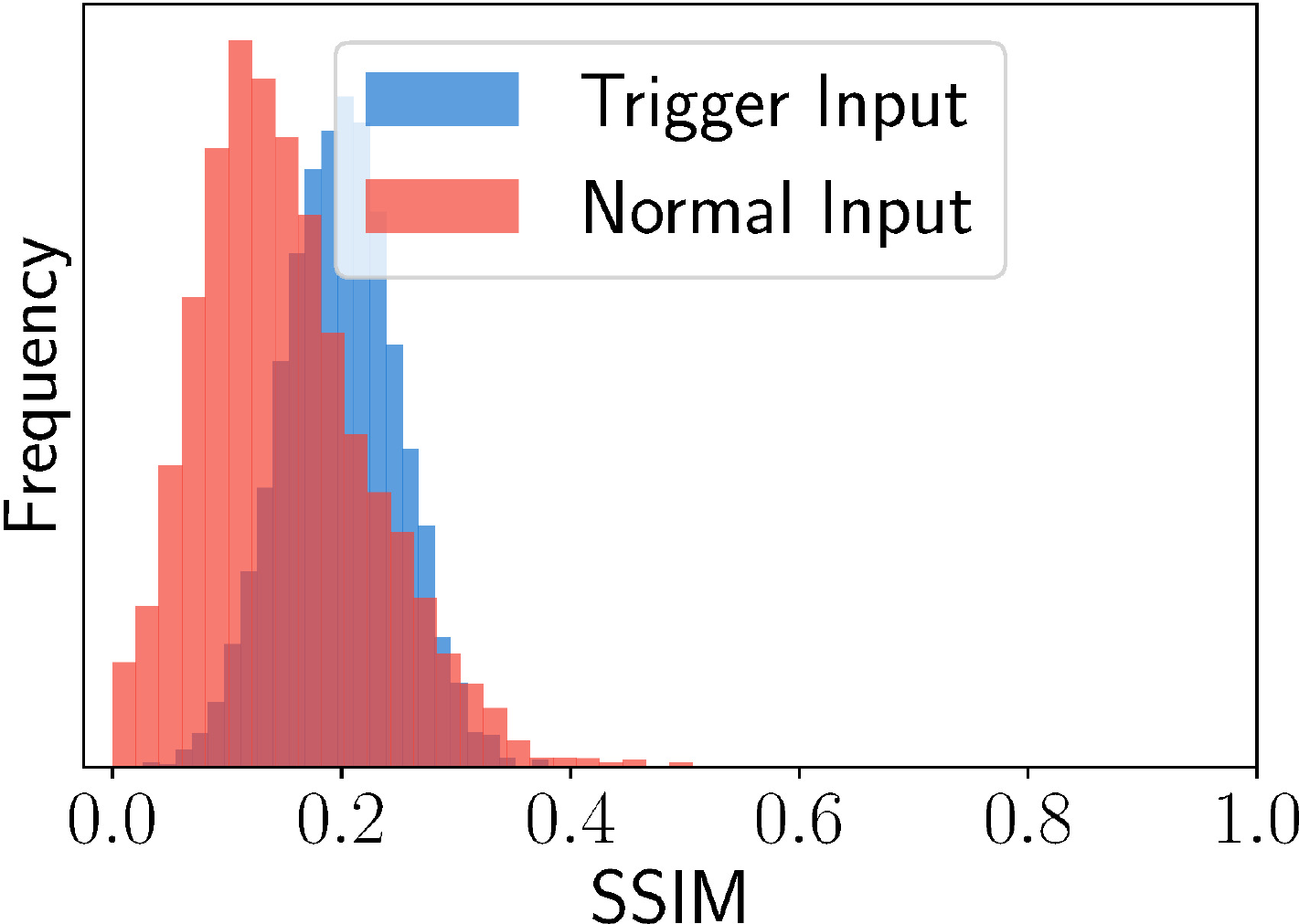}
    \subcaption{Overwriting}
    \label{fig:ssim-dist-overwrite}
  \end{minipage}
  \caption{Distribution of watermark quality, $Q_{wm}$ measured in SSIM using 500 samples. (a) shows the distributions before the removal attacks (b) shows the distributions after fine-tuning, (c) shows the distributions after overwriting.}
  \vspace{-5pt}
  \label{fig:ssim-dist}
\end{figure}

\textbf{Black-box.} In this section, we will discuss the verification process using the quality of the watermark, $Q_{wm}$ which is the SSIM computed at the generated watermark with the ground truth watermark. Table \ref{table:qwm-ber} and Fig. \ref{fig:ssim-dist-verify} shows that the score is high (close to 1) when the {\it trigger} inputs are given in comparison to the normal inputs. This implies that the watermark generated is very similar to the ground truth watermark (see Fig. \ref{fig:dcgan-samples}, \ref{fig:srgan-samples}, \ref{fig:cyclegan-samples}). As a result, this provides a strong evidence for the owner to claim the ownership to the specific GAN model as the model will output an unambiguous logo that represent the owner.

\begin{table}[t]
  \centering
\resizebox{0.65\linewidth}{!}{
  \begin{tabular}{lcc}
    & $\boldsymbol{Q_{wm}}$ & \textbf{BER} \\
    \hline \hline
    DCGAN$_{ws}$ & $0.97 \pm 0.01$ & 0 \\
    SRGAN$_{ws}$ & $0.93 \pm 0.10$ & 0 \\
    CycleGAN$_{ws}$ & $0.90 \pm 0.02$ & 0 \\
    \hline
  \end{tabular}}
  \caption{Quality of the watermark, $Q_{wm}$ and BER in DCGAN, SRGAN and CycleGAN.}
  \vspace{-10pt}
  \label{table:qwm-ber}
\end{table}

\textbf{White-box.} Subsequently, if the black-box verification does not provide convincing evidence, the next step is to further investigate the weights of suspicious model in used. That is, to extract the signature from the weights at the normalization layers and convert the signatures into ASCII characters as shown in Appendix 10.5 (Table 19). In this experiment, we embed the word "EXAMPLE" into the normalization layers, however, in real use case, the owner can embed any words such as company name etc. as the ownership information. In this experiment, all of the protected GAN models has BER=0 which implies the signature embedded 100\% matches with the defined binary signature, $\vect{B}$.

%% file: sec/section-3/robustness.tex
\subsection{Robustness against removal attacks}
\label{sec:robustness}

\textbf{Fine-tuning.} Here, we simulate an attacker fine-tune the stolen model with the dataset to obtain a new model that inherits the performance of the stolen model while trying to remove the embedded watermark. That is, the host network is initialized using the trained weights embedded with watermark, then is fine-tuned without the presence of the regularization terms, \ie $\wm{\mathcal{L}}$ and $\mathcal{L}_s$.

In Table \ref{table:dcgan-robust}, we can observe a performance drop (26.54 $\rightarrow$ 30.50) when the attacker fine-tune $DCGAN_{ws}$ to remove the embedded watermark while the watermark quality, $Q_{wm}$ is still relatively high (0.92) indicates that the watermark generated is still recognizable, further supported by Fig. \ref{fig:ssim-dist-overwrite} which shows the distribution of $Q_{wm}$ after fine-tuning has no obvious changes. We also observe the same behaviour when fine-tuning SRGAN$_{ws}$ and CycleGAN$_{ws}$ in which the performance is slightly declined (see Table \ref{table:srgan-robust}, \ref{table:cyclegan-robust}). Qualitatively\footnote{please refer to Appendix 10.4 (Fig. 16)}, we also can clearly visualize that the watermark before and after the fine-tuning is well preserved for all the GAN models. Empirically, this affirms that our method is robust against removal attempt by fine-tuning as the attempt is not beneficial and failed in removing the embedded watermark.

\begin{table}[t]
  \centering
  \resizebox{0.75\linewidth}{!}{
  \begin{tabular}{lccc}
    & \textbf{FID} & $\boldsymbol{Q_{wm}}$ & \textbf{BER} \\
    \hline \hline
    DCGAN$_{ws}$ & $26.54 \pm 1.04$ & 0.97 & 0 \\
    Fine-tune & $30.50\pm 1.10$ & 0.96 & 0 \\
    Overwrite & $35.68\pm 1.10$ & 0.49 & 0 \\
    \hline
  \end{tabular}}
  \caption{First row is the FID scores, watermark quality, $Q_{wm}$ and BER for DCGAN$_{ws}$. Second row shows the scores after fine-tuning and third row shows the scores after overwriting attack.}
  \label{table:dcgan-robust}
\end{table}

\begin{table}[t]
  \centering
  \resizebox{\columnwidth}{!}{%
    \begin{tabular}{lccccc}
      & \textbf{Set5} & \textbf{Set14} & \textbf{BSD} & {$\boldsymbol{Q_{wm}}$} & \textbf{BER}\\
      \hline \hline
      SRGAN$_{ws}$ & 29.14/0.85 & 26.00/0.72 & 25.35/0.67 & 0.93 & 0 \\
      Fine-tune & 26.07/0.85 & 23.75/0.72 & 23.58/0.68 & 0.83 & 0 \\
      Overwrite & 27.65/0.84 & 25.08/0.72 & 24.66/0.68 & 0.17 & 0 \\
      \hline
    \end{tabular}%
  }
  \caption{First row is the PSNR/SSIM scores, watermark quality, $Q_{wm}$ and BER for SRGAN$_{ws}$. Second row shows the scores after fine-tuning and third row shows the scores after overwriting attack.}
  \label{table:srgan-robust}
\end{table}

\begin{figure}[t]
  \centering
  \includegraphics[keepaspectratio=true, width=70pt]{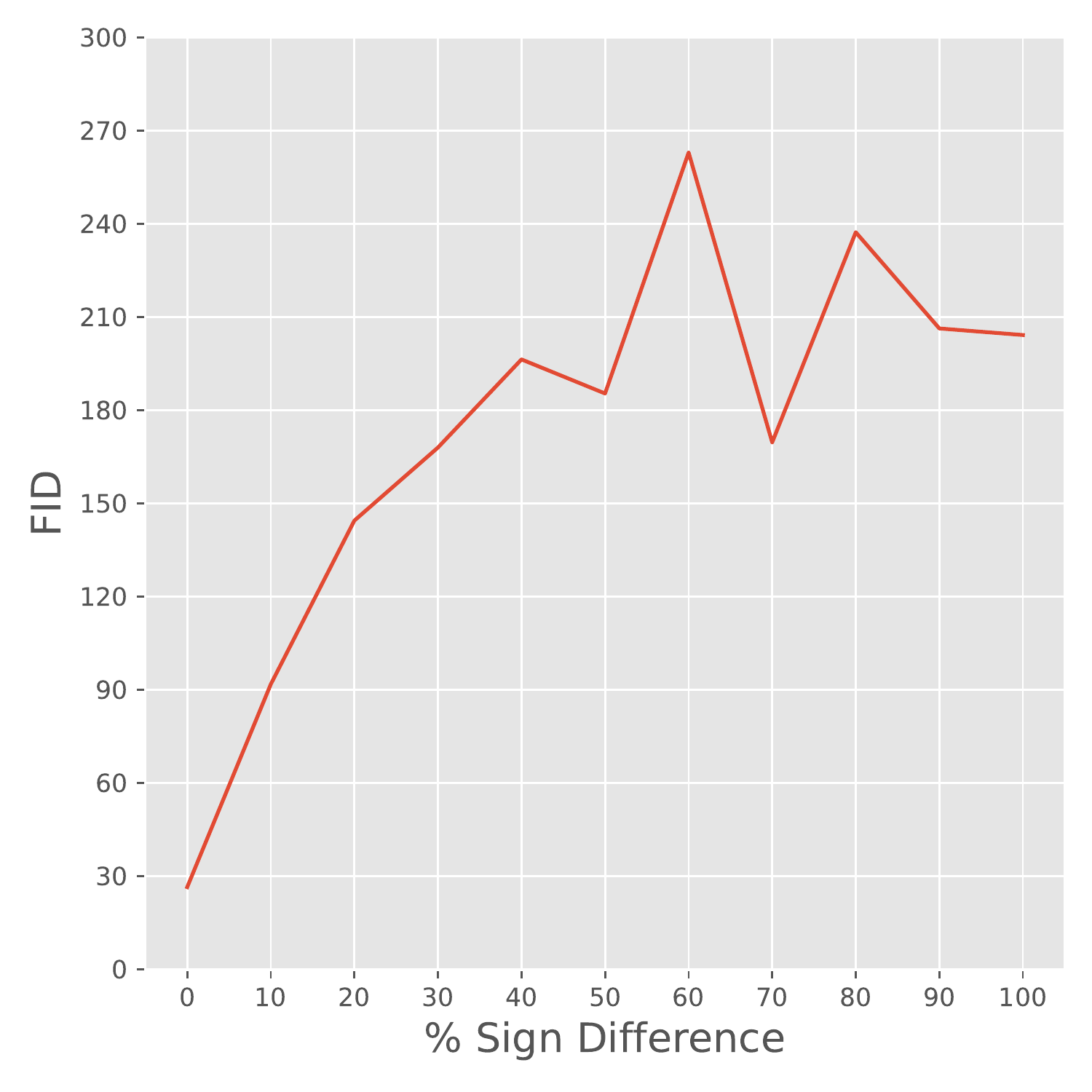}
  \includegraphics[keepaspectratio=true, width=70pt]{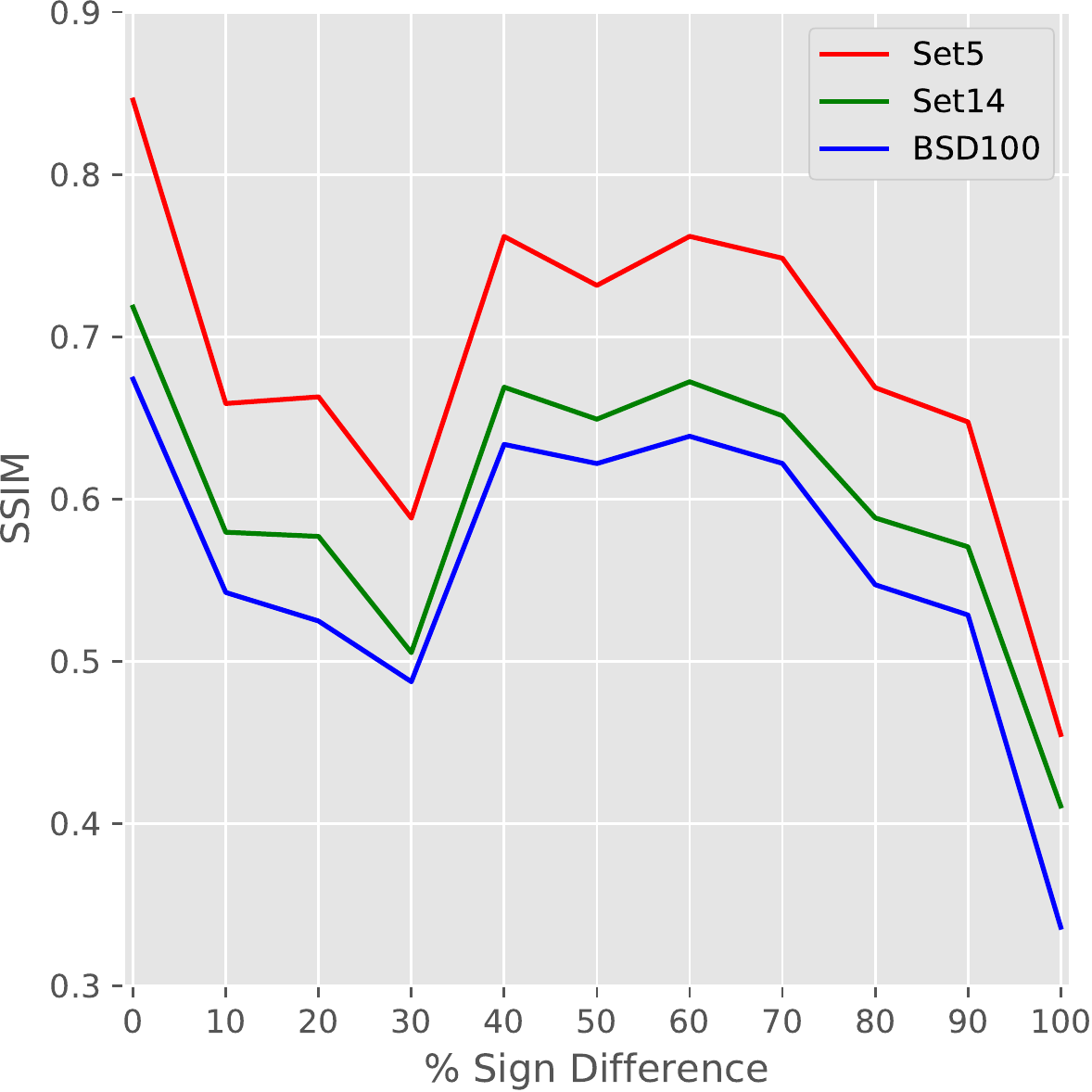}
  \includegraphics[keepaspectratio=true, width=70pt]{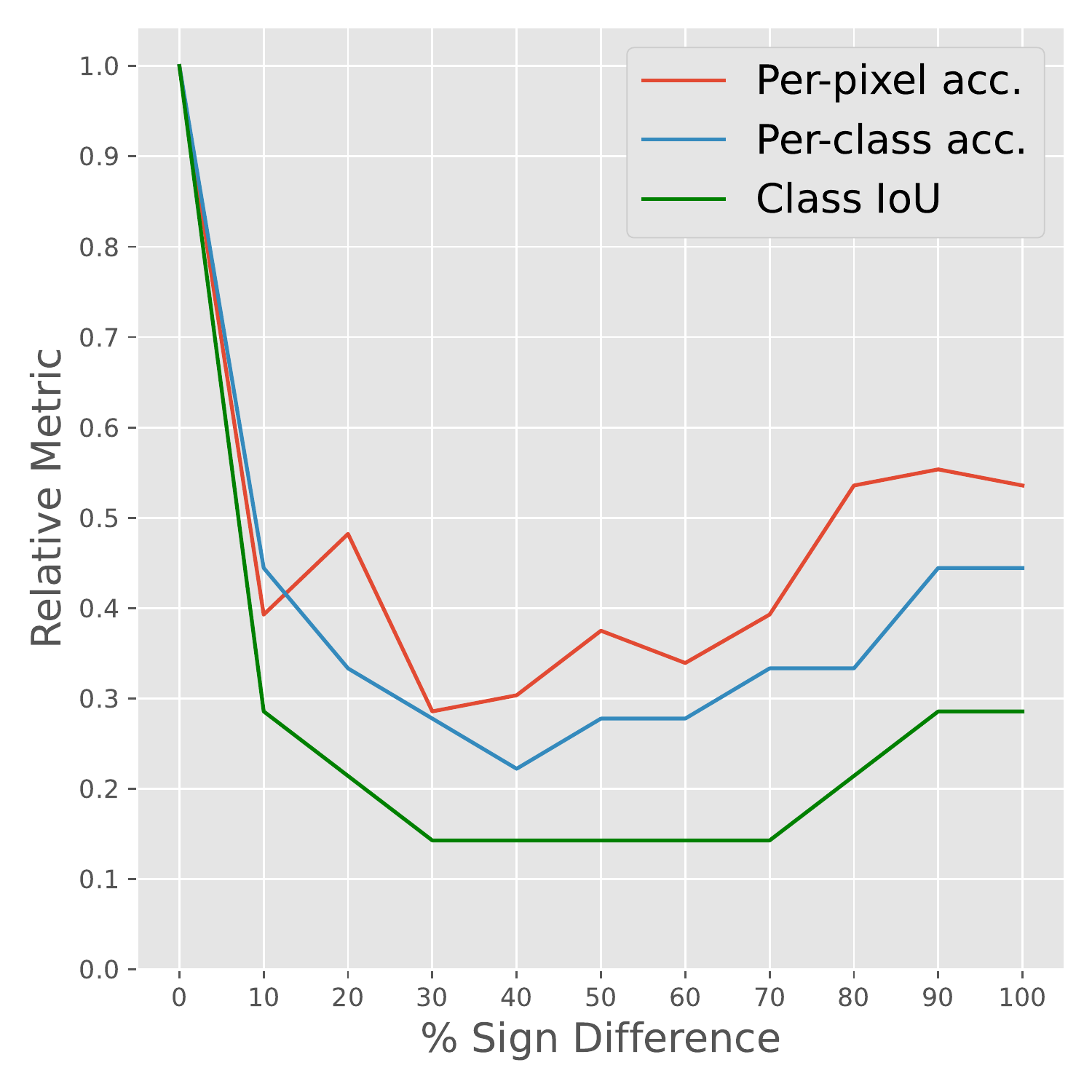}
  \caption{From left to right shows the performance of DCGAN, SRGAN, CycleGAN when different percentage (\%) of the $\text{sign}({\gamma})$ is being modified (compromised).}
  \vspace{-5pt}
  \label{fig:graph-flip-sign}
\end{figure}

\begin{table}[t]
  \centering
  \resizebox{\columnwidth}{!}{%
    \begin{tabular}{lccccc}
      & \textbf{Per-pixel acc.} & \textbf{Per-class acc.} & \textbf{Class IoU} & {$\boldsymbol{Q_{wm}}$} & \textbf{BER} \\
      \hline \hline
      CycleGAN$_{ws}$ & 0.58 & 0.19 & 0.13 & 0.90 & 0 \\
      Fine-tune & 0.55 & 0.18 & 0.14 & 0.85 & 0 \\
      Overwrite & 0.57 & 0.17 & 0.13 & 0.15 & 0 \\
      \hline
    \end{tabular}%
  }
  \caption{First row is the FCN-scores, watermark quality, $Q_{wm}$ and BER for CycleGAN$_{ws}$. Second row shows the scores after fine-tuning and third row shows the scores after overwriting attack.}
  \vspace{-10pt}
  \label{table:cyclegan-robust}
\end{table}


\textbf{Overwriting.} We also simulate the overwriting scenario where the attacker is assumed to embed a new watermark into our trained model using the same method as proposed. Table \ref{table:dcgan-robust}, \ref{table:srgan-robust}, \ref{table:cyclegan-robust} show the results of the attempt. Although we can notice the proposed method is being compromised (\ie~$Q_{wm}$ drops in all 3 GAN models), the performance has also worsened explicitly. However, if we ever met such condition, we can still claim the ownership by further investigate the normalization layers and retrieve the signature embedded into the weights since the signature remains intact in all sort of removal attacks (see next).

%% file: sec/section-3/resilience.tex
\subsection{Resilience against ambiguity attacks}
\label{sec:resilience}

Through Table \ref{table:dcgan-robust}, \ref{table:srgan-robust} and \ref{table:cyclegan-robust}, we can observe that the embedded signature remains persistent even after removal attacks such as fine-tuning and overwriting as the BER remains 0 throughout the experiments. Thus, we can conclude that enforcing the sign in defined polarity using sign loss is rather robust against diverse adversarial attacks.

\begin{figure}[t]
  \centering
  \begin{subfigure}[b]{\linewidth}
    \centering
    \includegraphics[keepaspectratio=true, width=35pt]{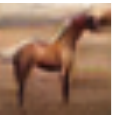}
    \includegraphics[keepaspectratio=true, width=35pt]{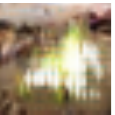}
    \includegraphics[keepaspectratio=true, width=25pt]{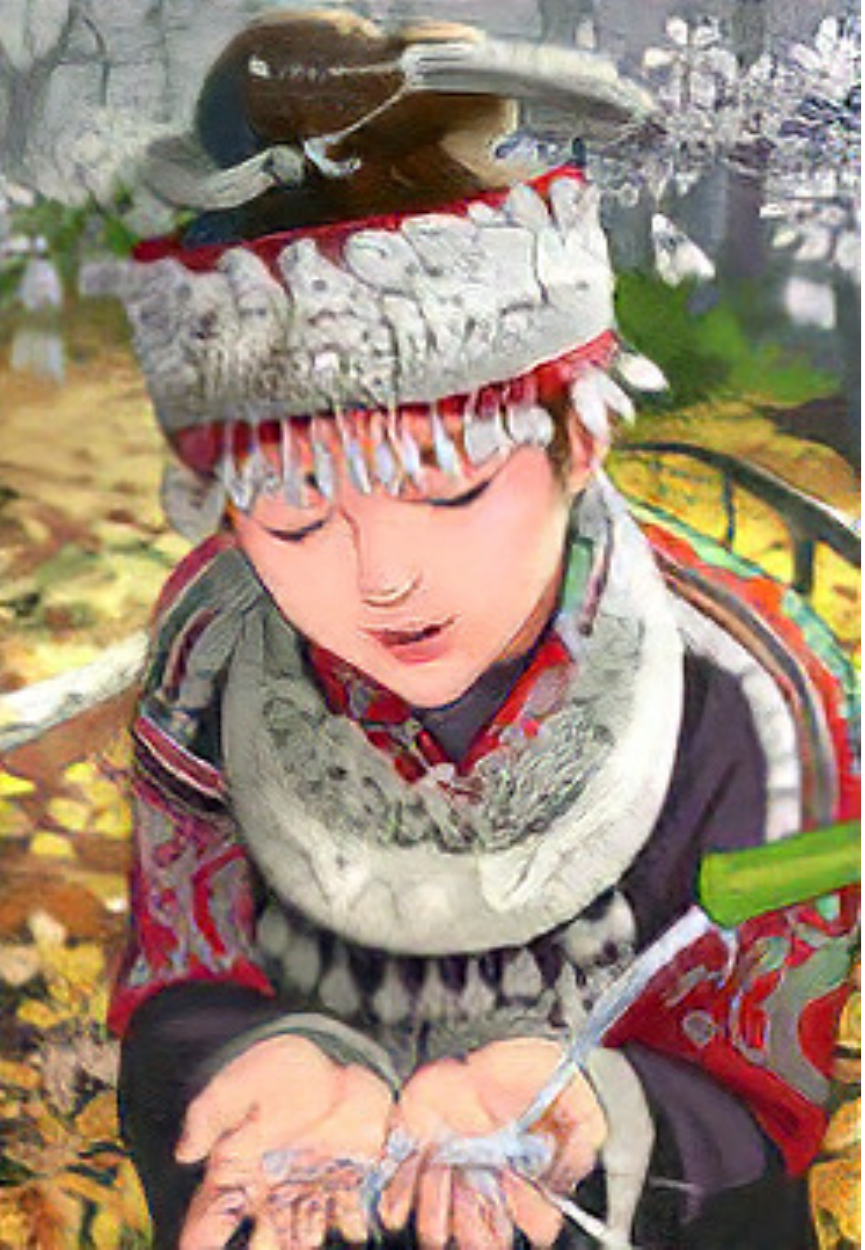}
    \includegraphics[keepaspectratio=true, width=25pt]{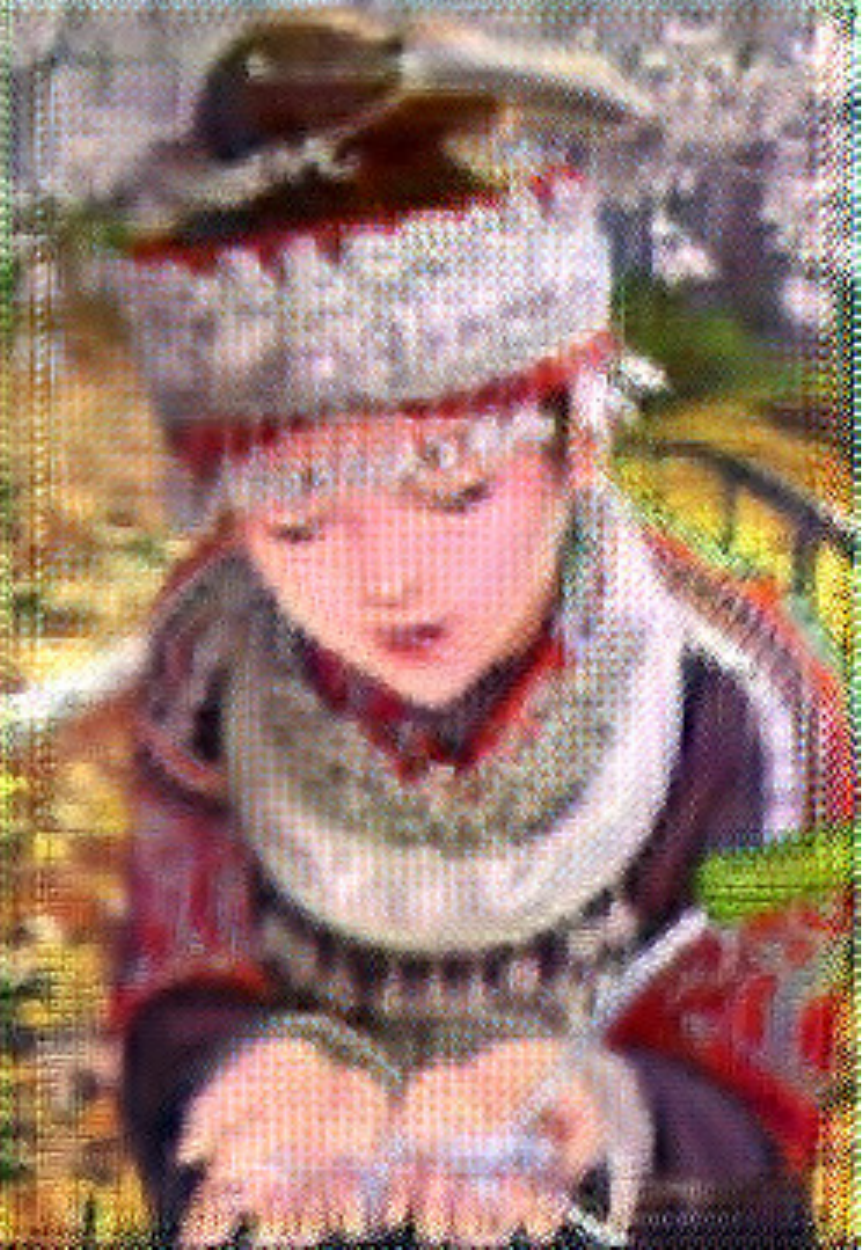}
     \includegraphics[keepaspectratio=true, width=35pt]{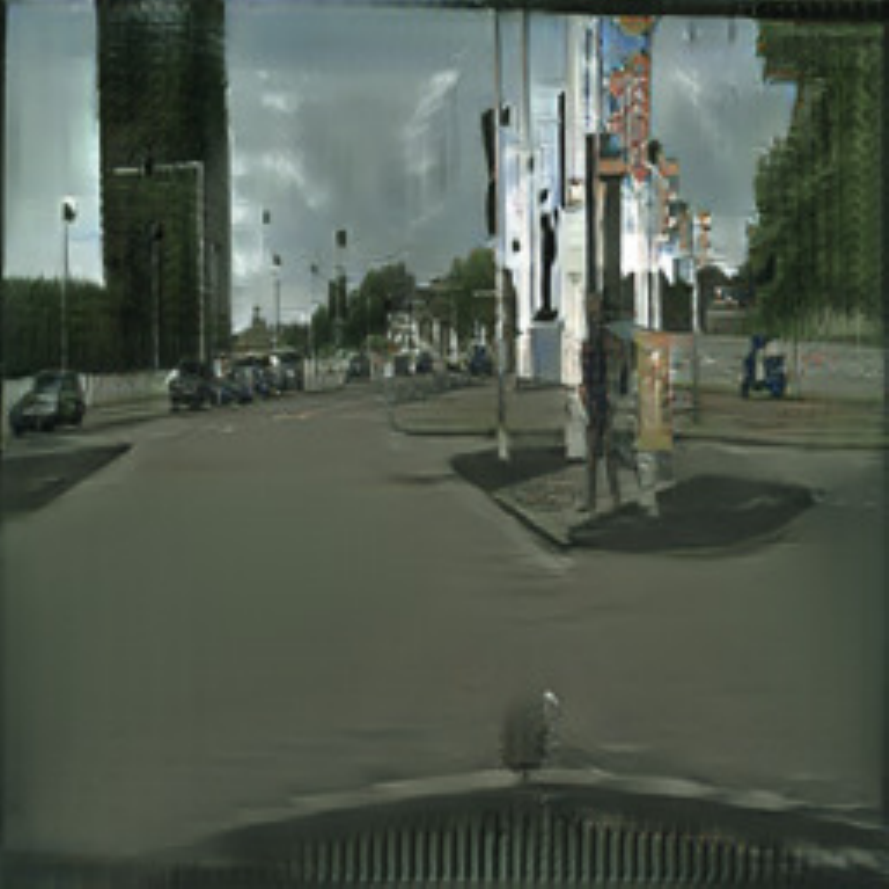}
    \includegraphics[keepaspectratio=true, width=35pt]{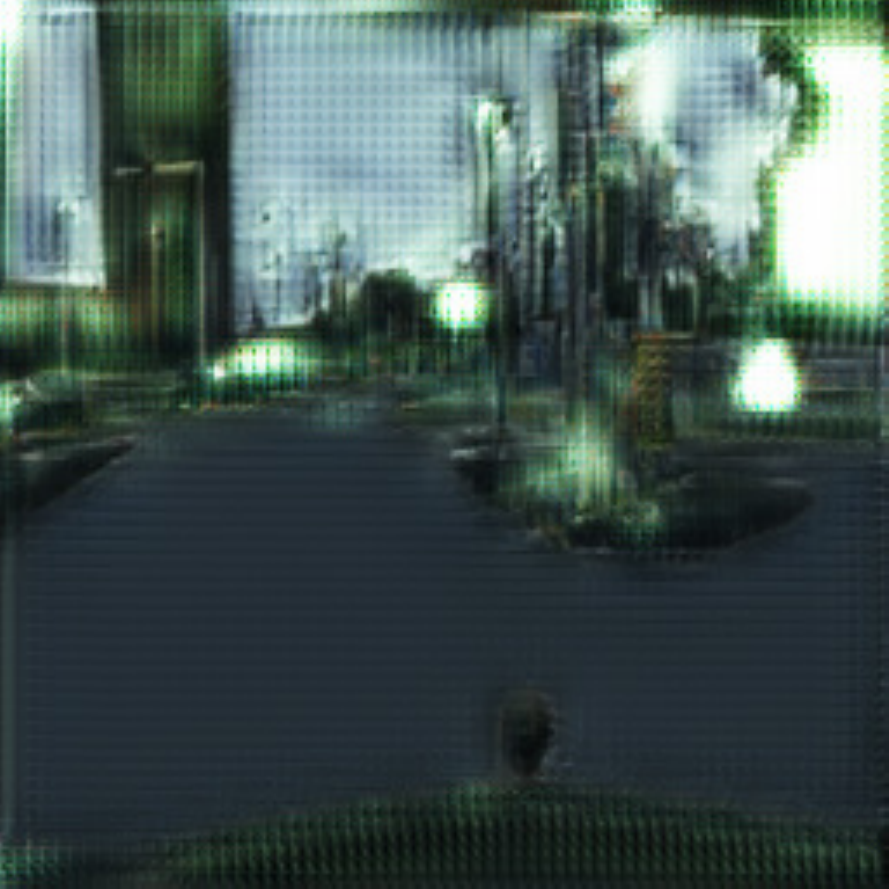}
  \end{subfigure}
  \caption{Image pairs from left to right is GAN$_{ws}$ with 0\% and 10\% of the total signs were being randomly toggle.}
  \vspace{-10pt}
  \label{fig:flip-sign-samples}
\end{figure}

We also simulated a scenario of an insider threat where the watermark and scale signs were exposed completely. As shown in Fig. \ref{fig:graph-flip-sign}, it shows that the FID of DCGAN$_{ws}$ increases drastically (from 26 $\rightarrow$ 91) and SSIM of SRGAN$_{ws}$ drops, despite only 10\% of the signs are modified. Qualitatively, Fig. \ref{fig:flip-sign-samples} clearly shows the quality of the generated images is badly deteriorated when the signs are compromised. This is same for SRGAN$_{ws}$ and CycleGAN$_{ws}$ where the quality of the generated SR-images and (label $\rightarrow$ photo) images are very poor in quality where obvious artefact is observed even the signature signs are modified at only 10\%.

In summary, we can deduce that the signs enforced in this way remain rather persistent against ambiguity attacks and offenders will not be able to employ new (modified) scale signs without compromising the GANs performance.

%% file: sec/section-3/ablation-study.tex
\subsection{Ablation Study}
\label{sec:ablation-study}

\subsubsection{Coefficient $\lambda$.}
The coefficient, $\lambda$ is multiplied to the reconstructive regularizing term, $\wm{\mathcal{L}}$ to balance between the original objective and the quality of generated watermark. We perform an ablation study and from Table \ref{table:ablation-lambda}, we show that when $\lambda$ is low (0.1), the FID score is at the lowest, meaning the GAN model has a very good performance in the original task. Oppositely, when $\lambda$ is set to very high (10.0), the quality of the watermark, $Q_{wm}$ is at the best, but the FID score is the lowest along the spectrum. However, qualitatively, it is hardly to visualize this. As a summary, there is a tradeoffs between GAN model performance and the watermarking quality. We find that $\lambda = 1.0$ is reasonable as the quality of watermark is relatively good without hurting the performance of the original task too much.

\begin{table}[t]
  \centering
  \resizebox{0.65\linewidth}{!}{
  \begin{tabular}{cccccc}
    $\lambda$ & 0.1 & 0.5 & 1.0 & 5.0 & 10.0 \\
    \hline \hline
    FID & \textbf{25.88} & 26.57 & 28.19 & 32.46 & 47.38 \\
    $Q_{wm}$ & 0.926 & 0.956 & 0.965 & 0.979 & \textbf{0.982} \\ \hline
  \end{tabular}} \\
  \vspace{+5pt}
  \includegraphics[keepaspectratio=true, width=30pt]{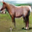}
  \includegraphics[keepaspectratio=true, width=30pt]{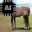}
  \includegraphics[keepaspectratio=true, width=30pt]{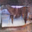}
  \includegraphics[keepaspectratio=true, width=30pt]{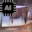}
  \includegraphics[keepaspectratio=true, width=30pt]{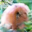}
  \includegraphics[keepaspectratio=true, width=30pt]{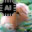}
  \caption{$\lambda$ vs. GAN performance measured in FID and quality of the generated watermark measured in SSIM. Image pairs from left to right is $\lambda$=0.1; $\lambda$=1.0 and $\lambda$=10.}
  \vspace{-5pt}
  \label{table:ablation-lambda}
\end{table}

\subsubsection{$n$ vs. $c$.}
This experiment investigates the effects of different $n$ and $c$ settings to the original DCGAN performance (measured in FID) and the quality of the generated watermark (measured in SSIM). We conclude that setting $n=5$ and $c=-10$ perform the best (See Table \ref{table:ablation-nc}) in terms of quality of both generated image and watermark, however, the choice can be vary depends on the situation. Notice that it performs the worst when setting $c=0$ and the performance increases when the magnitude of $c$ increases, moving away from 0. This effect explains the reason why the {\it trigger} input set must have a very different distribution from the training data. For DCGAN, the training input has a normal distribution of $\mu = 0$, and setting $c$ to 0 will not change the distribution, thus causing confusion between the normal input and {\it trigger} input.

\begin{table}[t]
  \centering
  \resizebox{0.65\linewidth}{!}{
  \begin{tabular}{l|cccccc}
    \backslashbox{$n$}{$c$}
    & -10 & -5 & 0 & +5 & +10 \\ \hline
    \\ \multirow{2}{*}{ 5} & 26.05 & 26.37 & 276.36 & 25.98 & 26.19
    \\  & 0.961 & 0.960 & 0.367 & 0.953 & 0.958
    \\ \multirow{2}{*}{10} & 28.35 & 27.49 & 332.88 & 26.18 & 27.11
    \\  & 0.958 & 0.953 & 0.338 & 0.956 & 0.956
    \\ \multirow{2}{*}{15} & 25.51 & 26.85 & 316.78 & 27.27 & 26.24
    \\  & 0.954 & 0.945 & 0.343 & 0.951 & 0.953
  \end{tabular}}
  \caption{Effect of $n$ and $c$ to model's performance in terms of FID (above) and the quality of generated watermark measured in SSIM (below).}
  \label{table:ablation-nc}
  \vspace{-10pt}
\end{table}

%

%% file: sec/discussion-and-conslusion.tex
\section{Discussion and Conclusion}

This paper illustrates a complete and robust ownership verification scheme for GANs in  black-box and white-box settings.  While extensive experiment results are conducted for three representative variants \ie~DCGAN, SRGAN and CycleGAN, the formulation lay down is generic and can be applied to any GAN variants with generator networks as the essential component.  Empirical results showed that the proposed method is robust against removal and ambiguity attacks, which aim to either remove existing watermarks or embed counterfeit watermarks. It was also shown that the performance of the model's original tasks (\ie image generation, super-resolution and style transfer) were not compromised. The importance of this work, in our view, can be highlighted by numerous disputes over IP infringements between giant and/or startup companies, which are now heavily investing substantial resources on developing new DNN models. It is our wish that the ownership verification for GANs will provide technical solutions in discouraging plagiarism and, hence, reducing wasteful lawsuit cases.
